\documentclass[%
 aip,
 fleqn,
 pof,
 amsmath,amssymb,
preprint,
]{revtex4-2}
\usepackage{graphicx}
\usepackage{dcolumn}
\usepackage{bm}
\usepackage[retainorgcmds]{IEEEtrantools}
\usepackage[utf8]{inputenc}
\usepackage[T1]{fontenc}
\usepackage{mathptmx}
\usepackage{etoolbox}
\usepackage{diffcoeff} 
\usepackage{amsmath}
\usepackage{float}
\usepackage{amssymb}
\usepackage{caption}
\usepackage{subcaption}
\usepackage{latexsym}
\usepackage{times}
\usepackage[pagewise]{lineno}
\usepackage{hyperref}
\usepackage{siunitx}
\DeclareSIUnit\atm{atm}
\usepackage[version=4]{mhchem}

\usepackage[normalem]{ulem}
\usepackage[xcolor]{changebar}

\definecolor{revOne}{HTML}{cc0000}
\definecolor{revTwo}{HTML}{cc9900}
\definecolor{revThree}{HTML}{238E23}
\definecolor{editor}{HTML}{0057fa}

\newcommand{\revise}[2]{\textcolor{#2}{#1}}  
\newcommand{\former}[2]{\textcolor{#2}{\sout{#1}}}
\renewcommand{\revise}[2]{{#1}}
\renewcommand{\former}[2]{}

\makeatletter
\def\@email#1#2{%
 \endgroup
 \patchcmd{\titleblock@produce}
  {\frontmatter@RRAPformat}
  {\frontmatter@RRAPformat{\produce@RRAP{*#1\href{mailto:#2}{#2}}}\frontmatter@RRAPformat}
  {}{}
}%
\makeatother
\begin{document}


\title[]{A single-domain approach for modeling flow in and around porous media applied to buoyant reacting plume formation and ignition}
\author{D.~Behnoudfar}
\author{K.~E.~Niemeyer}%
 \email{kyle.niemeyer@oregonstate.edu}
\affiliation{ 
School of Mechanical, Industrial, and Manufacturing Engineering, Oregon State University, Corvallis, Oregon, United States of America
}%


\begin{abstract}
Many processes involve mixed porous-solid fluid domains where fluid flow, heat transfer, and chemical reactions interact over disparate length scales, such as the combustion of multi-species solid fuels. 
Although many studies have concentrated on detailed physics within the fluid or porous phase, few consider both phases, in part due to the challenge in determining suitable boundary conditions between the regions, particularly in turbulent flows where eddies might penetrate the pores. 
Here, we apply a single-domain approach that eliminates the need for boundary conditions at the interface, and simulate scenarios involving porous solids and a surrounding fluid. Similar to large eddy simulation, the method averages properties over a small spatial volume---but over the entire domain. 
We focus on ignition and related interfacial phenomena. After verifying and validating the model, we examine the emission of buoyant reacting plumes from the surface of a heated solid and the near-field flow dynamics. 
The results indicate flow instabilities similar to Rayleigh--Taylor and Kelvin--Helmholtz phenomena. 
A combination of viscous and baroclinic torques triggers vorticity generation near the interface and its growth in the surrounding fluid region. 
Furthermore, we explore the effect of interface morphology, finding that geometrical characteristics such as asymmetry or gap size can alter ignition time and location, or even suppress it. 
Asymmetry-induced oscillations initially cause negative heat fluxes, which prevent the temperature from reaching the critical level necessary to trigger ignition. 
These behaviors could significantly influence the mixing of oxidizer and fuel, ignition processes, and fire propagation.
\end{abstract}

\maketitle
 
%

\section{Introduction\label{sec:introduction}}

Flow in mixed porous solid--fluid domains, possibly involving heat transfer and chemical reactions, arises in a variety of systems in engineering and nature. 
This work is particularly motivated by the physics of natural fires. 
The increase in the extent of areas globally impacted by wildfires has become a major environmental and societal concern in recent years. 
Similar interfacial phenomena occur in electrochemical devices, microbial activity at river bed sediments, flow in biological tissues, and micro-devices for rapid heat transfer, among others. 

The research community has access to established detailed models for simulating turbulent reactive flows \cite{savard2015, motheau2016, Fillo2020, desai2021} (e.g.) and semi-detailed models focusing solely on porous domains \cite{cfd, lautenberger2009}. 
However, problems of interest such as forest fires or solid-fuel combustion happen across both regions with strong coupling between the processes. 
From a fluid-dynamics perspective, the flow at the interface between the two regions can exhibit a variety of behaviors, depending on the flow regime and other conditions. 
Direct simulations that resolve flow inside the pores are often unfeasible since most porous media have complex geometries, for which the details are not fully known.
\revise{If such details are known, one modeling strategy couples the discrete element method with computational fluid dynamics to represent reacting flow over granular assemblies of particles~\cite{Mahiques2023}.}{revOne}
The other approach to the problem is using parameterized boundary conditions, incorporating the effect of wall permeability. 
For example, the method proposed by Beavers and Joseph \cite{beavers1967} for laminar flow over a permeable wall allows for a slip velocity parallel to the wall, with a slip coefficient that has to be specified, whereas the wall-normal velocity is set to zero; others extended this method for application in turbulent flow cases (e.g., Hahn et al.~\cite{hahn2002}). This type of boundary condition is based on the Stokes flow assumption close to the permeable wall \cite{saffman1971} and therefore can only be applied to turbulent flow cases when the ratio of the effective pore diameter to the length scale of the near-wall turbulent eddies is small \cite{hahn2002}. For larger values of this ratio, the eddies might penetrate the permeable wall, violating the Stokes flow assumption. Therefore, the penetration of turbulence in this case cannot be modeled simply using boundary conditions.

Single-domain formulations offer a way to avoid these issues. Early attempts
for such a methodology were based on the Darcy--Brinkman equation, after Brinkman's empirical correction to Darcy's law \cite{brinkman1949calculation}. The single-domain formulations can be mathematically derived by spatial averaging the transport equations. This method is similar to the approach used in large eddy simulation (LES) \cite{lesieur1996} where the flow is filtered over a small spatial volume. 
Applying this procedure to the porous region smooths inhomogeneities so that the porous medium can be considered as a continuum \cite{whitaker1998}. 
One advantage is that no boundary conditions are required for coupling the processes in the two regions; however, permeability, sub-filter-scale drag, and sub-filter-scale stress need to be modeled. 
Also, the model should allow for variable porosity (and its gradual increase) at the interface region since the averaging filter no longer fully falls in the homogeneous porous layer \cite{whitaker1998}. A few studies have applied these formulations for non-reacting turbulent flow problems \cite{bae2011effect, sadowski2023large} but they have not gone beyond deriving a volume-averaged momentum equation.

The ignition of solid fuel and the issuing of pyrolysis products into the surrounding gas has rarely been studied in detail from a modeling perspective. Prior numerical studies mostly assumed simplified boundary conditions for coupling gas-solid processes \cite{zhou2002}, which is insufficient for capturing the flow behavior in complicated situations. 
In the case of buoyant jets, instabilities can emerge within the flow that grow over time, impacting the mixing of gas species and flame structure \cite{jiang2000, meehan2022}. Such works have mostly focused on non-reacting buoyant instabilities where complexities
associated with chemical reactions and surface topography are absent. Flow in the vicinity of surfaces with roughness elements has been of interest in several fields, particularly aerospace \cite{bres2008three}, environmental science \cite{banerjee2020effects}, and meteorology \cite{li2024structure}. For example, plant and urban canopies (vegetation like forests and agricultural fields or cities with buildings and other structures) significantly influence turbulence and the resulting transport of momentum, heat, water vapor, and pollutants. These interactions can also affect the mixing of air and the intensity and spread of a wildfire.

In this work, we explore potential reacting instabilities and near-field flow dynamics at the interface of the porous solid and surrounding fluid regions using the above-mentioned detailed approach. The study also addresses the effect of surface morphology on ignition and attempts to answer how different configurations influence the location and time of ignition.
We first present the single-domain framework, including the LES equations extended to energy and species conservation, for modeling mixed porous-fluid domains in turbulent and non-turbulent flow regimes. We then verify this model using a non-reacting channel flow with a permeable wall and validate using a reactive-flow experiment of ignition of a wooden sphere in a tubular reactor.
Finally, we investigate phenomena up to the point of ignition in a buoyancy-driven flow above a porous fuel bed, examining the sources of flow instabilities.

\section{Methods\label{sec:methods}}
We consider a domain composed of a fluid-saturated porous region and the surrounding fluid, governed by a single set of transport equations. 
The model is partly based on the work of \.Zuk et al.~\cite{zuk2022} and relies on a mixture-theoretic framework in which both the solid matrix and the fluid phase coexist as immiscible thermodynamic entities \cite{kaviany2012}. 
Each control volume in the domain is characterized by its porosity ($\phi$), defined as the local volume fraction of fluid. 

\subsection{Governing equations} 
For the fluid phase, the model solves the volume-averaged conservation of mass, species, momentum, and thermal energy (in the entire domain) described with

\begin{align}
\diffp{}{t} \left(\phi \langle \rho \rangle^f\right) + \nabla \cdot \left(\phi \langle \rho \rangle^f \langle \textbf{u} \rangle\right) &= (1-\phi)\sum_{j=1}^{N_g} \langle \dot \omega^s_j \rangle^s \;, \label{eq:1} \\
\diffp{}{t} \left(\phi \langle \rho \rangle^f \langle Y_j \rangle \right) + \nabla\cdot \left(\phi \langle \rho \rangle^f \langle Y_j \rangle \langle \textbf{u} \rangle\right) &= \nabla\cdot(\phi \langle \rho \rangle^f D \nabla \langle Y_j \rangle) \nonumber\\
&+\> \nabla\cdot\left(\phi \langle \rho \rangle^f \langle \hat{Y_j}  \hat{\textbf{u}} \rangle \right) + \phi \langle \dot\omega_{j} \rangle^f + \left(1-\phi\right) \langle \dot\omega_{j}^s \rangle^s \;,  \label{eq:2} \\
\diffp{}{t} \left(\phi \langle \rho \rangle^f \langle \textbf{u} \rangle\right) + \nabla\cdot\left(\phi \langle \rho \rangle^f \langle \textbf{u} \rangle \langle \textbf{u} \rangle \right) + \nabla \langle p \rangle &= \nabla\cdot \langle \Bar{\Bar{\mathbb{\tau}}}\rangle - \nabla\cdot\left(\phi \langle \rho \rangle^f \langle \hat{\textbf{u}}  \hat{\textbf{u}} \rangle \right) +\> \phi \langle \rho \rangle^f \textbf{g} + \textbf{f} \;, \label{eq:3} \\
\diffp{}{t} \left(\phi \langle \rho \rangle^f \langle h \rangle \right) + \nabla\cdot(\phi \langle \rho \rangle^f \langle h \rangle \langle \textbf{u} \rangle) &= \nabla\cdot( \lambda \nabla \langle T \rangle) + \dot Q \nonumber\\
&-\> \nabla\cdot(\phi \langle \rho \rangle^f \langle \hat{h}  \hat{\textbf{u}} \rangle) -\> h_{fs} A_{fs} \left(\langle T \rangle - \langle T^s \rangle\right) + \diffp{\langle p \rangle}{t} \nonumber \\
&+\> (1-\phi) \sum_{j=1}^{N_g} \langle h_j \rangle \langle \dot\omega_{j}^s \rangle^s + \langle S^{f,\> \text{radiation}} \rangle\;, \label{eq:4} 
\end{align}
where $\textbf{u}$ is the volume fluid velocity vector, $T$ is fluid temperature, $T^s$ is solid temperature, $Y_j$ is mass fraction of gas species $j$, $\dot\omega_{j}$ is volumetric mass change rate of gas species $j$ resulting from homogeneous reactions, $\dot\omega_{j}^s$ is volumetric mass change rate of gas species $j$ resulting from heterogeneous reactions, $p$ is fluid pressure, 
$\rho$ is density, $D$ is fluid diffusion coefficient,
$h$ is fluid's mixture-averaged enthalpy, $h_j$ is the enthalpy of gas species $j$, $\lambda$ is fluid thermal conductivity, $\dot Q = -\phi \sum_{j=1}^{N_g}\langle \dot\omega_{j} \rangle^f h_{f,j}$ is heat release rate due to reactions, $N_g$ is the number of gas species, $h_{f,j}$ is the heat of formation of individual gas species $j$, $\textbf{g}$ is gravitational acceleration, and $S^{f,\> \text{radiation}}$ is the heat exchanged through radiation for the gas phase, evaluated using a participating media approach (Appendix~\ref{radiation} provides more details). 

The symbols $\langle \cdot \rangle$, $\langle \cdot \rangle^f$ and $\langle \cdot \rangle^s$ denote volume averaging operators defined as $\langle \cdot \rangle = 1/\Delta V\int_{\Delta V} (\cdot)\,dV$. For $\langle \cdot \rangle^f$ and $\langle \cdot \rangle^s$, the total averaging volume ($\Delta V$) is replaced with $\Delta V^f$ or $\Delta V^s$, which are the volumes of the fluid and solid, respectively.
$\hat{\textbf{u}}$, $\hat{h}$, and $\hat{Y_j}$ are the fluctuating components of velocity, enthalpy and mass fraction, respectively, due to averaging defined as  $\hat{\textbf{u}}= \textbf{u} - \langle\textbf{u}\rangle$, $\hat{h}= h - \langle h \rangle$ and  $\hat{Y_j}= Y_j - \langle Y_j \rangle$. 

$\Bar{\Bar{\tau}}$ is the stress tensor following Stokes hypothesis:
\begin{equation}
    \langle \Bar{\Bar{\tau}}\rangle = \mu^*\left(\nabla \langle\textbf{u}\rangle+{\nabla\langle\textbf{u}\rangle}^T\right)-\frac{2}{3}\left(\nabla \cdot \langle\textbf{u}\rangle\right)\Bar{\Bar{I}}\;,
\end{equation}
where $\mu^*$ is the effective viscosity \cite{le2006}; in this work, we assume $\mu^* = \mu$  for simplicity. For highly porous materials ($\phi > 0.7$), the $\mu^*/\mu$  ratio is not significantly greater than one, especially in the region close to the porous layer--fluid interface \cite{kaviany2012}.

$\textbf{f}$ represents the drag force that the solid exerts on the fluid phase:
\begin{equation}
    \textbf{f}=\frac{1}{\Delta V}\int_{\Delta A}(p+\Bar{\Bar{\tau}})\cdot\textbf{n}\,dA \;,
\end{equation}
where $\Delta A$ is the interfacial area between the fluid and solid phases in $\Delta V$ and $\textbf{n}$ is the normal vector. The closure for this term is discussed in the next section.

Darcy's law or the Navier--Stokes equations are retrieved from Eq.~\eqref{eq:3} by setting the porosity to 0 or 1, respectively. 
The spatially averaged energy equation for the porous medium is modeled using the two-medium treatment allowing for a thermal
non-equilibrium between the phases \cite{kaviany2012}. 
The term $h_{fs} A_{fs} (T - T^s)$ approximates spatially averaged convective heat transfer between the fluid and solid inside the porous region, where $h_{fs}$ is the heat transfer coefficient and $A_{fs}$ is the interfacial surface area between fluid and solid. 
Using the unity Schmidt and Lewis number assumptions, diffusion coefficient and thermal conductivity are approximated from Sutherland's transport model for viscosity evaluation \cite{sutherland1893}. 

For the solid phase, the model solves a separate set of conservation equations, as described below:
\begin{align}
\diffp{}{t} \left((1-\phi) \langle \rho \rangle^s\right) &= (1-\phi)\sum_{i=1}^{N_s} \langle \dot \omega^s_i \rangle^s \;, \label{eq:6} \\
\diffp{}{t} \left((1-\phi) \langle \rho \rangle^s \langle Y_i \rangle \right) &= (1-\phi)\sum_{i=1}^{N_s} \langle \dot \omega^s_i \rangle^s \;, \label{eq:7} \\
\diffp{}{t} \left((1-\phi) \langle \rho \rangle^s \langle h^s \rangle \right) &= \nabla\cdot( \lambda^s \Bar{\Bar{A}}\nabla \langle T^s \rangle) 
-\> (1-\phi) \sum_{i=1}^{N_s} h_{f,i} \langle \dot\omega_{i}^s \rangle^s 
+\> h_{fs} A_{fs} \left(\langle T \rangle - \langle T^s \rangle\right) \nonumber\\
&-\> (1-\phi) \sum_{j=1}^{N_g} \langle h_j \rangle \langle \dot\omega_{j}^s \rangle^s + \langle S^{s,\> \text{radiation}} \rangle
 \;, \label{eq:solid_energy}
\end{align}
where $N_s$ is the number of solid species, $\dot\omega_{i}^s$ is volumetric mass rate of change for solid species $i$ resulting from heterogeneous reactions, $Y_i$ is the mass fraction of solid species $i$, $h^s$ is the solid's mixture-averaged enthalpy, $\lambda^s$ is the solid's mixture-averaged thermal conductivity, $\Bar{\Bar{A}}$ is the anisotropy tensor of the solid matrix, and $S^{s,\> \text{radiation}}$ is the heat exchanged through radiation for the solid phase (see Appendix~\ref{radiation}). 

\subsection{LES model for mixed porous-pure fluid flow\label{subsec:les}}

The LES equations for a mixed porous-fluid medium are obtained by applying a low-pass spatial filter on the general transport equations. A common filter (equivalent in its effect to the finite volume discretization) is the top-hat filter, which physically means the volume average of a physical quantity in a grid cell.
The volume-averaged Navier--Stokes equations for a Newtonian fluid flowing in a stationary matrix~\cite{le2006} were presented as Eq.~\eqref{eq:3}. The integral term ($\textbf{f}$) in this equation corresponds to the microscopic momentum exchange of the fluid with the solid matrix and is commonly modeled using the correlation form $\textbf{f} = - \mu \Bar{\Bar{K}}^{-1}[\Bar{\Bar{I}}+\Bar{\Bar{F}} (Re_p)]\cdot \langle\textbf{u}\rangle$, where $\Bar{\Bar{K}}$ is Darcy permeability tensor (which is a function of porosity) and $\Bar{\Bar{F}}$ is the Forchheimer correction tensor \cite{breugem2006influence, wood2020}. 
The other non-closed term, $\nabla\cdot\left(\phi \langle \rho \rangle^f \langle \hat{\textbf{u}}  \hat{\textbf{u}} \rangle \right)$, appears because velocity fluctuates inside the control volume and thus differs from its averaged value. 
The quantity $\Bar{\Bar{\tau}}_{SGS}=-\left(\phi \langle \rho \rangle^f \langle \hat{\textbf{u}}  \hat{\textbf{u}} \rangle \right)$  is a part of the sub-filter scale stress and, according to LES techniques, is commonly modeled as 
$-\left(\phi \langle \rho \rangle^f \langle \hat{\textbf{u}}  \hat{\textbf{u}} \rangle \right)= \mu_t\left(\nabla\langle\textbf{u}\rangle + \left(\nabla\langle\textbf{u}\rangle\right)^T\right)$  where $\mu_t$ is turbulence eddy viscosity. After applying the mentioned models and approximations, Eq.~\eqref{eq:3} becomes
\begin{eqnarray}
\diffp{}{t} \left(\phi \langle \rho \rangle^f \langle \textbf{u} \rangle\right) &+& \nabla\cdot\left(\phi \langle \rho \rangle^f \langle \textbf{u} \rangle \langle \textbf{u} \rangle \right) + \nabla \langle p \rangle = \nonumber\\
&& [\mu +\mu_t]\left(\nabla\langle\textbf{u}\rangle + (\nabla\langle\textbf{u}\rangle)^T\right) +\> \phi \langle \rho \rangle^f \textbf{g} - \mu \Bar{\Bar{K}}^{-1}[\Bar{\Bar{I}}+\Bar{\Bar{F}} (Re_p)]\cdot \langle\textbf{u}\rangle  \;. \label{eq:6} 
\end{eqnarray}
In a laminar flow regime, $\mu_t$  and $\Bar{\Bar{F}}$  can be neglected in the above equation.  
Here, we use the one-$k$-equation-eddy model \cite{yoshizawa1986} to parameterize  $\mu_t$:
\begin{equation}
\mu_t = C_s k^{1/2} \Delta \;,
\end{equation}
where $C_s$ is the model constant, $k$ is sub-grid kinetic energy obtained by solving its transport equation as described by Yoshizawa~\cite{yoshizawa1986}, and $\Delta$ is a sub-grid length scale, taken to be the cubic root of the computational cell volume: $\Delta = \left(\Delta x \Delta y \Delta z \right)^{1/3}$.

Similarly, the turbulent heat flux on the sub-grid scale is modeled as $-\left (\phi \langle \rho \rangle^f \langle \hat{h}  \hat{\textbf{u}} \rangle\right) = \lambda_t \nabla \langle T \rangle$ 
where $\lambda_t$ is the heat transfer eddy conductivity and is estimated based on the Reynolds analogy, according to which $\lambda_t = \mu_t c_p / Pr_t$, where $c_p$ is the fluid heat capacity and $Pr_t$ is the turbulent Prandtl number. 

\subsection{Numerical Methods\label{subsec:numerical}}

To discretize the advective terms in the momentum equation, we use a blend of central differencing (75\%) and linear upwind (25\%) schemes to interpolate the velocity at cell faces. 
For the advective terms in the energy and species conservation equations, we use a central differencing scheme that limits towards upwind in regions of rapidly changing gradient according to the procedures in the total variation diminishing (TVD) schemes \cite{harten1997high}.
The diffusive terms are discretized using standard central differencing along with linear interpolation for the diffusivity.
The solver in this work is based on the \texttt{reactingFOAM} solver of OpenFOAM~\cite{openfoam} that uses a sequential splitting method in which the reactive terms are integrated separately. 
We use an extrapolation-based stiff ordinary differential equation (ODE) solver (SEULEX)~\cite{hairer1991ii} to integrate the solid-phase reactive terms and either a L-stable, stiff embedded Rosenbrock ODE solver of order (2)3~\cite{sandu1997benchmarking} or a backward Euler scheme to integrate the gas-phase reactive terms. 
The solution algorithm starts by solving the governing equations for the solid phase; next, it enters the PISO-SIMPLE loop to solve the coupled fluid-phase conservation equations, as described by \.{Z}uk et al.~\cite{zuk2022}.

\section{Results and discussion\label{sec:results}} 

In this section, we first describe the outcomes of the model applied to two scenarios: a non-reacting channel flow with a permeable wall, for which direct numerical simulation data are available, and a reactive flow experiment involving the ignition of a wooden sphere in a tubular reactor. 
We then explore the phenomena occurring up to the ignition point in a buoyancy-driven flow within a system comprising a porous fuel bed in unconfined air.

\begin{figure}[h!]
\centering
\includegraphics[width=0.5\textwidth]{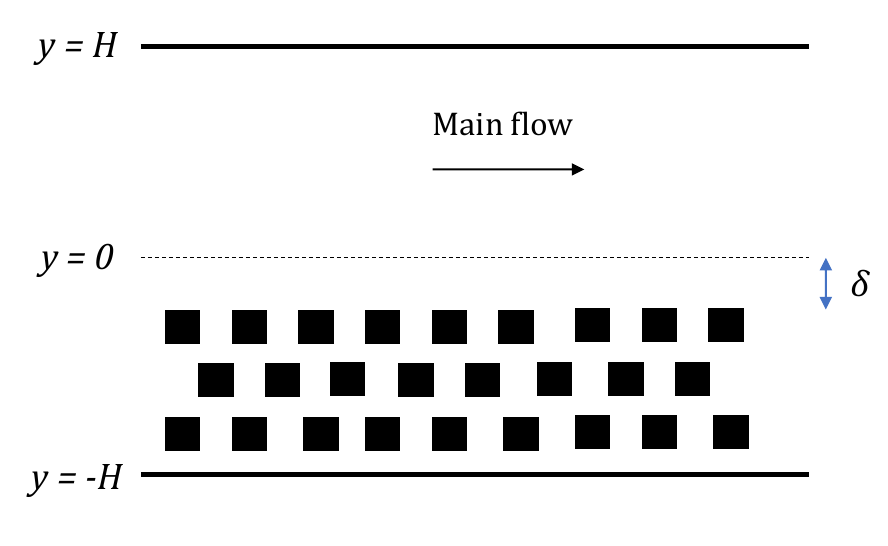}
\caption{\footnotesize Schematic of the domain for flow simulations in a channel with a permeable wall.}
\label{fig:channel}
\end{figure}

\subsection{Flow in a channel with a permeable wall\label{subsec:channel}} 

For these simulations, the setup consists of a two-dimensional plane channel with a permeable bottom wall (Figure~\ref{fig:channel}). 
The domain of size \SI{3}{\centi\meter}$\times$\SI{2}{\centi\meter} is discretized using a \revise{uniform}{revOne} rectangular grid of 300$\times$200, \revise{resulting in a grid spacing of $0.01H$}{revOne}. 
\revise{The grid uniformity is necessary for avoiding the commutation error involved in LES filtering}{revOne}.
The flow domain is divided into three regions: the homogeneous or channel region between $y=0$ to $y=H$ where porosity equals one, the interface region with thickness $\delta$ (between $y=0$ and $-\delta$) where porosity varies as a function of height, and the homogeneous porous region between $y= -\delta$ and $y=-H$ where porosity has a constant value of  $\phi_c = 0.875$. 
In the interface region, the porosity varies according to the following relationship (since the averaging filter no longer fully falls in the homogeneous porous layer)~\cite{breugem2006influence}:
\begin{eqnarray}
\phi(y) &=& -6(\phi_c -1)\left (\frac{y}{\delta}\right)^5 -15(\phi_c -1)\left (\frac{y}{\delta}\right)^4 -\> 10(\phi_c -1)\left (\frac{y}{\delta}\right)^3 +1 \;. \label{eq:7}
\end{eqnarray}
The permeability tensor is isotropic ($\Bar{\Bar{K}} = K\Bar{\Bar{I}}$) and modeled as~\cite{breugem2006influence}
\begin{equation}
 K = \frac{[1-(1-\phi)^{1/3}]^3[1+(1-\phi)^{1/3}]}{11.4(1-\phi)}d_p^2 \;,
\label{eq:perm}
\end{equation}
where $d_p$ ($=0.05H$) is the pore dimension. 
The domain is bounded by two solid walls at  $y=H$  and  $y=-H$, at which the no-slip and no-penetration boundary conditions are imposed. 
Periodic boundary conditions are imposed in the horizontal direction.

\begin{figure}[htbp]
\centering
\includegraphics[width=0.5\textwidth]{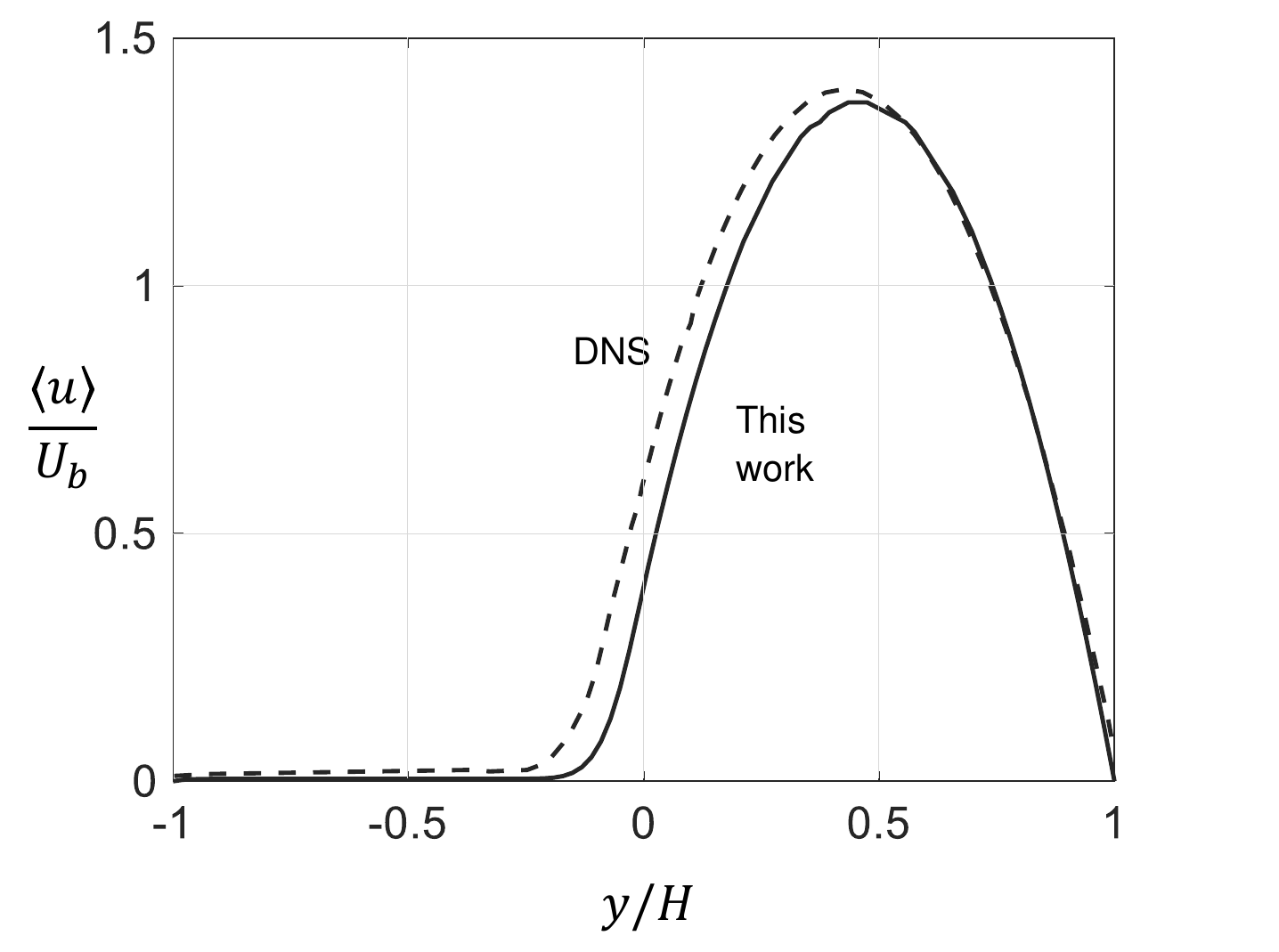}
\caption{\footnotesize Profile of Reynolds- and volume-averaged streamwise velocity, normalized by bulk velocity $U_b$, in laminar flow case.}
\label{u_lam}
\end{figure}

\begin{figure}[htbp]
\centering
\includegraphics[width=0.5\textwidth]{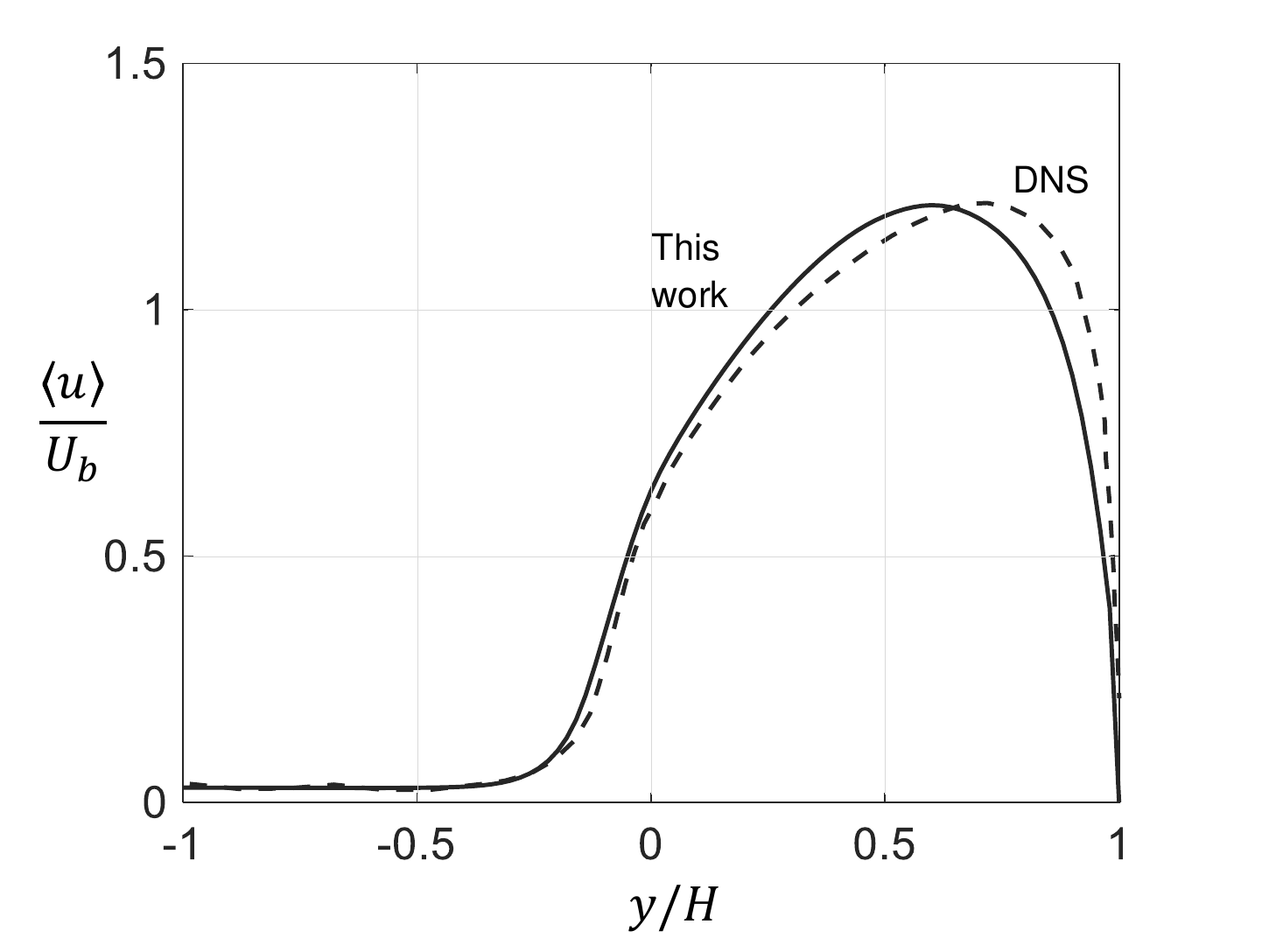}
\caption{\footnotesize Profile of Reynolds- and volume-averaged streamwise velocity, normalized by bulk velocity $U_b$, in turbulent flow case.}
\label{u_turb}
\end{figure}

Figures~\ref{u_lam} and \ref{u_turb} compare the profiles of calculated Reynolds- and volume-averaged velocity with those from the DNS of Breugem et al.~\cite{breugem2006influence} in which the porous region consists of a three-dimensional Cartesian grid of cubes and the flow is fully resolved at the pore scale. 
\revise{We recreate laminar and  turbulent flow simulations with bulk Reynolds numbers ($Re \equiv \rho U_b H/\mu $) of 1 and 5500, respectively.}{revOne}
The model slightly underpredicts the velocity near the interface in the laminar flow case but shows good overall agreement with the DNS for both flow regimes, with errors within 13\% of DNS values for most data points. For the turbulent flow case, the model can capture the skewness with the position of maximum velocity, which corresponds to the larger shear stress at the permeable wall than at the top wall. 
This results from the flow being dominated by relatively large-scale vortical structures near the interface. 
Similar vortical structures have been detected in experiments of flow over plant canopies that originate from a Kelvin--Helmholtz type instability \cite{finnigan2000}.

\subsection{Ignition of a wooden sphere in a tubular reactor\label{subsec:sphere}} \addvspace{10pt}

\begin{figure}[htbp]
\centering
\includegraphics[width=0.3\textwidth]{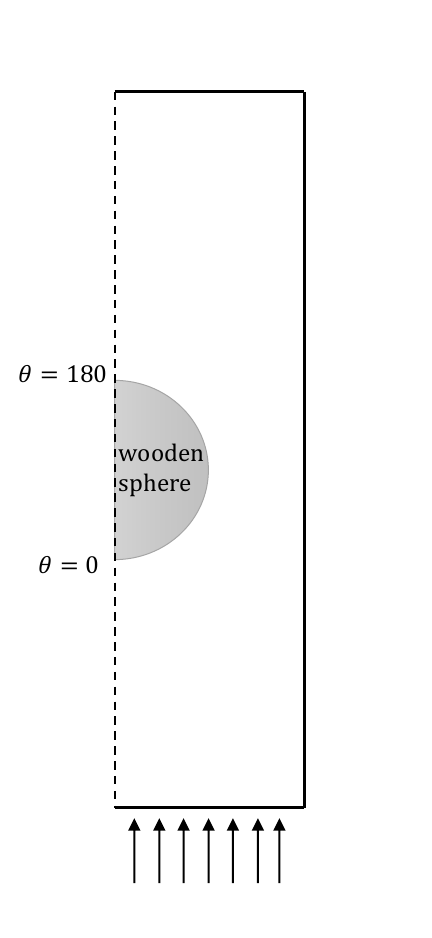}
\caption{\footnotesize Schematic of the domain for the simulation of wooden sphere ignition.}
\label{fig:furnace}
\end{figure}

To test the model's performance on reactive systems, we simulate two physical experiments~\cite{sphere} in which hot air with a temperature of \SI{873}{\kelvin} and Reynolds number of 965 enters a combustion chamber (with an inner diameter of \SI{100}{\milli\meter} and height of \SI{250}{\milli\meter}) from below, providing the heat for the ignition of a wooden \former{oak}{revOne} sphere of diameter \SI{50}{\milli\meter}.
\revise{We considered separate simulations for oak and pine wooden spheres, with densities of \SI{780}{\kilo\gram\per\meter^3} and \SI{440}{\kilo\gram\per\meter^3}, respectively.}{revOne}
The two-dimensional computational domain (Figure~\ref{fig:furnace}) covers half of the chamber spanning from the centerline to the side wall and is discretized using a grid size of 75$\times$300. 

Adiabatic and no-slip boundary conditions are imposed at the side wall and axial symmetry conditions at the normal direction and the centerline. 
For the chemical kinetics of pyrolysis, we use the detailed biomass combustion model of Debiagi et al.~\cite{debiagi} with 46 species and 28 reactions. 
The gas-phase chemical kinetic model is based on GRI-Mech 3.0 \cite{gri} with 45 species and 324 reactions (with \ce{NO_x} chemistry removed). 
The initial composition of reference species representing the oak wood is based on American black oak (\textit{Quercus velutina}) from the CRECK Biomass Database \cite{debiagi}: 48.4\% cellulose, 21.9\% hemicellulose, 4.3\% H-rich lignin, 15.9\% O-rich lignin, 3.1\% triglyceride, 2.5\% tannin, and 4.0\% moisture content.
\revise{For the pine wood, the initial composition is based on ponderosa pine (\textit{Pinus ponderosa}) from the same database: 40.2\% cellulose, 29.7\% hemicellulose, 1.1\% H-rich lignin, 15.9\% O-rich lignin, 3.2\% C-rich lignin, 7.4\% triglyceride, 2.4\% tannin, and 4.0\% moisture content.}{revOne}


\begin{figure*}
\centering
\begin{subfigure}[b]{.4\textwidth}%
\centering\captionsetup{width=\linewidth}%
\includegraphics[width=\linewidth]{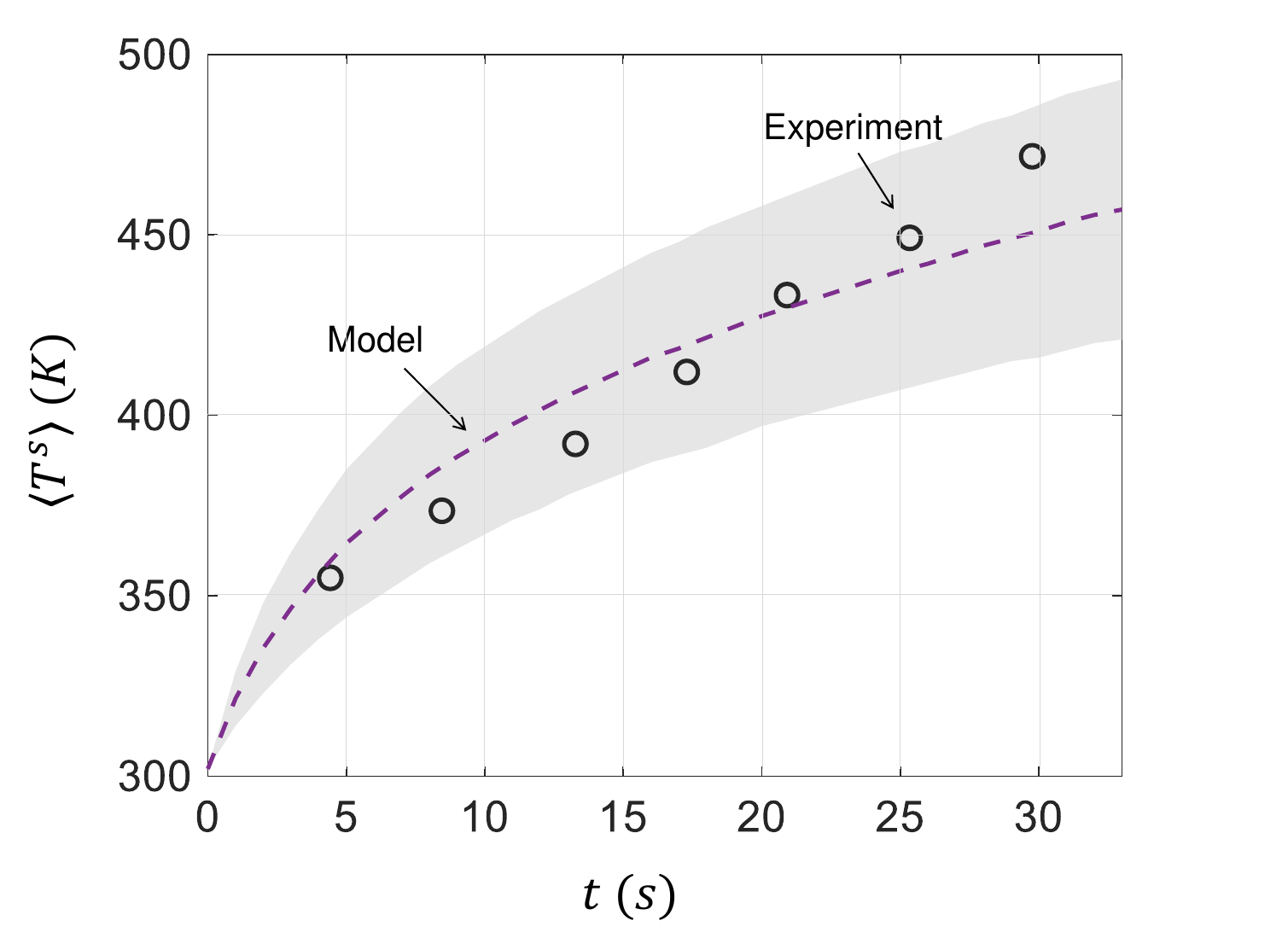}%
\caption{oak}
\label{oak}
\end{subfigure} 
~
\begin{subfigure}[b]{.365\textwidth}%
    \centering\captionsetup{width=\linewidth}%
\includegraphics[width=\linewidth]{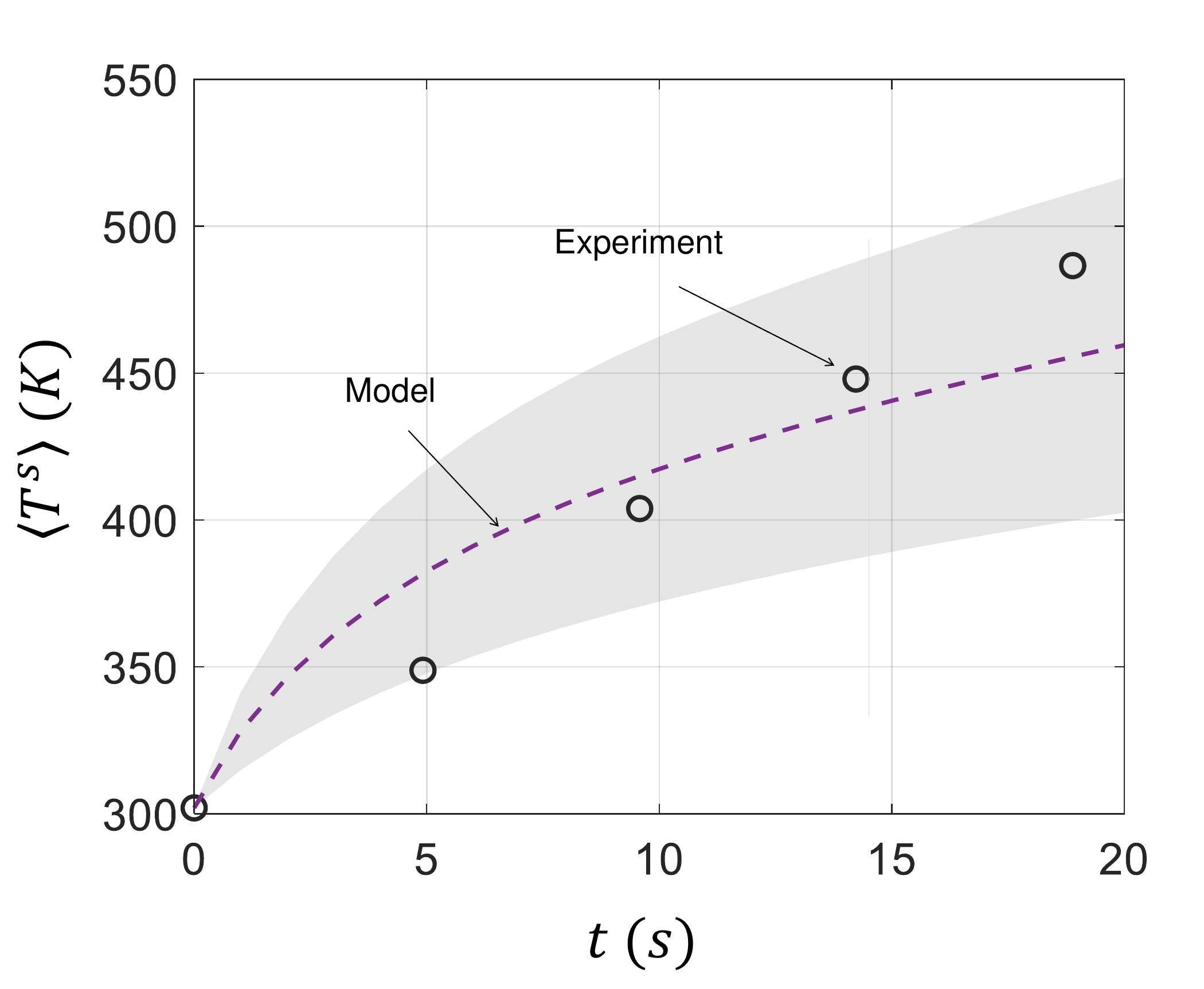}%
\caption{pine}
\label{pine}
\end{subfigure}

\caption{\small Temperature evolution at the surface of the sphere \revise{for the two wood types}{revOne}. The dashed line denotes the average calculated temperature in the region between the meridian angles of  270$^{\circ}$ and 360$^{\circ}$, as indicated by the shaded area.}
\label{temp_oak}
\end{figure*}

Figure~\ref{temp_oak} presents the evolution of temperature at the surface of the sphere \revise{for the two wood types}{revOne}. 
Since the description of the experiment did not include detailed information about the angle at which the probe was located on the surface, an exact temperature comparison is not possible; however, the calculated average temperatures are within 5\% of the reported experimental values, with similar resulting trends. 
The predicted ignition delay times \revise{are \SI{32.3}{\second} and \SI{20.1}{\second} for the oak and pine cases, respectively, within 3\% and 6\% of the experimental measurements.}{revOne}

\begin{figure}[htbp]
\centering
\includegraphics[width=0.5\textwidth]{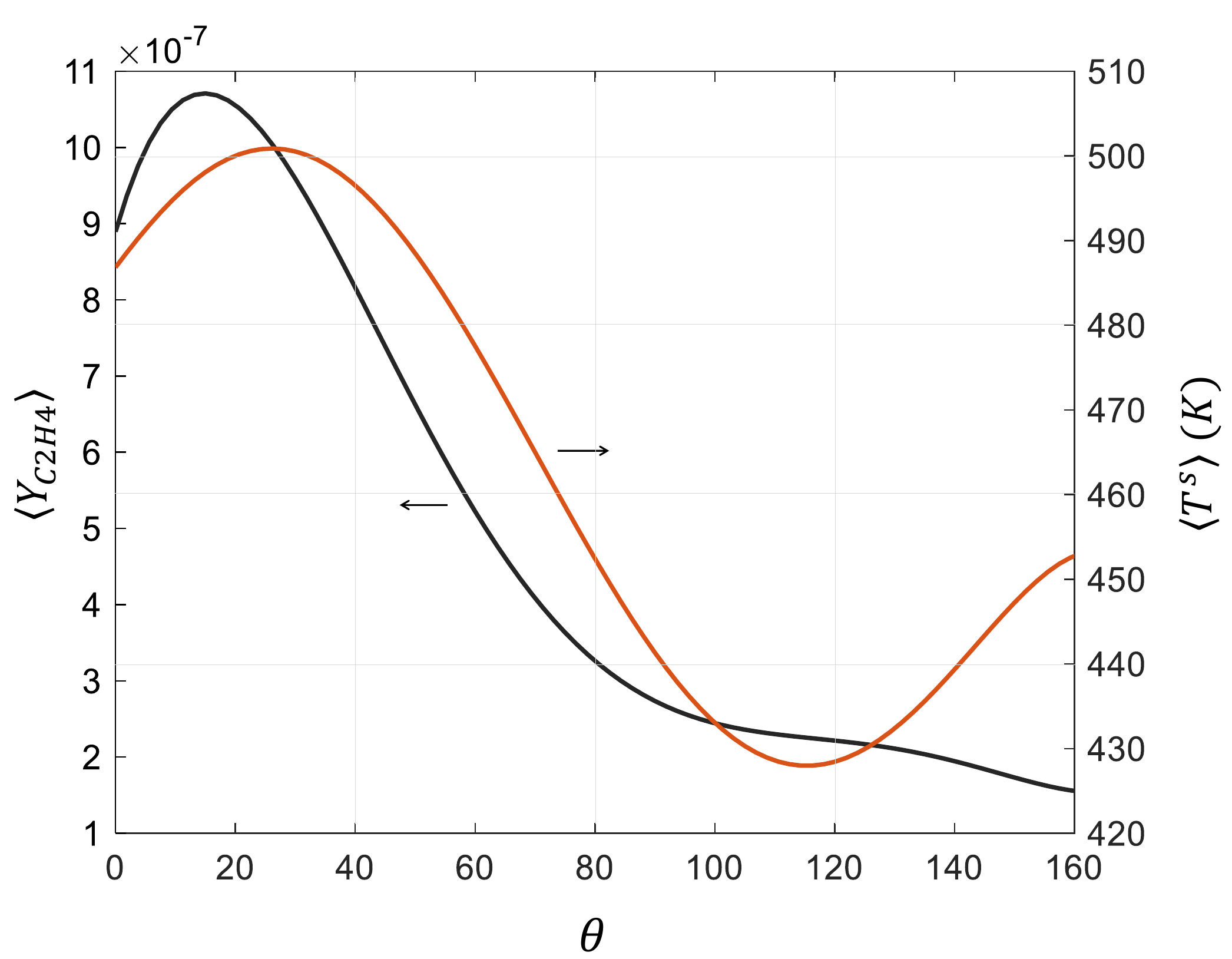}
\caption{\footnotesize Variations of solid temperature and \ce{C2H4} mass fraction across the surface as a function of meridian angle ($\theta$). $\theta = 0$ is set at the lowest point on the sample.}
\label{variations}
\end{figure}

Figure~\ref{variations} shows the variations of temperature and ethylene (one of the important pyrolysis products, acting as fuel for gas-phase combustion) mass fraction at the onset of ignition; the predicted peak locations confirm the experimental result that ignition happens at a location in the lower region of the sphere. The flow separation and formation of a wake region close to the upper half of the sphere degrades the heat and mass transfer.

\subsection{Flow instabilities during solid fuel combustion in quiescent air\label{subsec:instabilities}} 

\begin{figure}[htbp]
\centering
\includegraphics[width=0.5\textwidth]{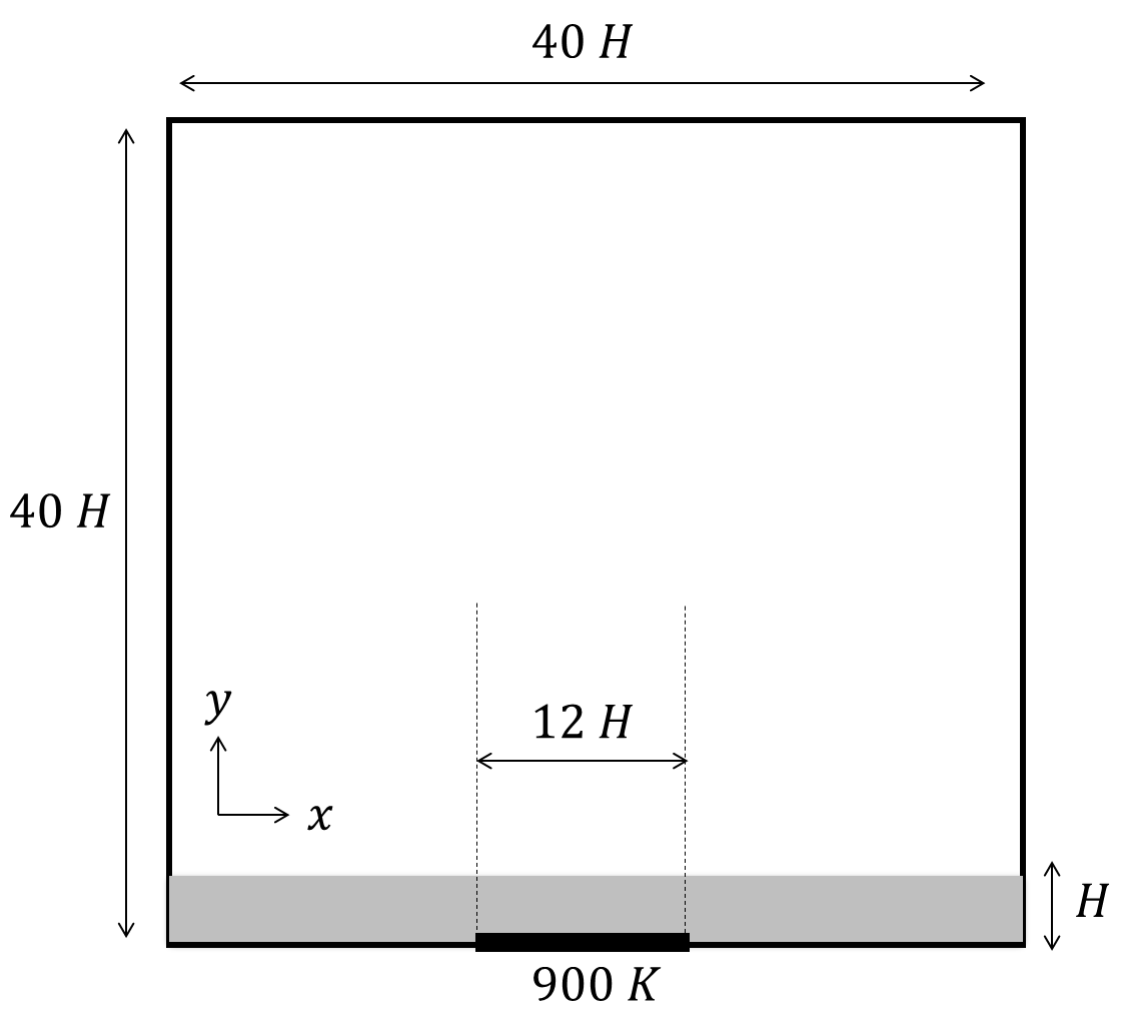}
\caption{\footnotesize The computational domain for the combustion of the heated porous layer in quiescent air. A constant temperature of \SI{900}{\kelvin} is imposed at center region of the lower boundary, indicated with a thick black line.}
\label{domain}
\end{figure}

In this section, we examine the formation of buoyant reacting plumes, a situation that can occur in the burning of solid fuels or natural fires when a vegetation layer is heated from below in the absence of a forced flow. 
We consider a two-dimensional square domain of side length $40H$, with a porous region at the bottom, as depicted in Fig.~\ref{domain}. 
The grid size is 400$\times$400; we performed tests to ensure the domain size was sufficiently large and the flow was not affected by the boundaries. 
All boundaries are impermeable and adiabatic except for the lower boundary of the porous layer where temperature (\SI{900}{\kelvin}) is imposed at the center region.
We assume the porous layer has a chemical composition similar to that of pine wood and the same structure as the one in Section~\ref{subsec:channel}. 
Since the focus of this analysis is not accurately capturing the ignition time, for simplicity we use a one-step pyrolysis model of pine wood \cite{lautenberger2009}, however, the model is generally capable of incorporating detailed chemical kinetics, as demonstrated in the previous section. 
\revise{We also neglect radiation heat transfer in these simulations, as it is not our focus here and Appendix~\ref{radiation} shows that it contributes relatively little to heat transfer for these problems.}{revTwo}

\begin{figure*}[htbp]
\centering
\includegraphics[width=.9\linewidth]{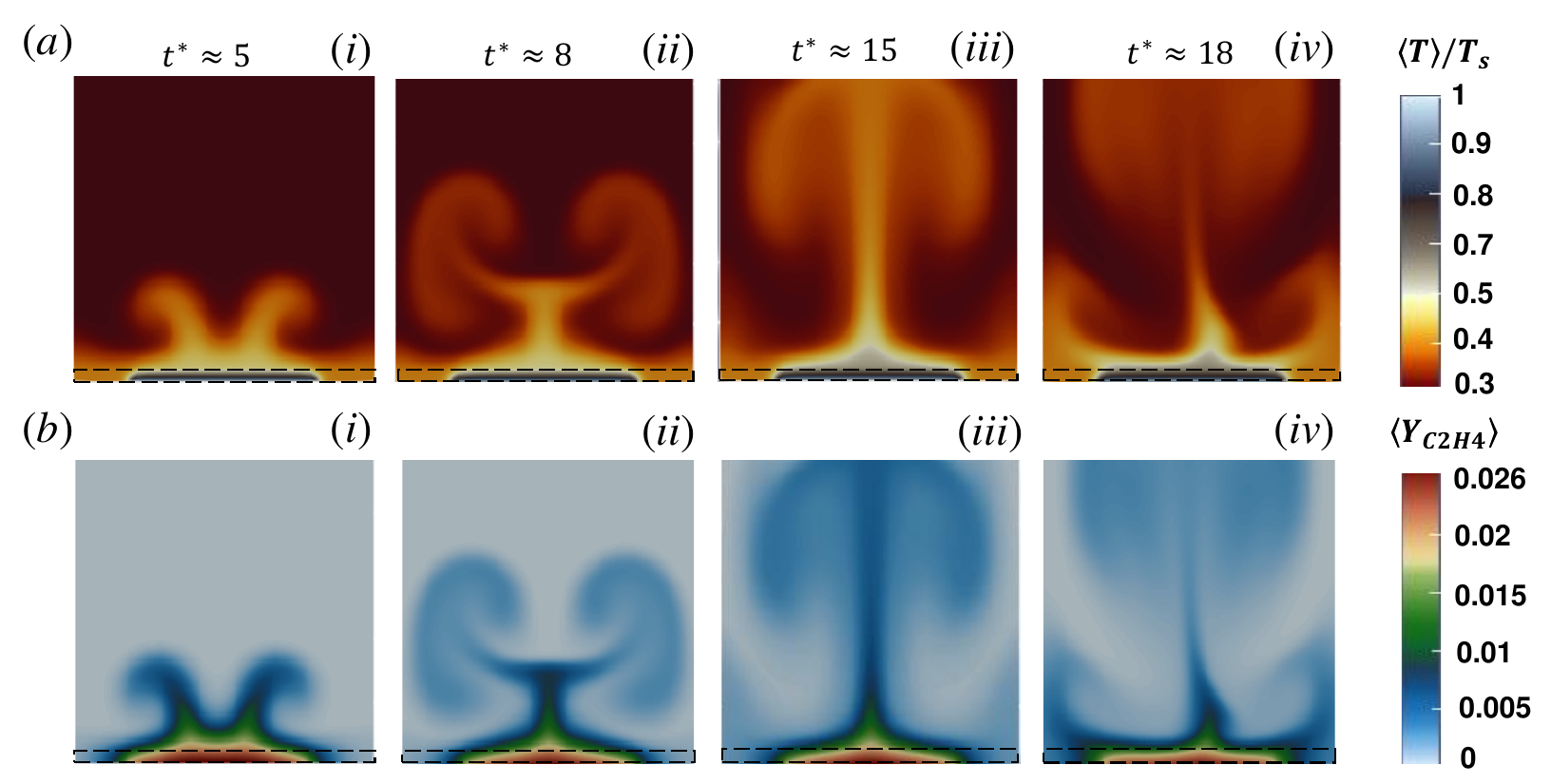}
\caption{\footnotesize Fields of fluid temperature (a) and mass fraction of \ce{C2H4} (b) as a function of nondimensional time $t^*=t\sqrt{g/(40H)}$. $T_s$ is the temperature of the bottom boundary (900 K).}
\label{snapshots}
\end{figure*}

Figure~\ref{snapshots} shows the evolution of variables of interest (gas temperature and mass fraction of fuel). 
In the initial stages, the boundary between the hot and cold fluids begins to form a bubble-spike; subsequently, it evolves into symmetrical mushroom-like flow structures before losing symmetry and reaching a state of chaotic behavior.
Some of these patterns resemble the characteristics of the classical Rayleigh--Taylor flow instability. 
This situation occurs when the density discontinuity between the two fluids is unstable to small perturbations, which grow in time due to the accumulation of vorticity. 
To determine the origins of these observations, we next analyze the vorticity dynamics.

\begin{figure*}[htbp]
\centering
\includegraphics[width=0.9\linewidth]{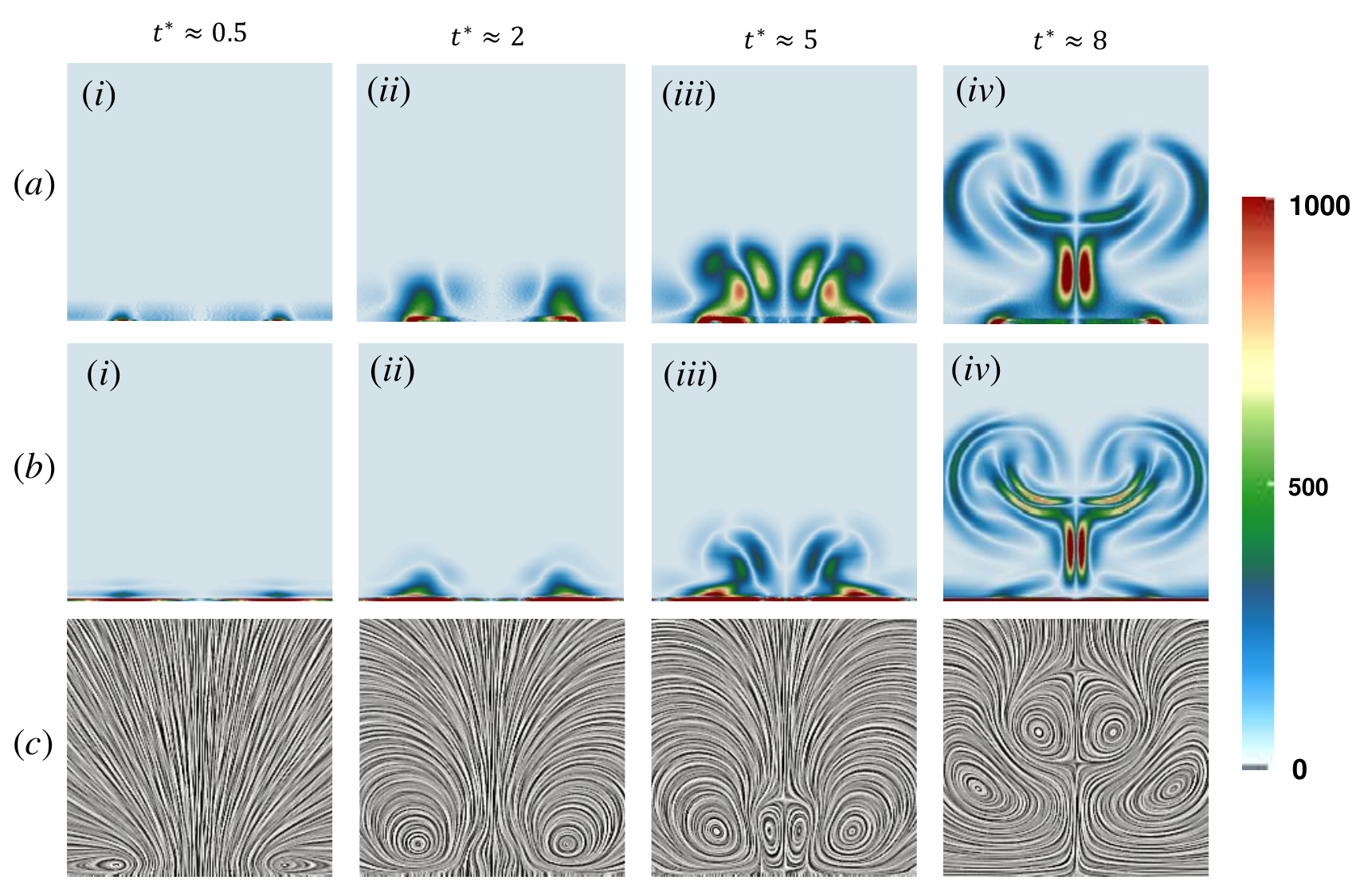}
\caption{\footnotesize Fields of vorticity production budgets: (a) baroclinic torque, (b) viscous effects, and (c) the corresponding velocity streamlines above the interface.}
\label{vorticity}
\end{figure*}

\begin{figure}[htbp]
\centering
\includegraphics[width=0.5\textwidth]{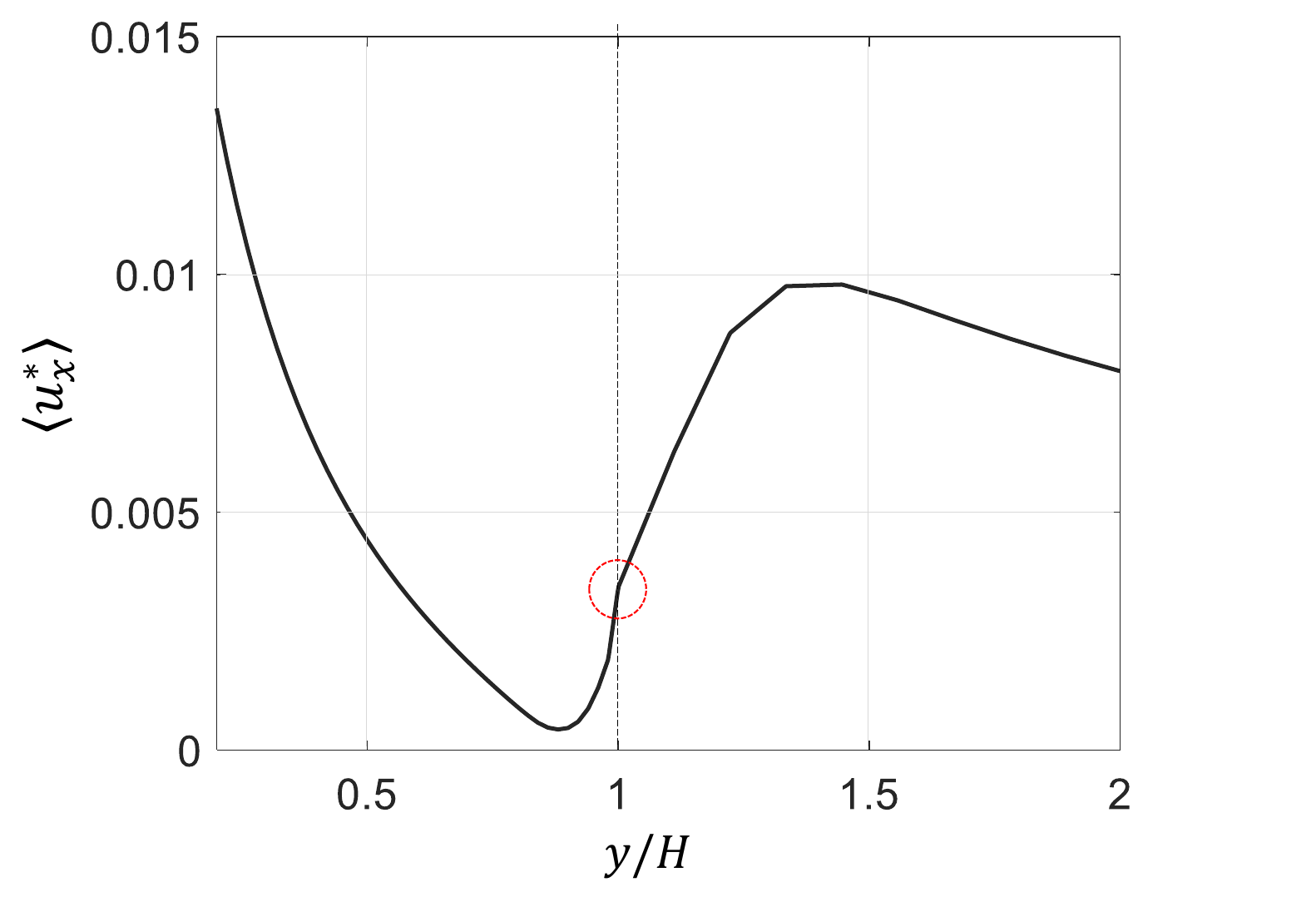}
\caption{\footnotesize Inflection point in the nondimensional velocity profile near the interface at $t^*=0.2$ and horizontal position of $x=26 H$. $\langle u^*_x \rangle = \langle u_x \rangle/(40H) \sqrt{g/(40H)}$.}
\label{infelction}
\end{figure}

The vorticity transport balance (based on the momentum equation) is
\begin{eqnarray}
\frac{D \bm{\omega}}{Dt} &=& (\bm{\omega}\cdot\nabla)\langle\textbf{u}\rangle-\bm{\omega}(\nabla \cdot \langle\textbf{u}\rangle) + \frac{1}{\langle\rho\rangle^2}\nabla\langle\rho\rangle\times \nabla \langle p \rangle
+\> \nabla \times \left (\frac{\nabla\cdot\langle\Bar{\Bar{\tau}}^*\rangle}{\langle\rho\rangle}\right) + \nabla\times\left(\frac{\textbf{f}}{\langle\rho\rangle}\right) \;,
\label{eq:vorticity}
\end{eqnarray}
where $\bm{\omega}$ is the vorticity, $\frac{D }{Dt}$ is the material derivative operator, and $\Bar{\Bar{\tau}}^*=\Bar{\Bar{\tau}}+\Bar{\Bar{\tau}}_{SGS}$. 
The last three terms on the right-hand side of Eq.~\eqref{eq:vorticity} can contribute to the formation of instabilities: baroclinic torque, viscous effects, and the torque arising from the drag force exerted by the solid matrix, respectively. However, in our computations, the last term is negligible compared to the other two terms (partly due to the high porosity near the interface) and is therefore ignored in this study. The first and second terms represent vortex stretching and dilatation, which we do not focus on here. 
These terms become important after the initial stages and during flame development, as Fillo et al.~\cite{fillo2022assessing} showed. 
They analyzed the enstrophy transport balance in turbulent premixed flames and found vortex stretching and viscous effects to be the primary source and sink terms, respectively.  

\begin{figure}[htbp]
\centering
\includegraphics[width=0.5\textwidth]{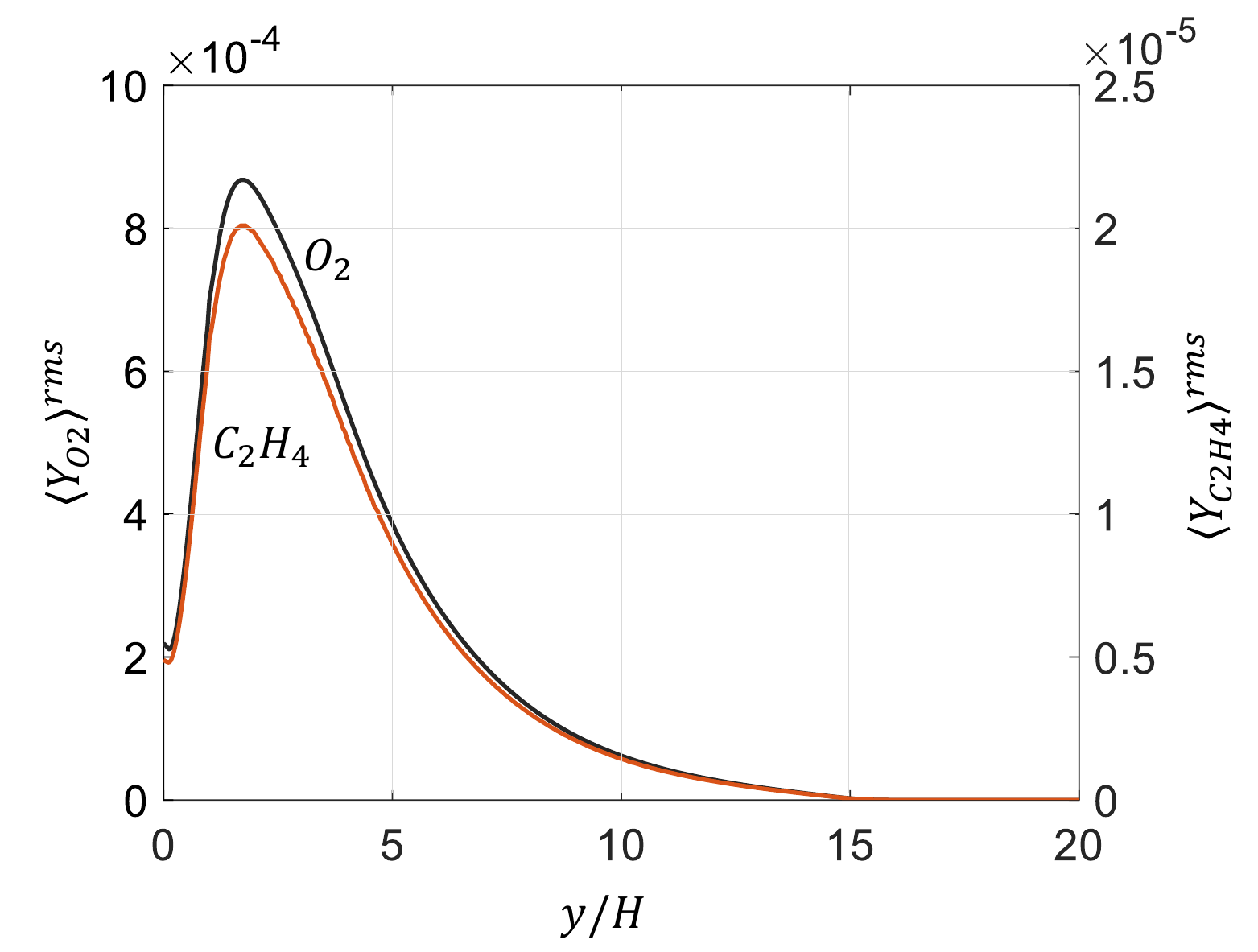}
\caption{\footnotesize Volume-averaged root-mean-square mass fractions of \ce{C2H4} and \ce{O2} along the centerline at $t^*\approx 10$.}
\label{rms}
\end{figure}

\begin{figure}
\centering
\includegraphics[width=0.5\linewidth]{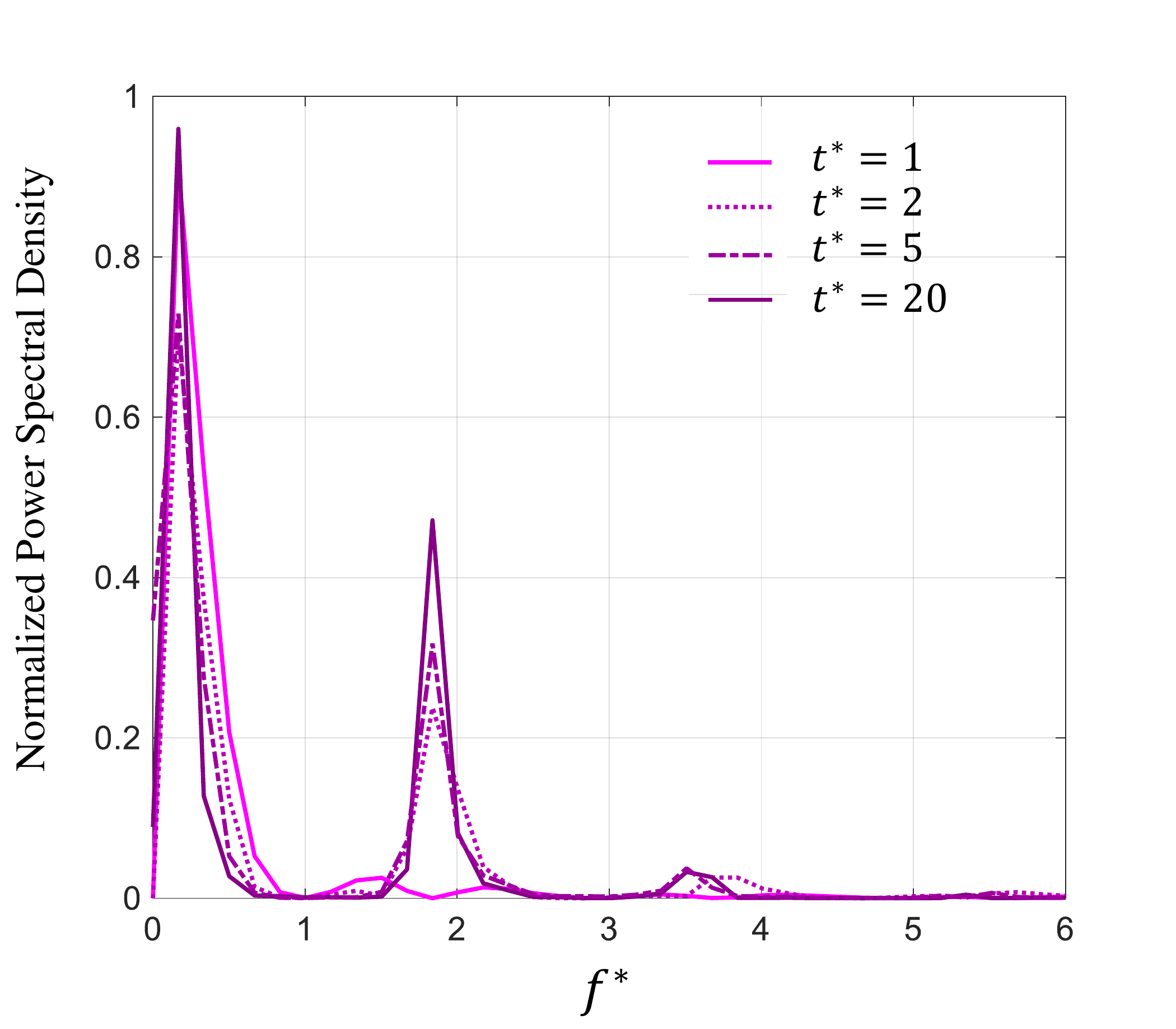}
\caption{\revise{\small Normalized power spectral densities (PSDs) of vertical velocity $\langle u_y \rangle$ with respect to nondimensional frequency ($f^*$), taken from four time periods of $[0 \ \  t^*]$.}{revTwo}}
\label{fig:freq}
\end{figure}

In the early stages, Figure~\ref{vorticity} shows mild vorticity generation at the corners of the interface between the porous and non-porous regions; density stratification leads to flow disturbances at the base of the porous layer due to the baroclinic term. 
The entrainment of ambient air from the sides then creates a shear layer, which appears to trigger the formation of vortices at the interface due to the viscous term (see the velocity streamlines). 
As these structures grow, another set of vortices starts to form at the interface center and rise over time, shaping the aforementioned flow structures. Overall, a combination of the baroclinic and viscous terms, associated with Rayleigh--Taylor and Kelvin--Helmholtz flow instabilities, appears to drive the formation of observed vortical structures. Other studies \cite{breugem2006influence, finnigan2000} have also suggested that the inflexional velocity profile at the interface of porous and free fluid regions could give rise to instabilities of the Kelvin--Helmholtz type. Figure~\ref{infelction} confirms that the mean horizontal velocity exhibits an inflection point just below the interface (before the formation of vortices), substantiating the above hypothesis. The presence of these vortical structures and their effect on variations of fuel and oxidizer mass fractions is also reflected in root-mean-square (rms) mass fractions ($Y^{\text{rms}}=\sqrt{\langle Y'Y'\rangle}$) as demonstrated in Figure~\ref{rms}: the peak in $Y^{\text{rms}}$ plots, near the interface ($y/H =1$), indicates the strong presence of vortices in that region. 

\revise{To provide further quantitative insights into the flow instabilities, Figure~\ref{fig:freq} presents frequency spectra of vertical velocity in the form of power spectral density (PSD). 
This tool is commonly used to measure the frequencies at which large-scale vortices are shed during the development of buoyant plumes \cite{meehan2022}. 
We obtain the PSD via singular value decomposition of a time series of vertical velocity taken from two-dimensional slices of data defined by the bounding box $x/L=[0.4 \ \  0.6], \  y/L = [H \ \  12H]$ and performing a fast Fourier transform on the second mode.
The presence of multiple peaks in the PSD suggests multiple interacting modes---the spectra broaden as the vortices form and potentially merge. 
In the time period of $[0 \ \  1]$, the peak at $f^* \approx 0.2$ is the only dominant feature present, corresponding to the early stages of vortices formation near the interface as seen in column $(i)$ of Figure~\ref{vorticity}. 
As the plume evolves, a second peak at $f^* \approx 2$ starts to grow after $t^* =1$, connected to the stages of bubble-spike formation and interface roll-up seen in Figures~\ref{snapshots}--\ref{vorticity}.}{revTwo}

The phenomena observed here share features with those associated with buoyant jets. For example, Meehan et al.~\cite{meehan2022} observed similar structures at low Reynolds numbers and low density ratios in the injection of helium into the quiescent air. The study suggested that the combination of Richardson and Reynolds numbers can determine the instability regime including laminar, transitional, and turbulent. An important difference here is the significant role of the porous medium and its structure in shaping the flow behaviors, in addition to the above parameters.

\subsection{Effects of interface morphology  \label{subsec:topology}} 

To demonstrate the model's applicability to more complicated situations, we revisit the previous case (formation of buoyant reacting plumes) by altering the shape of the boundary between the porous and free fluid regions. The new interface contains some surface-mounted rectangular features that could represent urban (or vegetative) canopies at a large scale. 

\begin{figure}[htbp]
\centering
\includegraphics[width=0.9\linewidth]{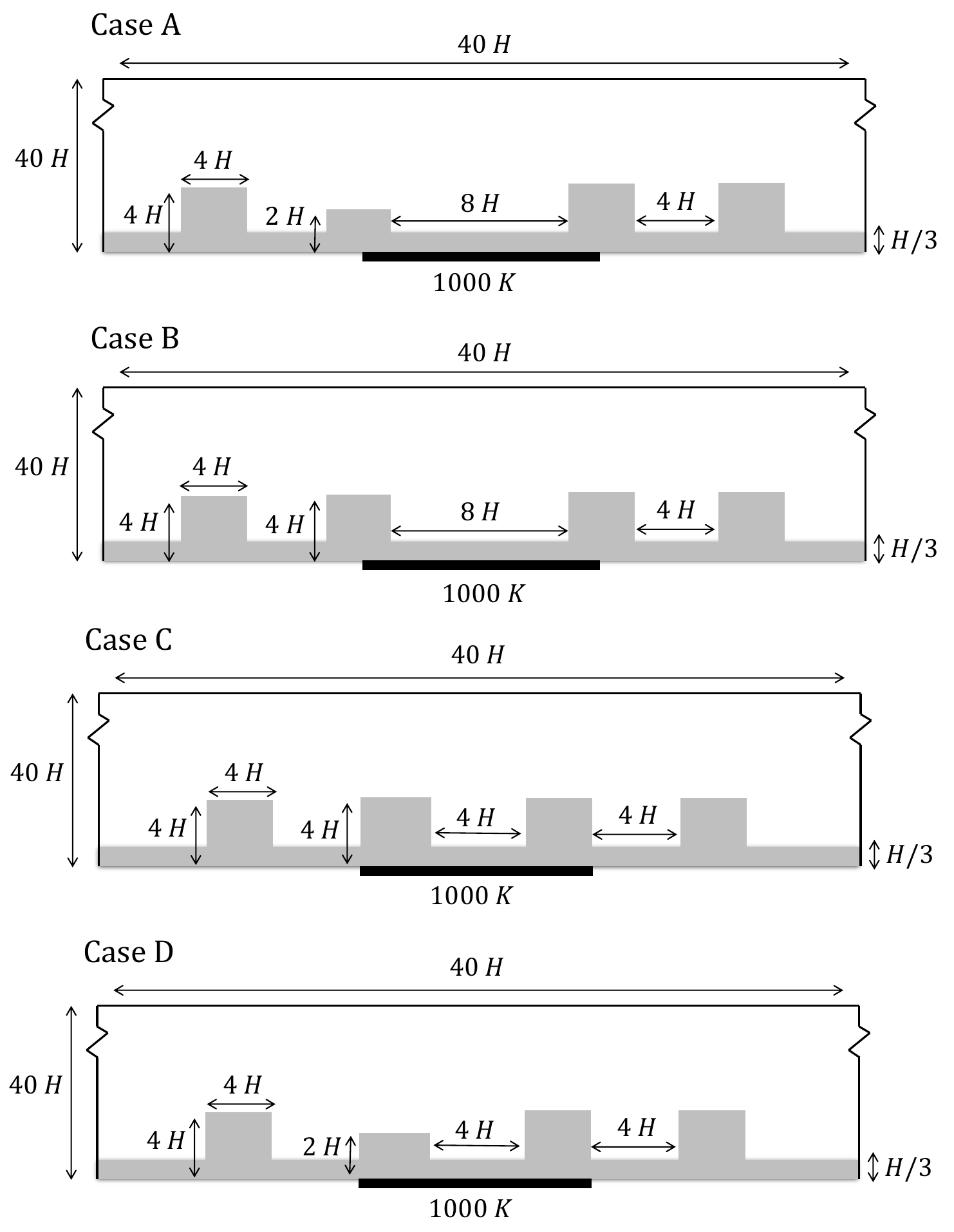}
\caption{\small Computational domains with surface-mounted geometrical features.}
\label{domain2}
\end{figure}

We consider four cases with either different spacing between the two middle rectangles or different heights of the left middle rectangle, as depicted in Fig.~\ref{domain2}. The conditions are the same as in Section~\ref{subsec:instabilities}, except for the temperature at the bottom boundary ($T_s= \SI{1000}{\kelvin}$). 

\begin{figure*}[htb]
\centering
\includegraphics[width=0.9\linewidth]{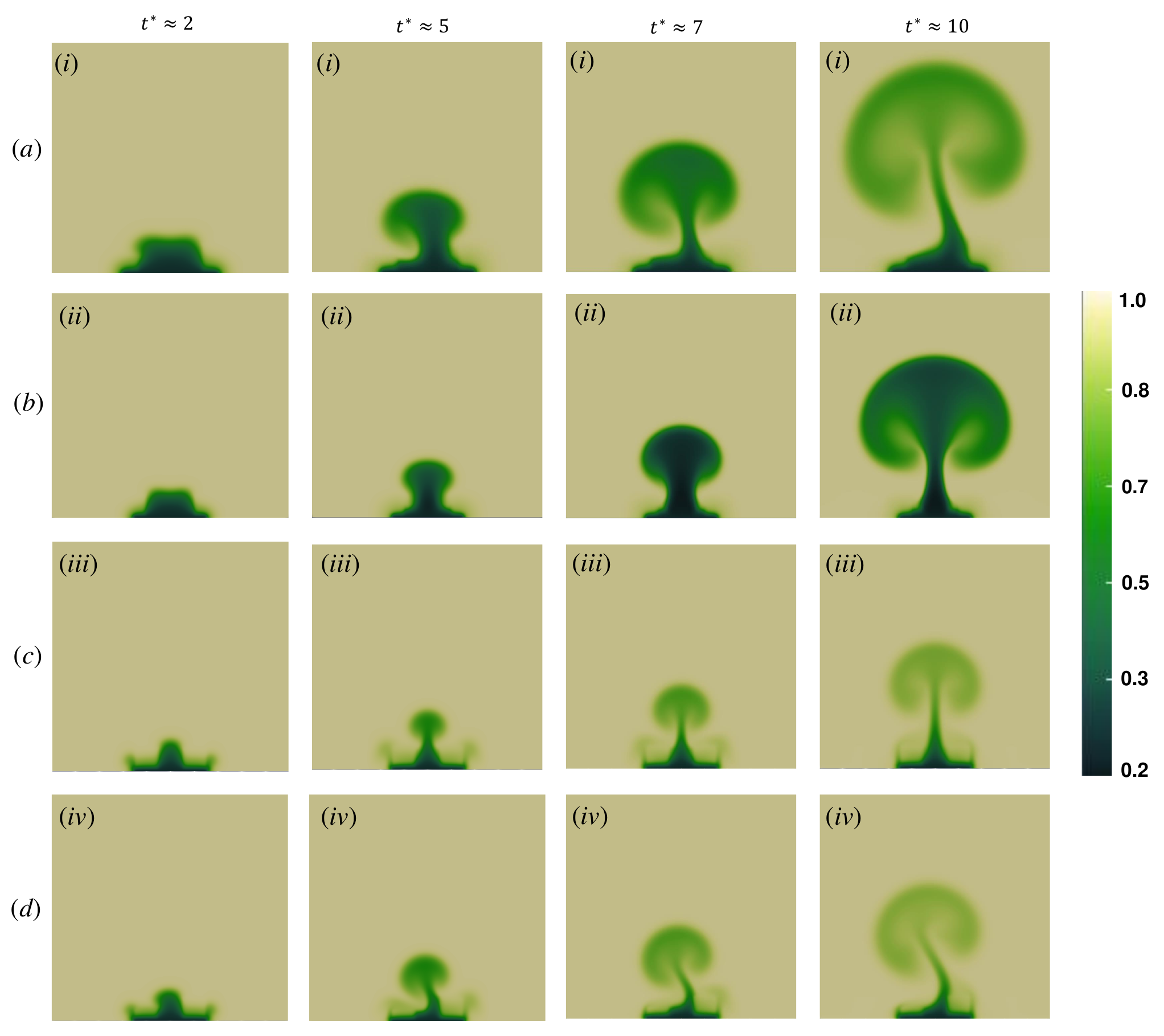}
\caption{\small Evolution of density field  $\langle \rho \rangle/\langle \rho_0 \rangle$ normalized by the initial density value of the ambient air for cases A (a), B (b), C (c), and D (d). }
\label{fig:density_field}
\end{figure*}

Figure~\ref{fig:density_field} shows the evolution of density for all four cases. The presence of the walls in the formed cavity region restricts the propagation of flow instabilities in the horizontal direction. In case A, the shorter height of the left wall reduces these restrictions on the left side of the plume core, leading to greater vortex growth on that side and the development of an asymmetric flow field. On the other hand, the presence of walls triggers the formation of more vortices due to Kelvin--Helmholtz instability near the wall region, which contributes to the flapping motion of the plume, discussed in the next section.

The four cases also exhibit different combustion behaviors: ignition events occur in cases A and B (with wider middle cavity), while they do not in cases C and D (with a narrower cavity). The asymmetry in case A considerably delays the time of the ignition event and changes its location. These observations can have important implications for the management of fire at large or small scales, therefore, this part of the study aims to explain these contrasting behaviors, specifically the differing ignition times in the first two cases and the absence of ignition in the last two. Since geometry is the only differing element in these experiments, the resulting flow and its interactions with the reactions must be the key to answering the questions.

\begin{figure*}[htb]
\centering
\includegraphics[width=0.9\linewidth]{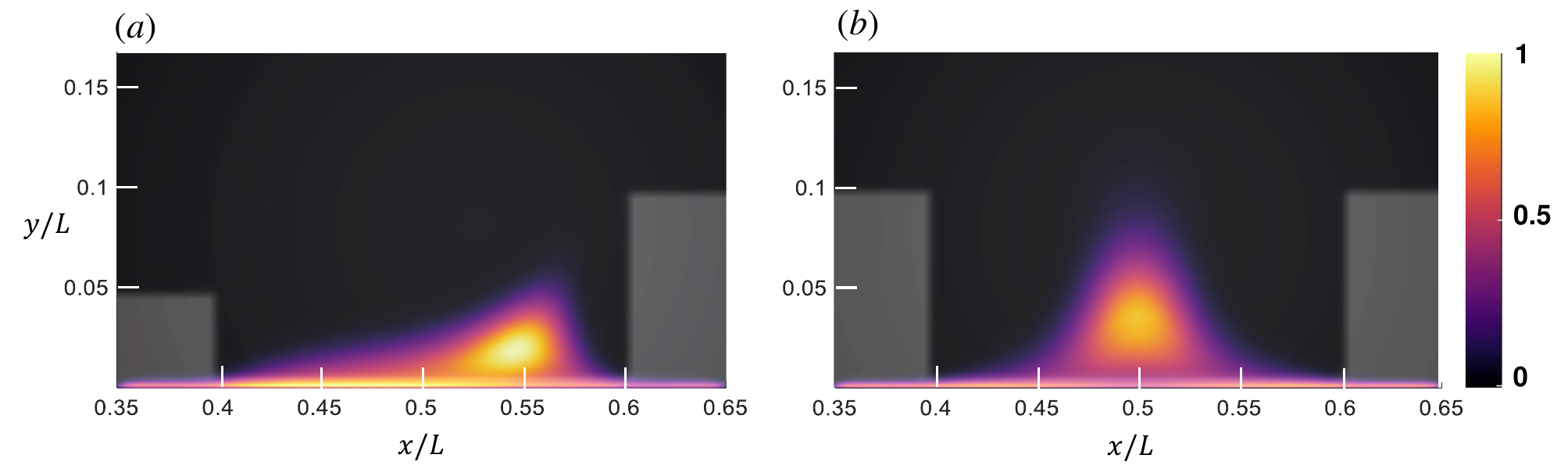}
\caption{\small Fields of chemical heat release rate, normalized by its maximum value ($\langle \dot Q \rangle/\dot Q_m$) at $t^*\approx 20$ for case A (a) and at $t^*\approx 4.5$ for case B (b).}
\label{fig:qdot_field}
\end{figure*}

\begin{figure*}[htbp]
\centering
\begin{subfigure}[b]{.35\textwidth}%
\centering\captionsetup{width=\linewidth}%
\textbf{Case A}\par\medskip
\includegraphics[width=\linewidth]{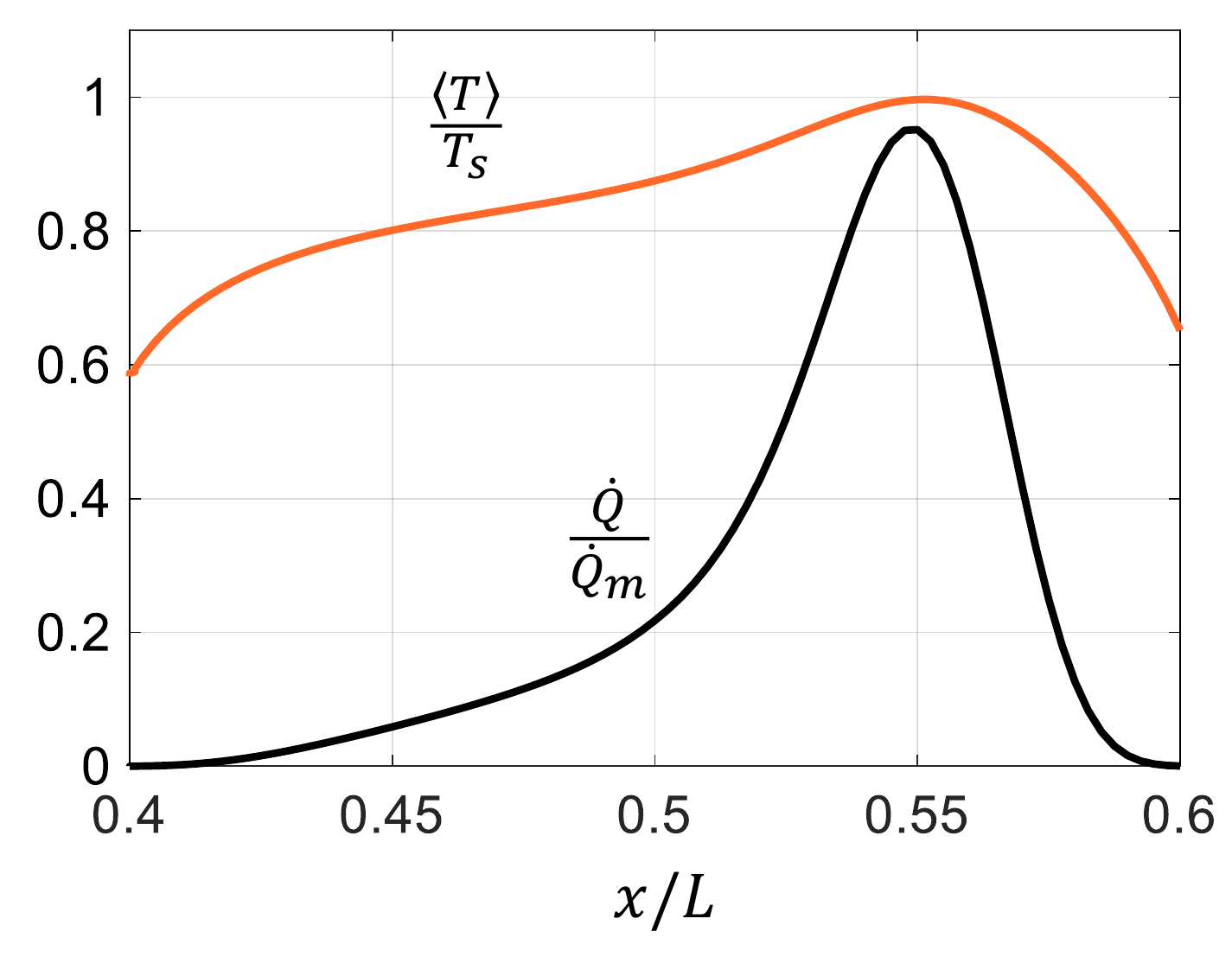}%
\caption{Temperature and chemical heat release rate}
\label{fig:fig12a}
\end{subfigure} \hspace{0.8cm}
~
\begin{subfigure}[b]{.35\textwidth}%
    \centering\captionsetup{width=\linewidth}%
     \textbf{Case B}\par\medskip
\includegraphics[width=\linewidth]{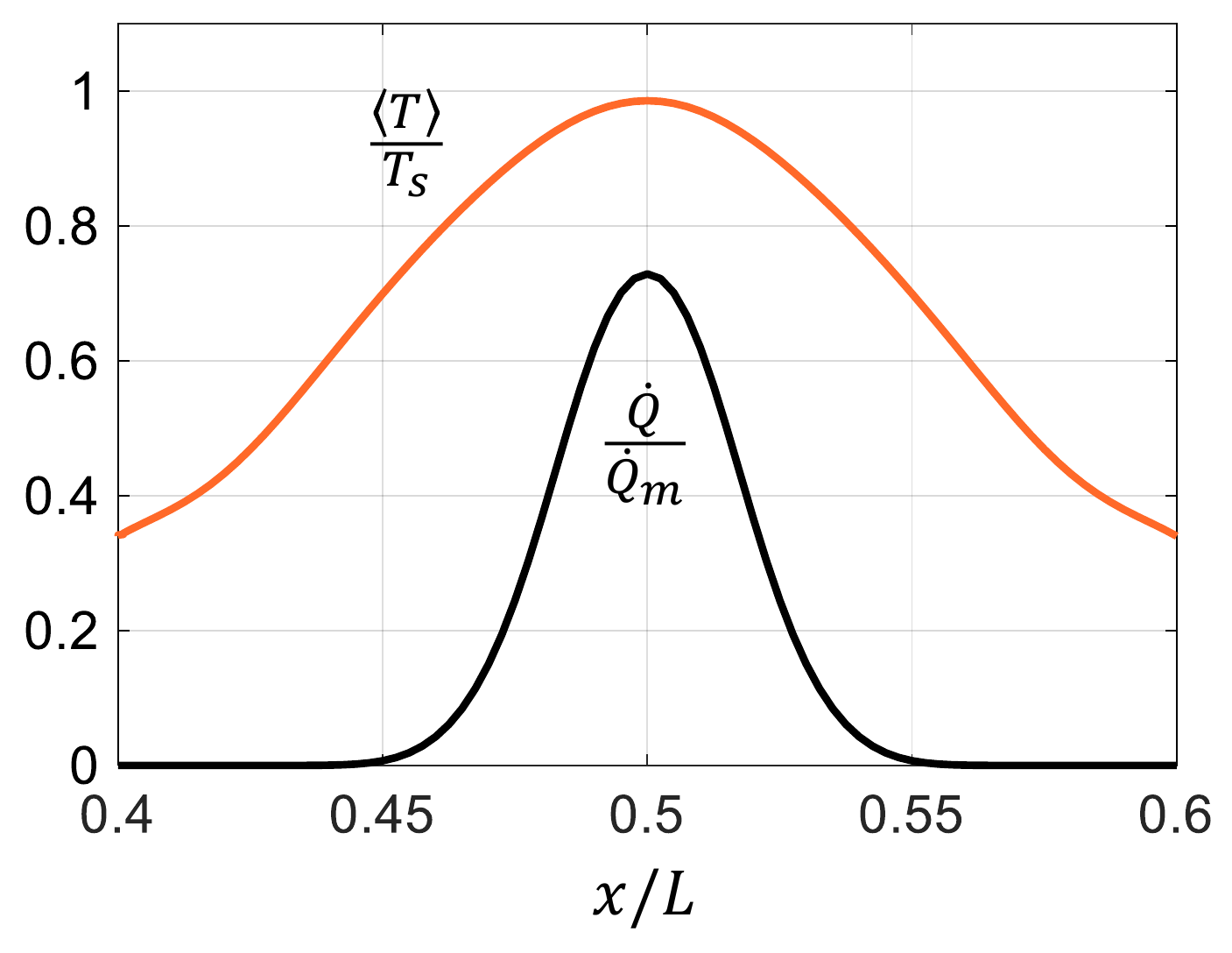}%
\caption{Temperature and chemical heat release rate}
\label{fig:fig12b}
\end{subfigure}

\begin{subfigure}[b]{.4\textwidth}%
    \centering\captionsetup{width=\linewidth}%
\includegraphics[width=\linewidth]{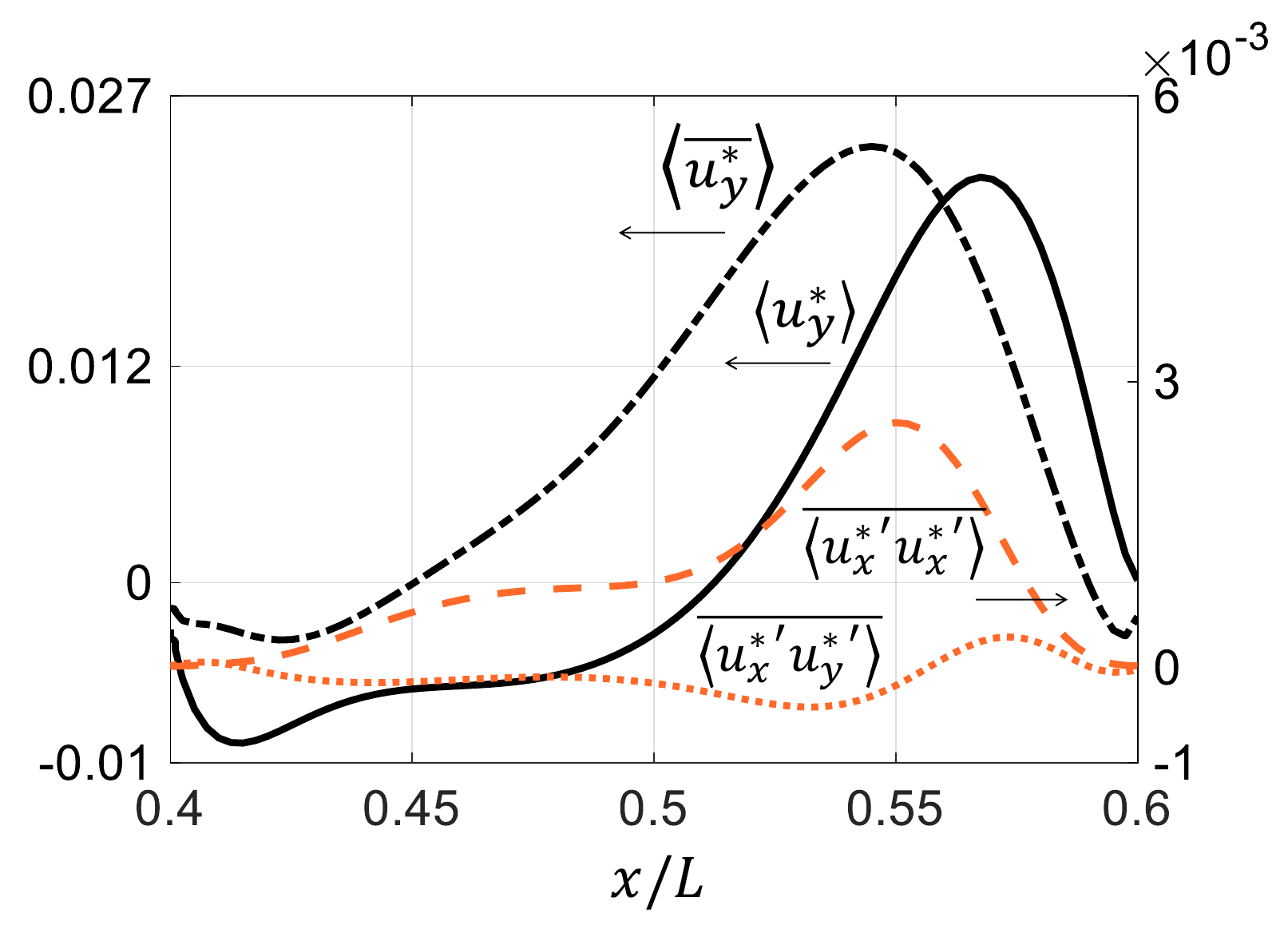}%
\caption{Instantaneous and average velocity and Reynolds stresses }
\label{fig:fig12c}
\end{subfigure}
~
\begin{subfigure}[b]{.4\textwidth}%
\centering\captionsetup{width=\linewidth}%
\includegraphics[width=\linewidth]{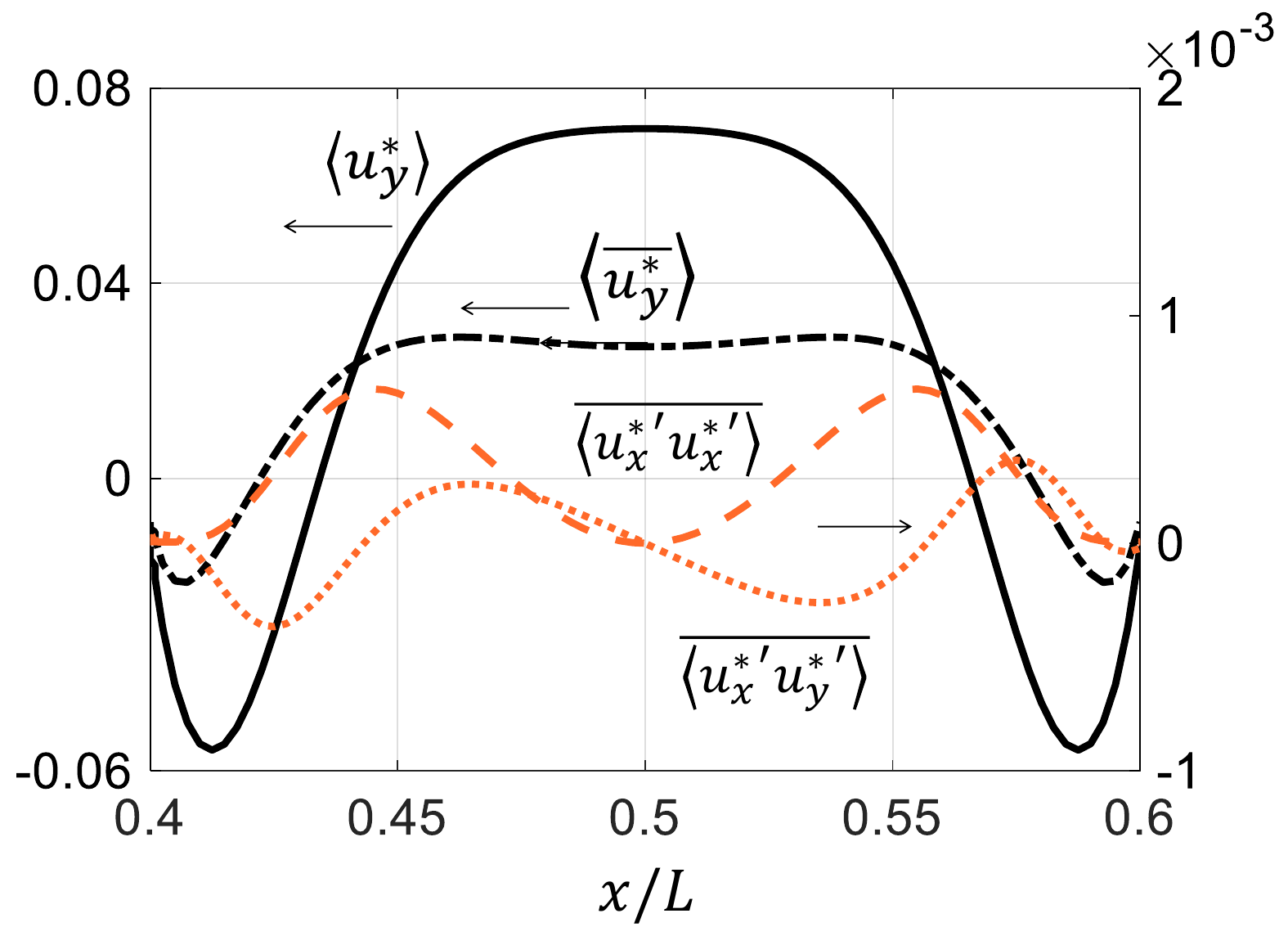}%
\caption{Instantaneous and average velocity and Reynolds stresses }
\label{fig:fig12d}
\end{subfigure}

\begin{subfigure}[b]{.38\textwidth}%
\centering\captionsetup{width=\linewidth}%
\includegraphics[width=\linewidth]{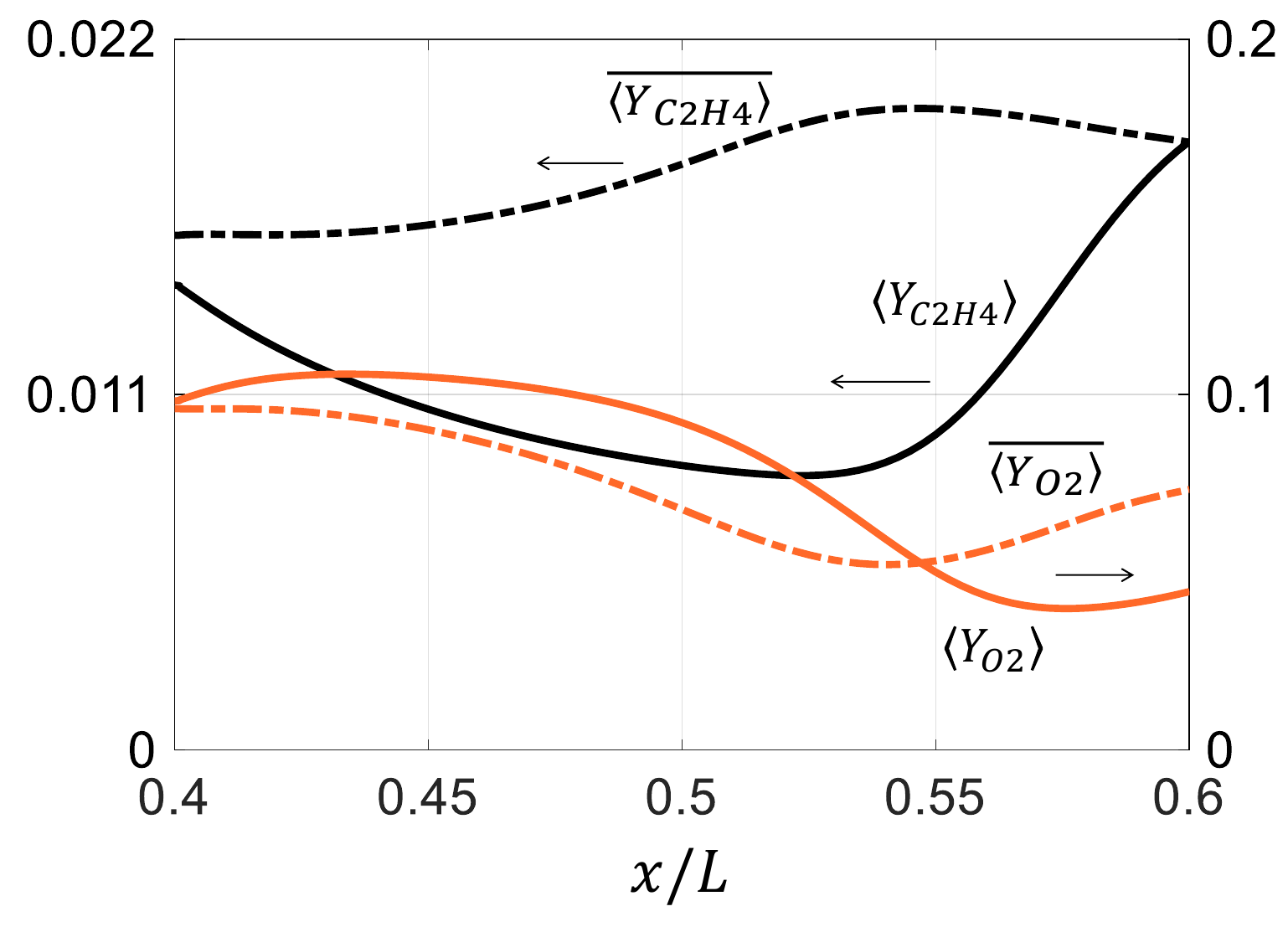}%
\caption{Instantaneous and average mass fractions of \ce{C2H4} and \ce{O2}}
\label{fig:fig12e}
\end{subfigure} \hspace{0.48cm}
~
\begin{subfigure}[b]{.39\textwidth}%
    \centering\captionsetup{width=\linewidth}%
\includegraphics[width=\linewidth]{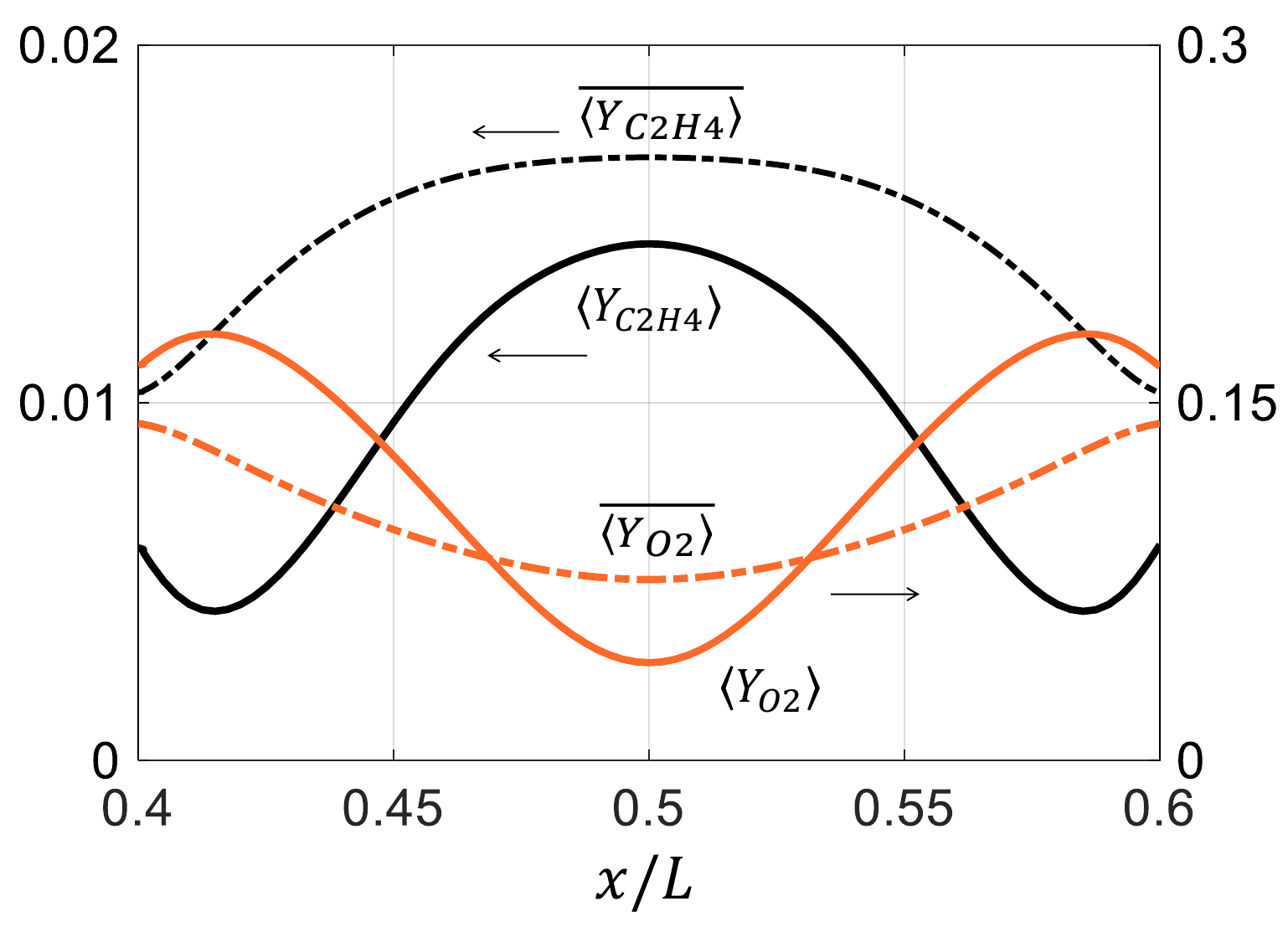}%
\caption{Instantaneous and average mass fractions of \ce{C2H4} and \ce{O2}}
\label{fig:fig12f}
\end{subfigure}

\caption{\small Spatial variability at the time of ignition across the horizontal line $y/L=0.025$ for case A and $y/L=0.045$ for case B}
\label{fig:spatial_variability}
\end{figure*}

\begin{figure}
\centering
\begin{subfigure}[b]{0.5\textwidth}%
\centering
\captionsetup{width=\linewidth}%
\includegraphics[width=\linewidth]{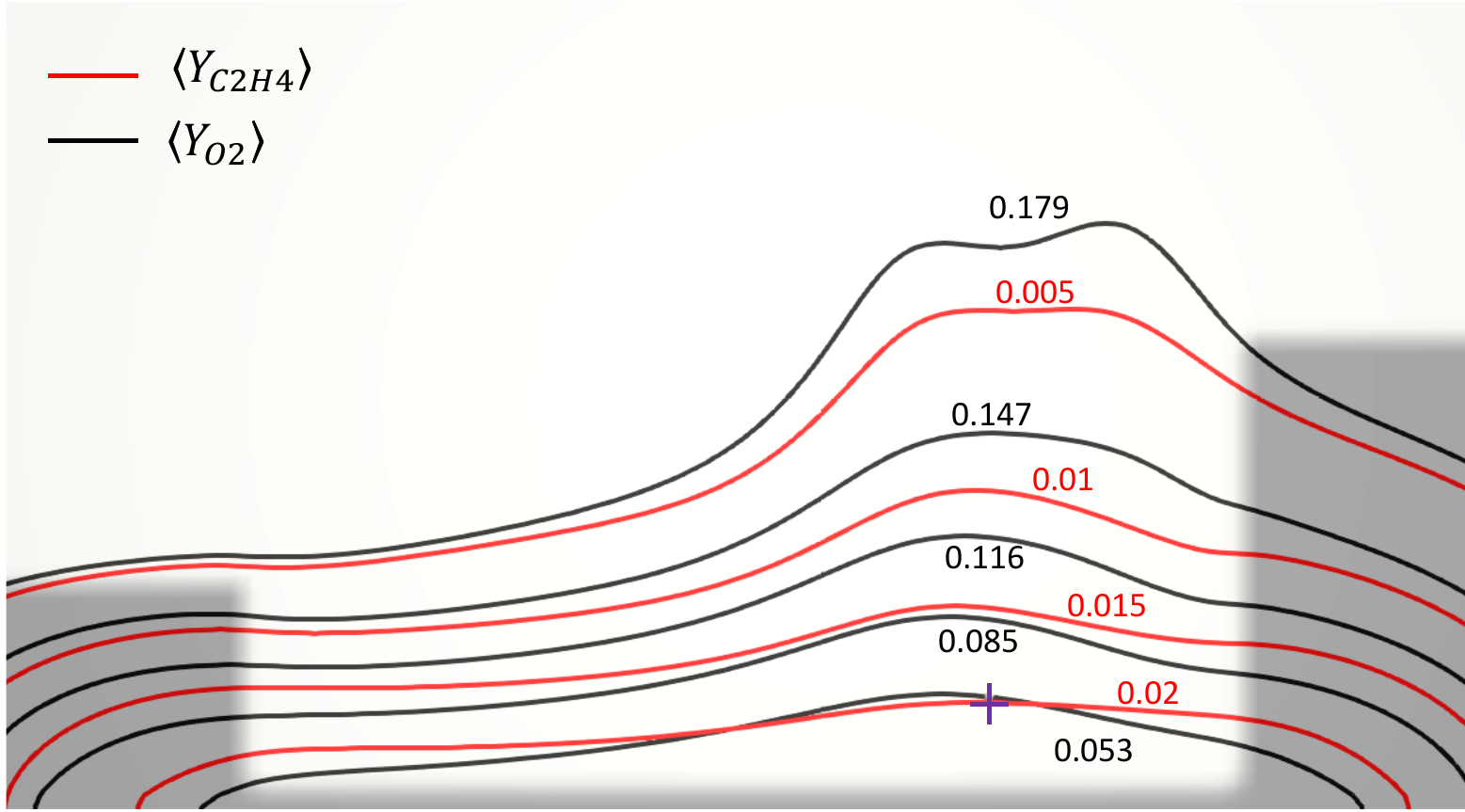}%
\caption{Case A}
\label{fig:contours_a}
\end{subfigure} 

\begin{subfigure}[b]{0.5\textwidth}%
\centering
\captionsetup{width=\linewidth}%
\includegraphics[width=\linewidth]{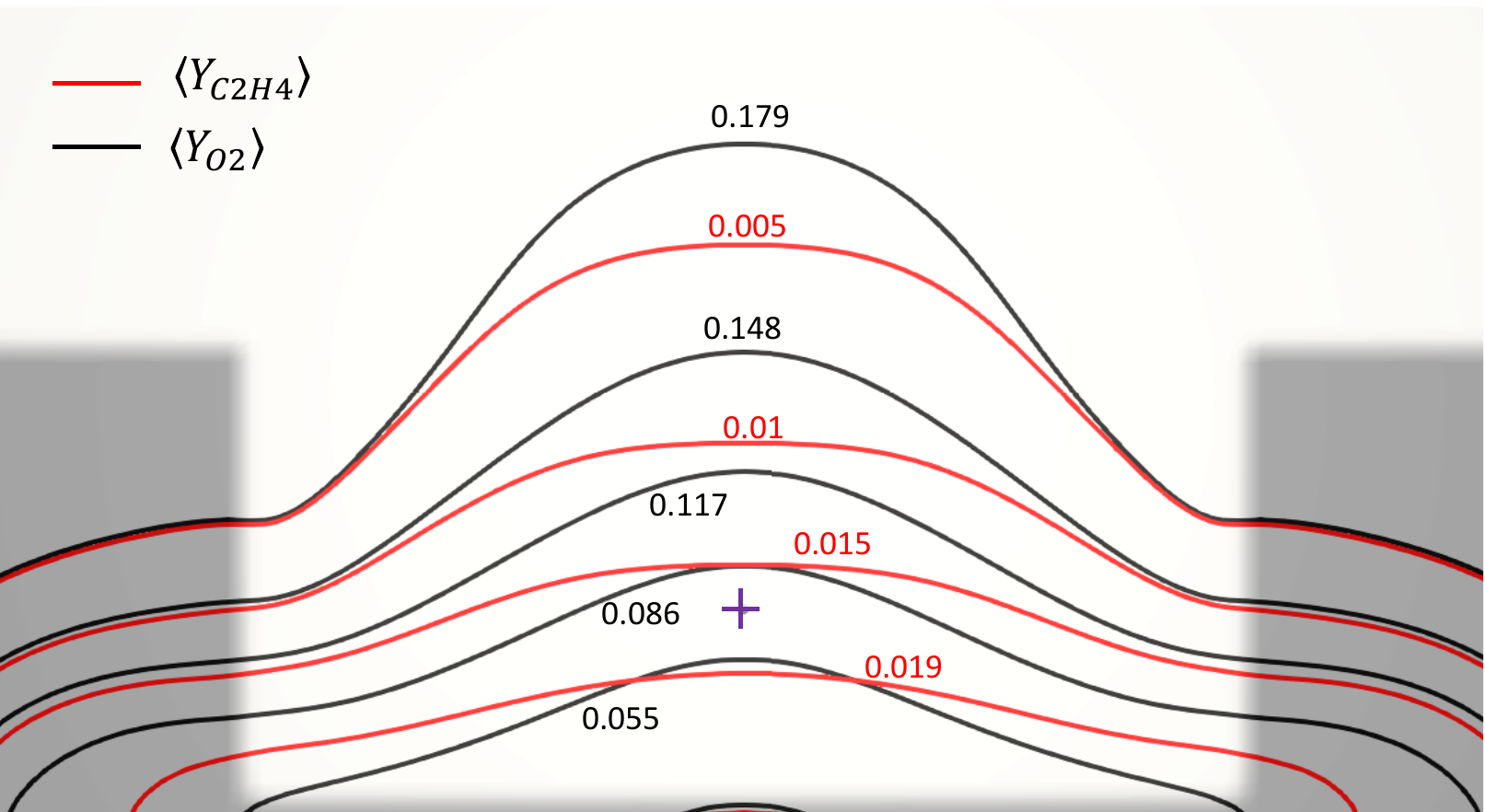}%
\caption{Case B}
\label{fig:contours_b}
\end{subfigure}

\caption{\small Contours of species mass fractions at the time of ignition. The plus sign indicates the ignition location.}
\label{fig:contours}
\end{figure}

\paragraph{Spatial variability and ignition location.} Examining the heat release rate due to reactions ($\dot Q$) at the ignition time (Figure~\ref{fig:qdot_field}), the two cases differ in that, in case B, ignition occurs at the centerline, while in case A, it is shifted to the side and closer to the surface. Figure~\ref{fig:spatial_variability} shows the variations of instantaneous and time-averaged velocities and species mass fractions across the horizontal line passing the ignition point. Interestingly, temperature and heat release rate appear to correlate with time-averaged quantities, and even more so with the fluctuating characteristics of the flow (resolved Reynolds stresses). The mass fractions of the fuel and oxidizer seem to control the vertical location of the ignition (Figure~\ref{fig:contours}). As seen in the figure, in both cases, the ignition location falls somewhere within the band corresponding to the same mass fraction ranges of fuel (\ce{C2H4}) and oxidizer (\ce{O2})---in case A, the flow field deforms this band, pushing it towards the side.

\begin{figure*}[htbp]
\centering
\begin{subfigure}[b]{.45\textwidth}%
\centering\captionsetup{width=\linewidth}%
\includegraphics[width=\linewidth]{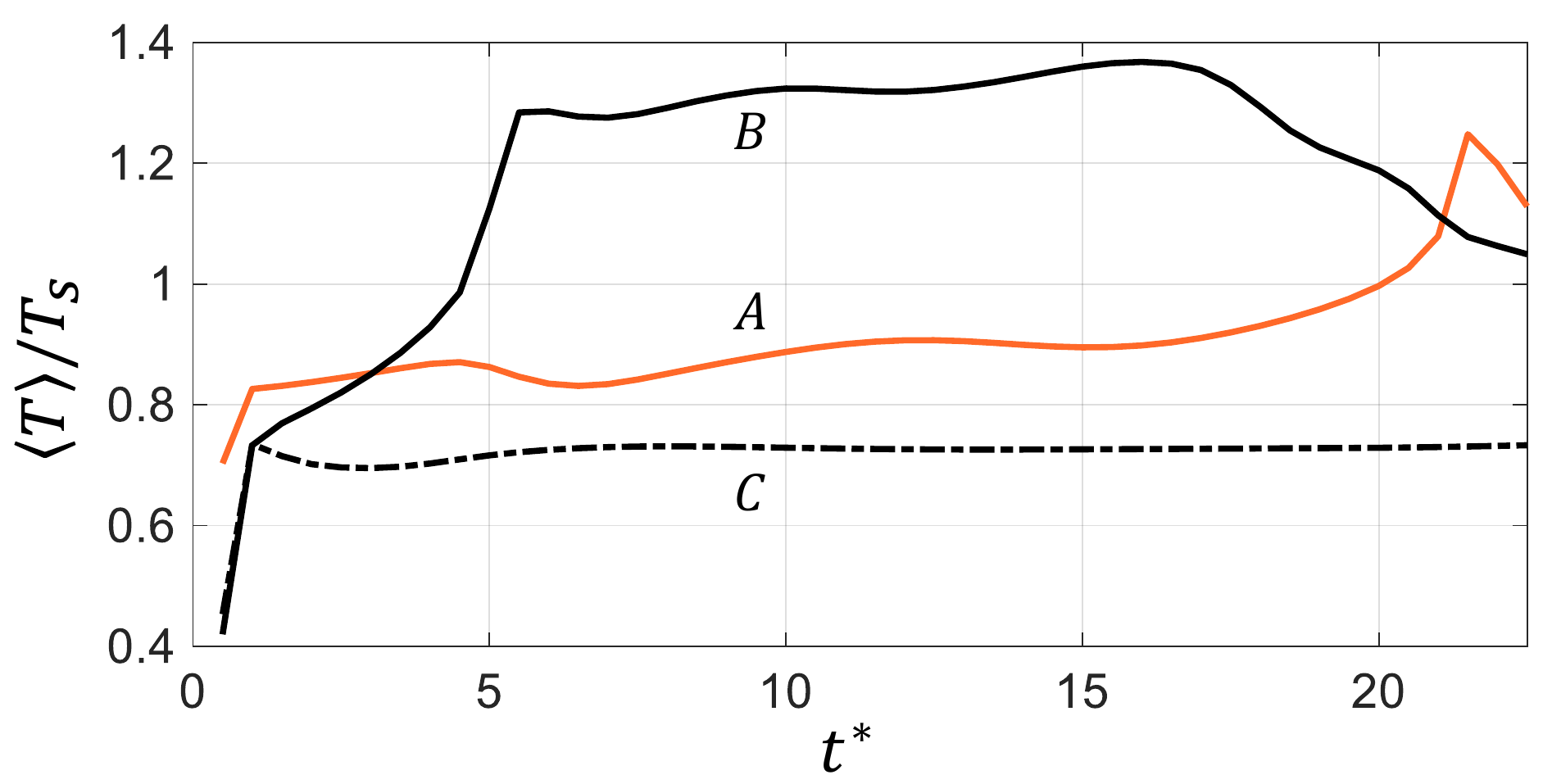}%
\caption{Temperature}
\label{fig:temporal_temp}
\end{subfigure} 
~
\begin{subfigure}[b]{.45\textwidth}%
    \centering\captionsetup{width=\linewidth}%
\includegraphics[width=\linewidth]{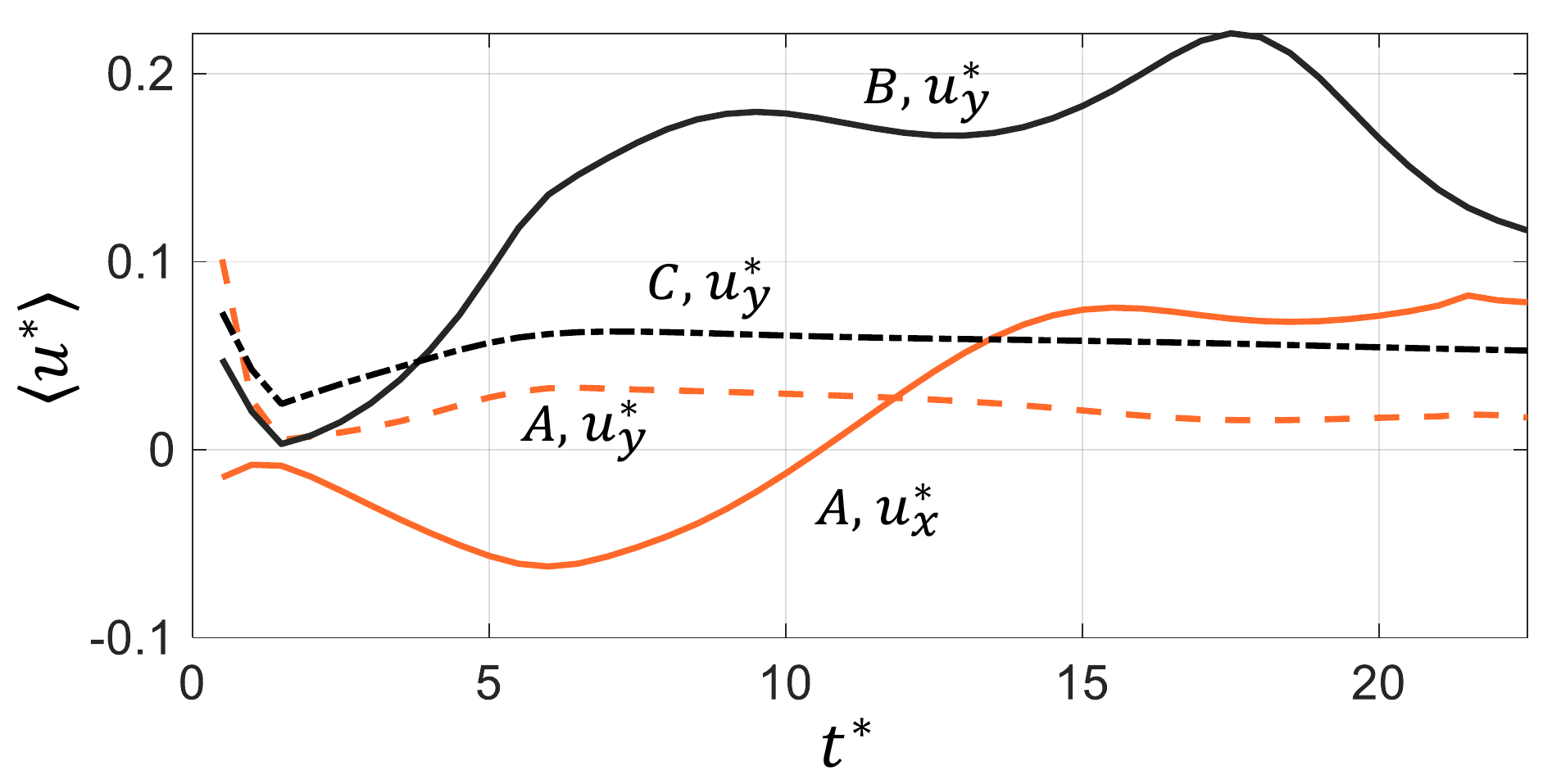}%
\caption{Velocity}
\label{fig:temportal_velocity}
\end{subfigure}

\begin{subfigure}[b]{.45\textwidth}%
    \centering\captionsetup{width=\linewidth}%
\includegraphics[width=\linewidth]{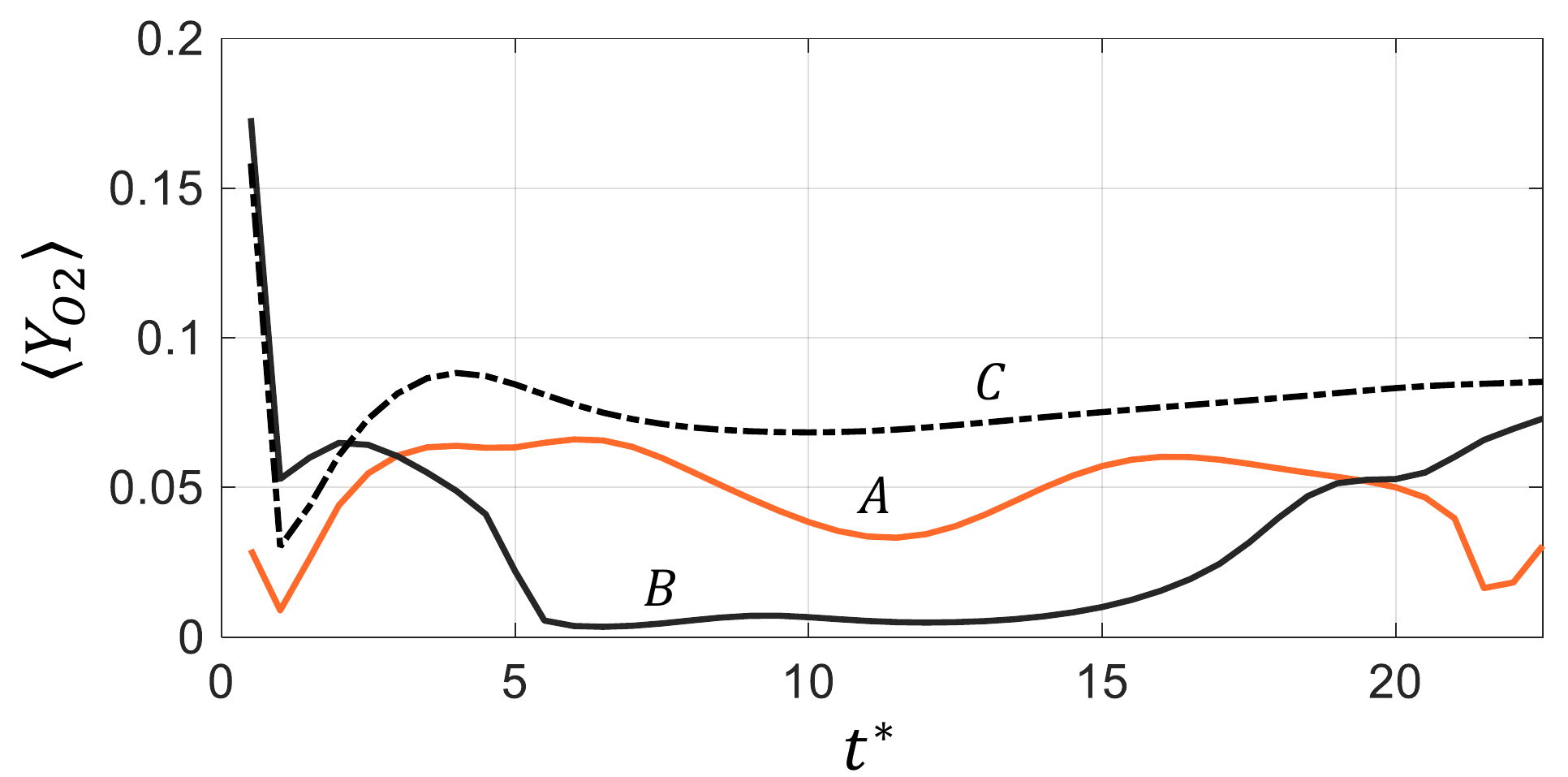}%
\caption{\ce{O2} mass fraction}
\label{fig:temporal_o2}
\end{subfigure}
~
\begin{subfigure}[b]{.45\textwidth}%
    \centering\captionsetup{width=\linewidth}%
\includegraphics[width=\linewidth]{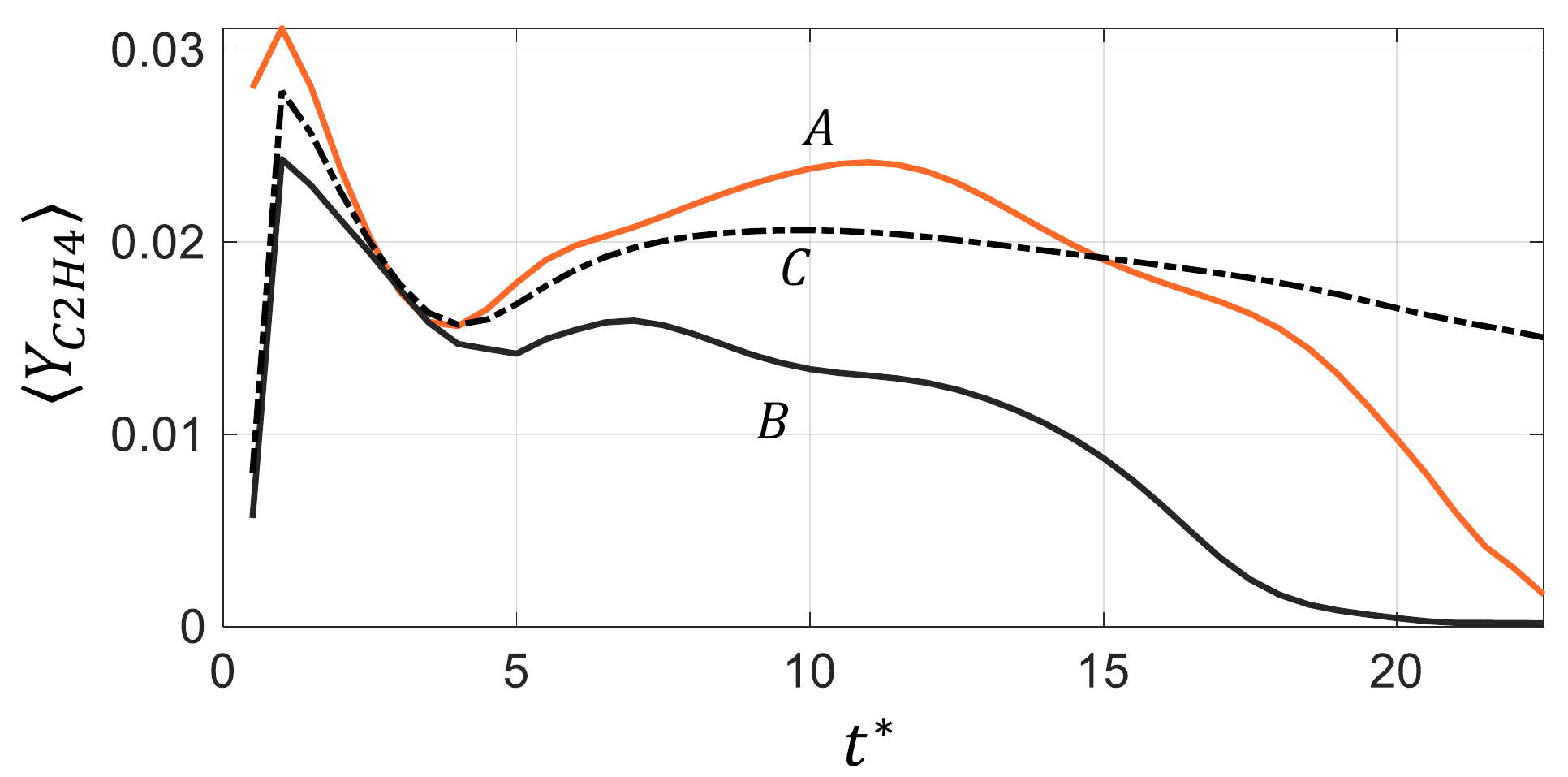}%
\caption{\ce{C2H4} mass fraction}
\label{fig:temporal_c2h4}
\end{subfigure}

\begin{subfigure}[b]{.45\textwidth}%
    \centering\captionsetup{width=\linewidth}%
\includegraphics[width=\linewidth]{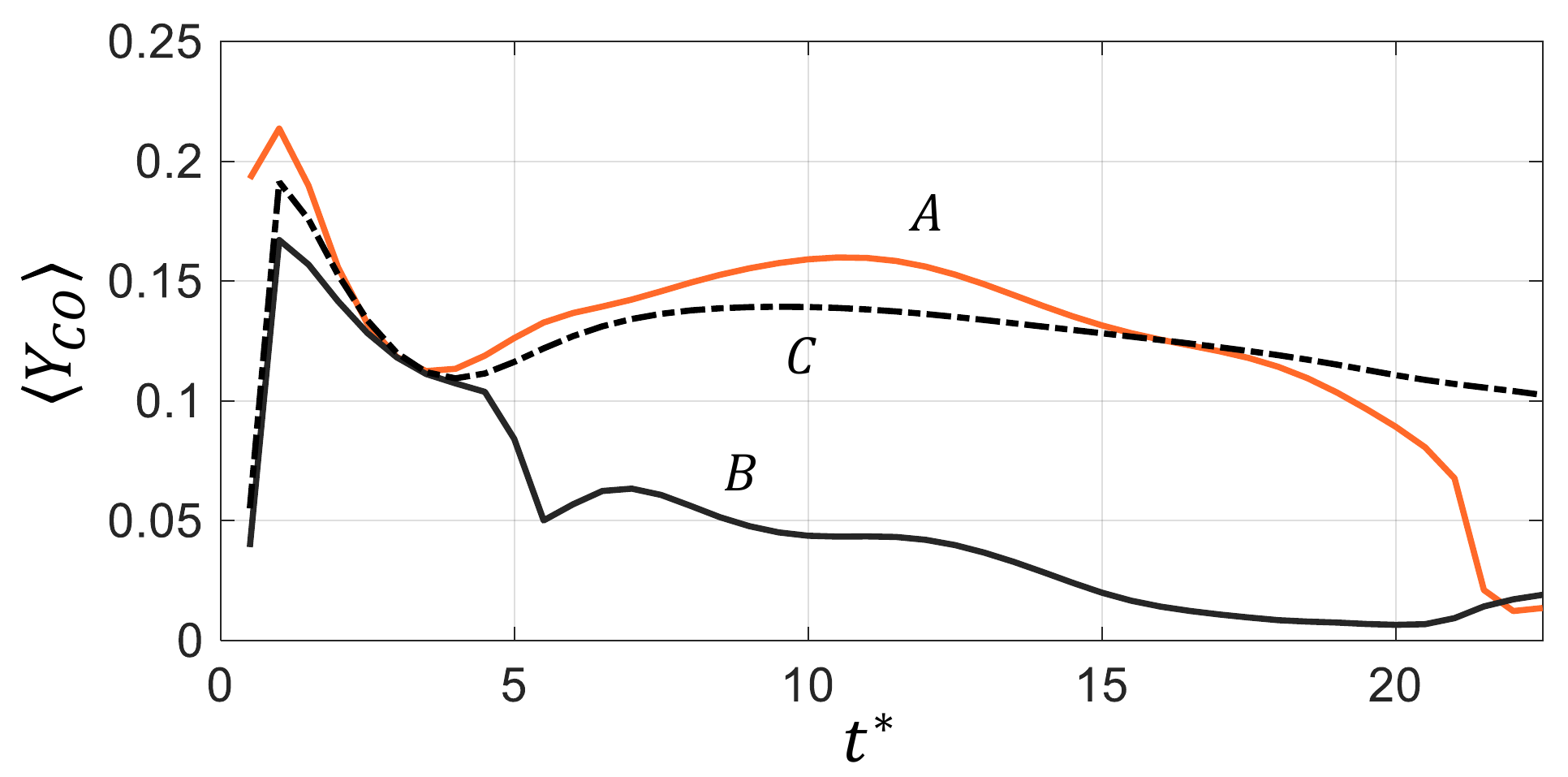}%
\caption{\ce{CO} mass fraction}
\label{fig:temporal_co}
\end{subfigure}
~
\begin{subfigure}[b]{.45\textwidth}%
    \centering\captionsetup{width=\linewidth}%
\includegraphics[width=\linewidth]{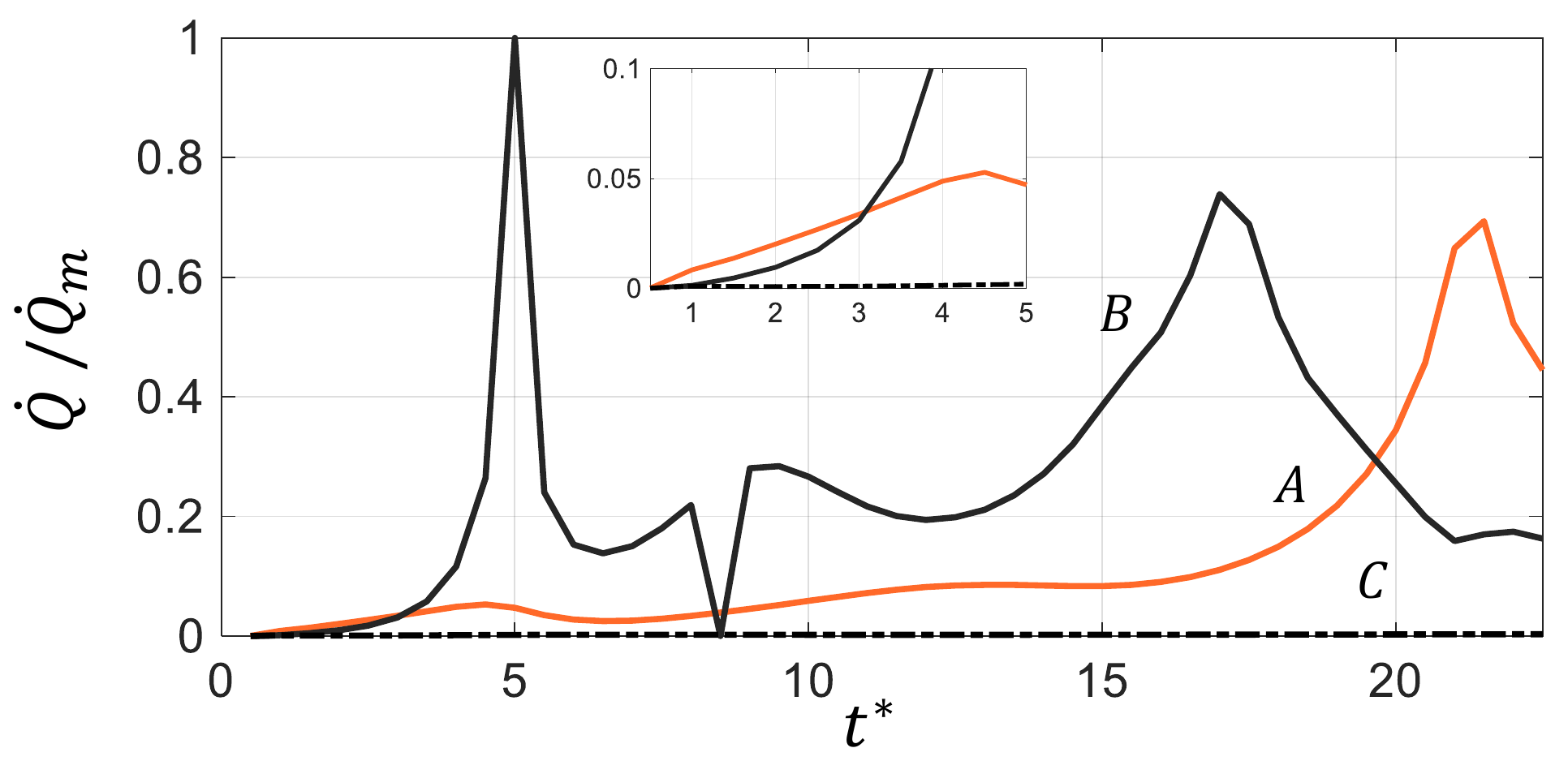}%
\caption{Chemical heat release rate}
\label{qdot}
\end{subfigure}

\caption{\small Evolution of variables at ($x/L=0.55, y/L=0.025$) for case A and at ($x/L=0.5, y/L=0.045$) for cases B and C}
\label{fig:temporal_variability}
\end{figure*}

\begin{figure*}[htbp]
\centering
\includegraphics[width=0.9\linewidth]{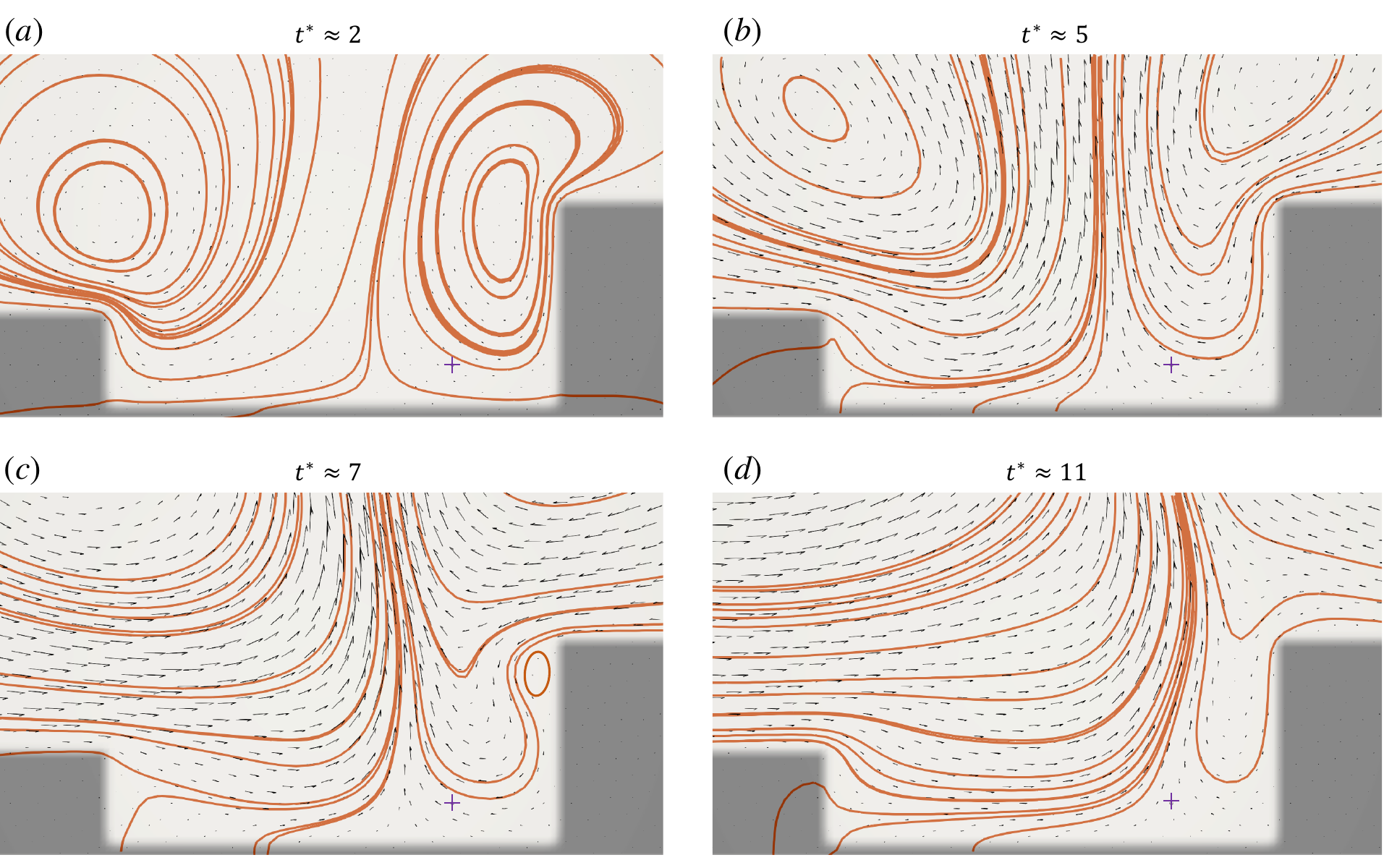}
\caption{\small Fields of velocity vectors and the associated streamlines at different times for case A. The plus sign denotes ignition location.}
\label{fig:streamlines}
\end{figure*}

\begin{figure*}[htbp]
\centering
\begin{subfigure}[b]{.45\textwidth}%
\centering\captionsetup{width=\linewidth}%
\includegraphics[width=\linewidth]{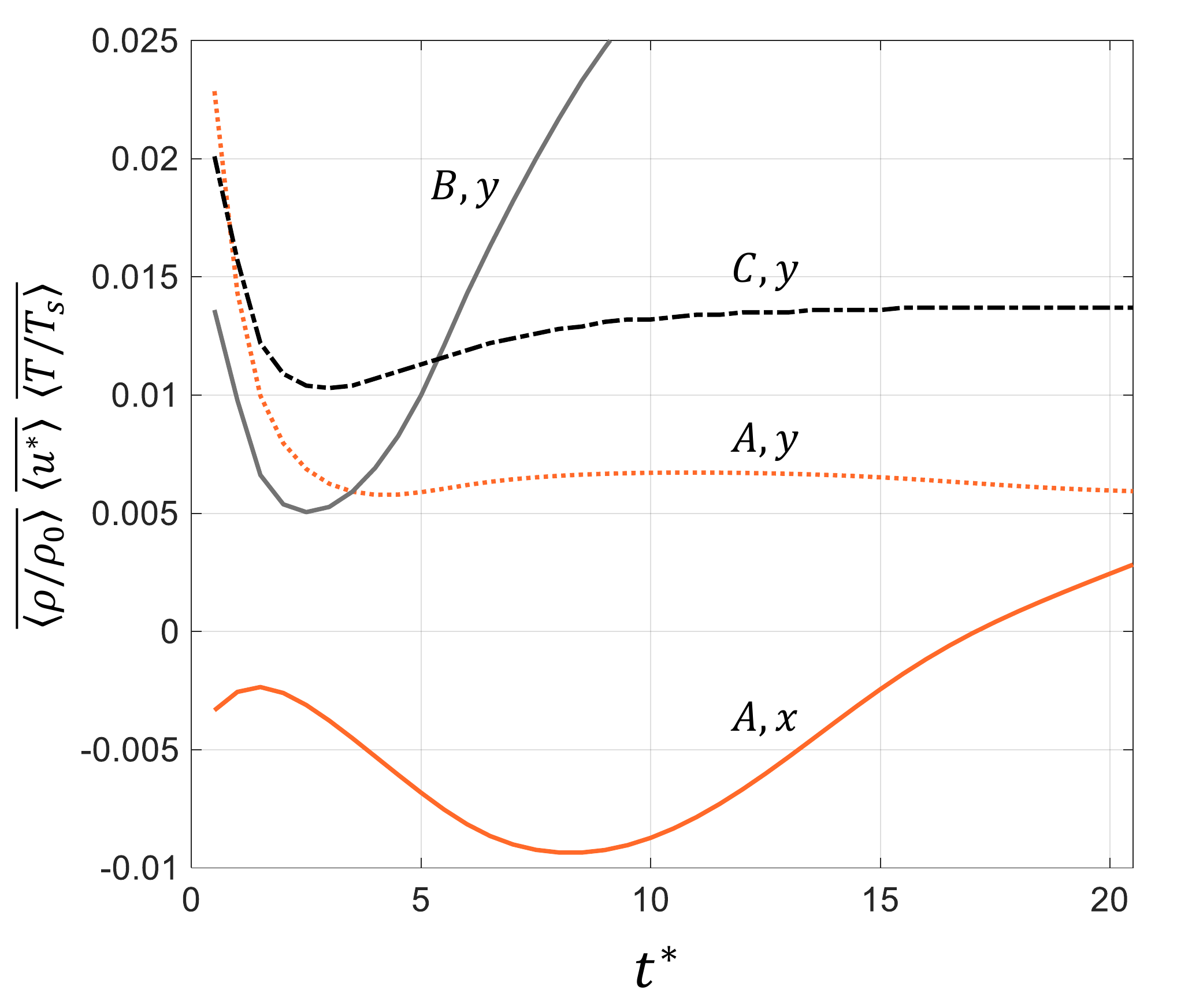}%
\caption{}
\label{fig:temp_umtm}
\end{subfigure} 
~
\begin{subfigure}[b]{.45\textwidth}%
    \centering\captionsetup{width=\linewidth}%
\includegraphics[width=\linewidth]{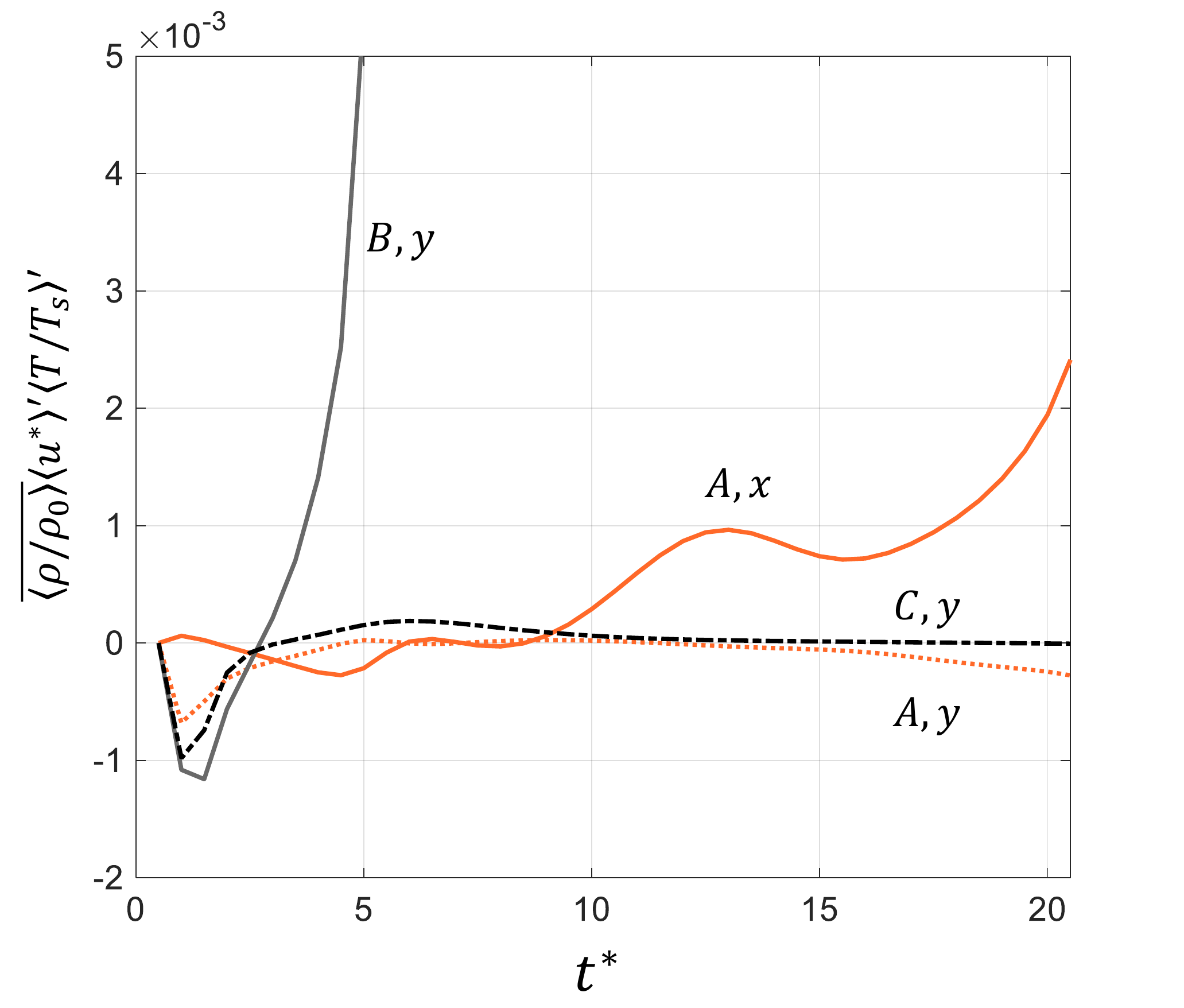}%
\caption{}
\label{fig:temp_tpup}
\end{subfigure}

\caption{\small Evolution of $\overline{\langle \rho/\rho_0\rangle} \Bar{\langle\textbf{u}^*\rangle}\overline{\langle T/T_s\rangle}$ and $\overline{\langle \rho/\rho_0\rangle} \langle\textbf{u}^*\rangle' \langle T/T_s\rangle'$ at ($x/L=0.55, y/L=0.025$) for case A and at ($x/L=0.5, y/L=0.045$) for cases B and C. $x$ and $y$ specify the velocity component that the fluxes are based on.}
\label{fig:tpup}
\end{figure*}

\begin{figure*}[htbp]
\centering
\begin{subfigure}[b]{0.45\textwidth}%
\centering\captionsetup{width=\linewidth}%
\includegraphics[width=\linewidth]{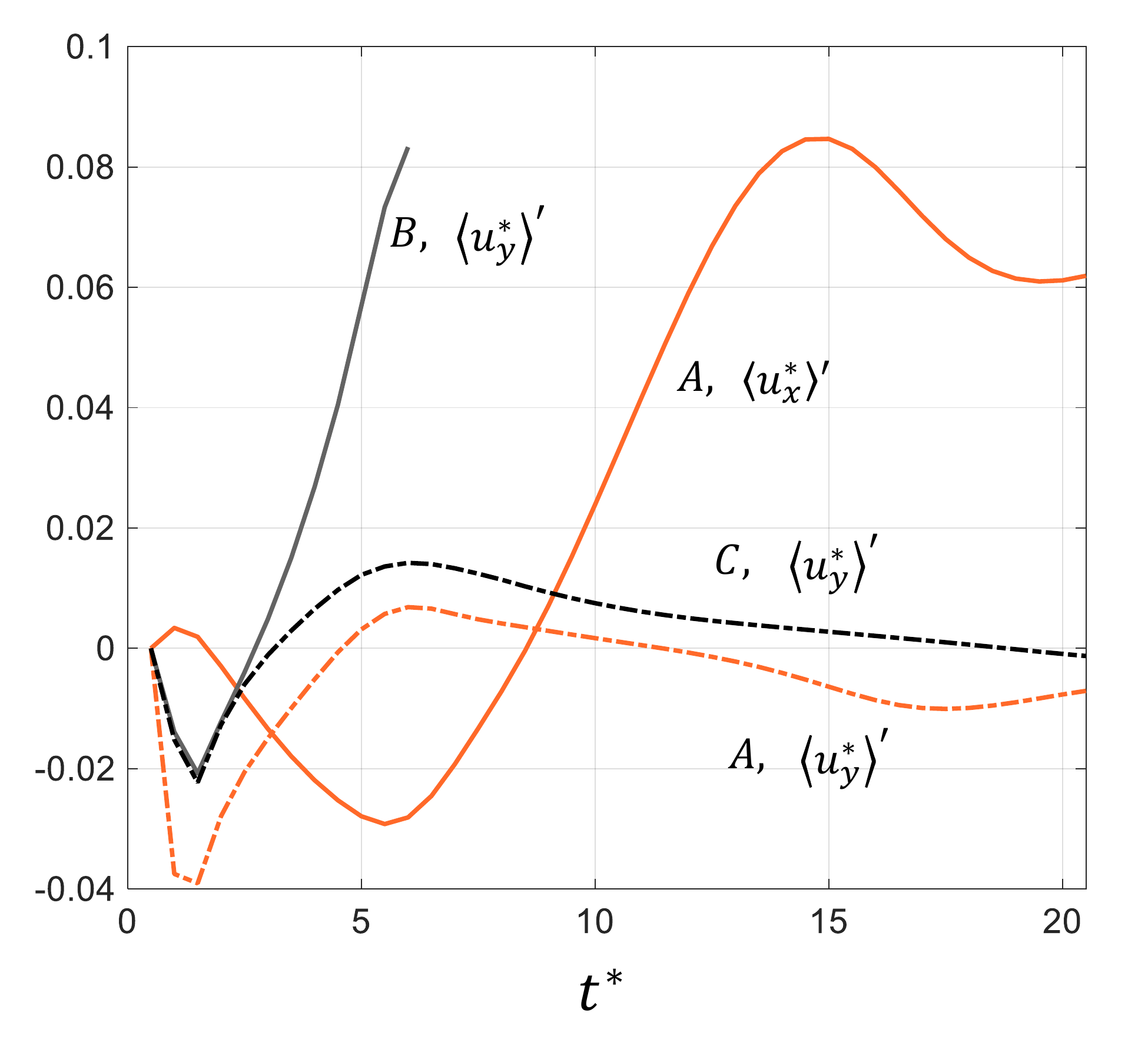}%
\caption{}
\label{}
\end{subfigure} 
~
\begin{subfigure}[b]{.45\textwidth}%
    \centering\captionsetup{width=\linewidth}%
\includegraphics[width=\linewidth]{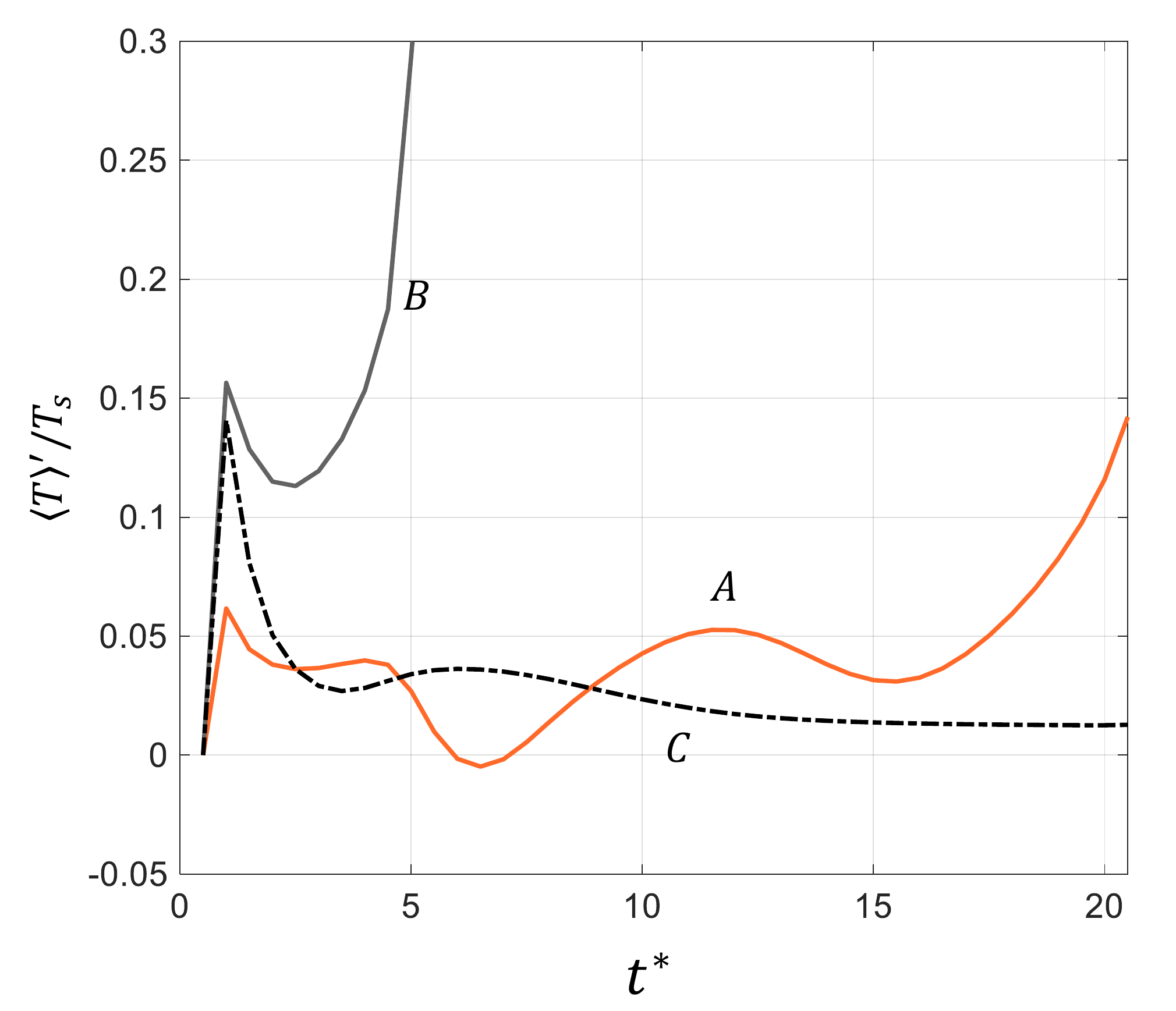}%
\caption{}
\label{}
\end{subfigure}

\caption{\small Evolution of (a) $\langle\textbf{u}^*\rangle'$ and (b) $\langle T\rangle'$ at ($x/L=0.55, y/L=0.025$) for case A and at ($x/L=0.5, y/L=0.045$) for cases B and C }
\label{fig:tp&up}
\end{figure*}

\begin{figure}[htbp]
\centering
\includegraphics[width=.5\linewidth]{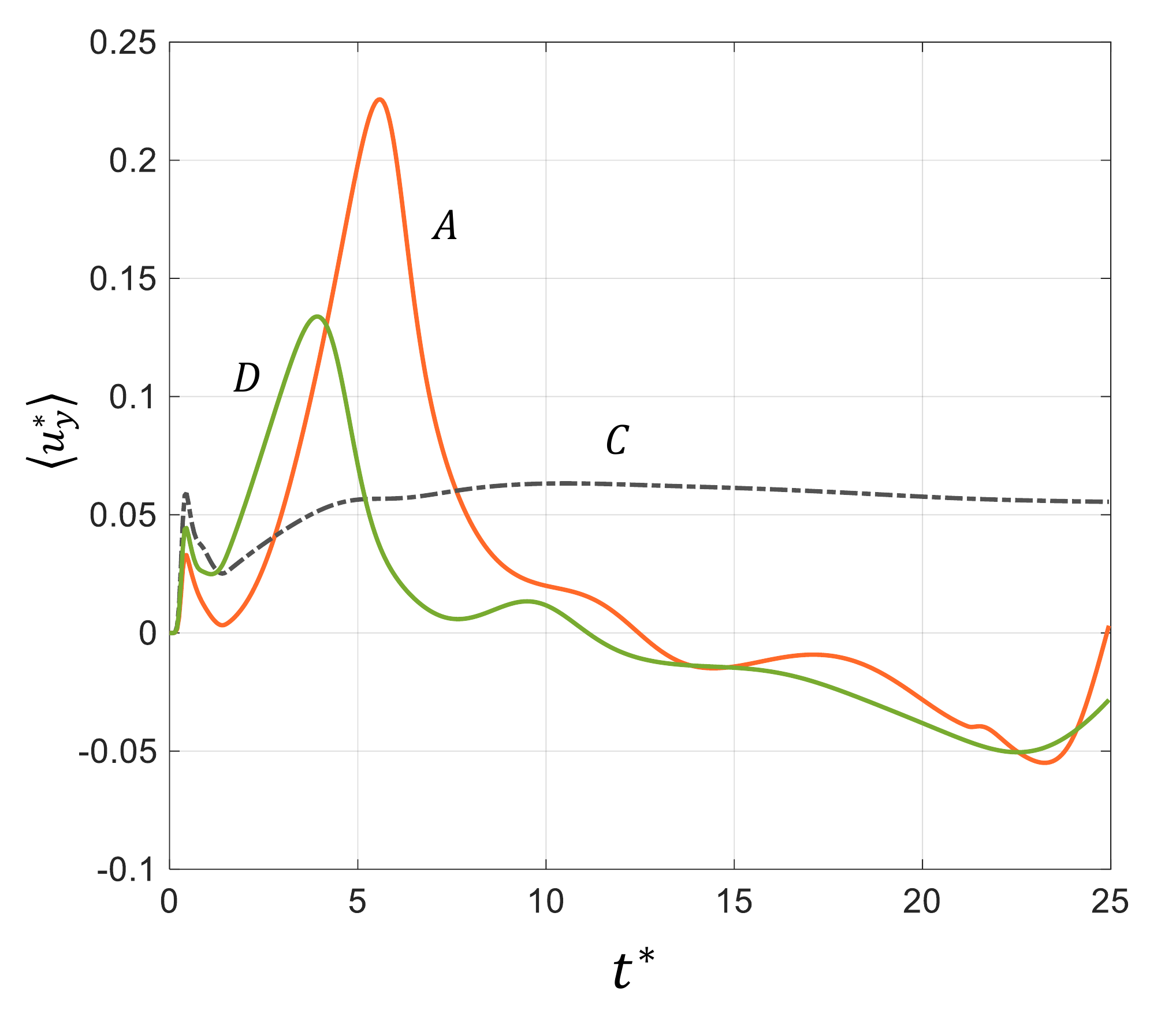}
\caption{\small Evolution of $\langle u_y^* \rangle$ at ($x/L=0.5, y/L=0.08$).}
\label{cavity_size}
\end{figure}

\paragraph{Temporal variability.} Figure~\ref{fig:temporal_variability} shows the time series of temperature, velocity, and mass fractions of \ce{C2H4}, \ce{CO}, and \ce{O2} at the ignition locations of cases A and B (and at the same location for case C). Case A exhibits the most oscillatory behavior, as seen in the velocity profiles, while case C results in more uniform profiles. In Case A, the velocity and temperature appear to oscillate at approximately the same frequency, although their oscillations are not synchronized. The same is true for the oscillations of \ce{O2} mass fraction and velocity, but these are more synchronized. Figure~\ref{fig:streamlines} presents a closer view of flow behavior during an early time window. As mentioned before, the imbalance in the vortex growth causes the plume to initially lean to one side; this is countered by the formation of vortices near the opposite wall, and the upward displacement of these vortices shifts the balance to the other side again. The result is a flapping motion, which manifests as oscillations in the horizontal velocity ($\langle u_x \rangle$). This effect has also been previously observed by other researchers \cite{cetegen1997measurements, meehan2022, cetegen2000experiments}.

The point with the sharp rise in $\dot Q$ (and $T$) indicates the time of ignition. This occurs at $t^*\approx 4.5$ for case B. Around this time, the temperature for case A begins to drop, preventing it from reaching the critical temperature necessary for triggering the release of high heat due to reactions. Since $\dot Q$ and species mass fractions all show an increasing trend at this point (case A), the heat loss must be due to convective fluxes (as conduction and dissipation are considered negligible). To investigate this, we take a closer look at the convective dynamics. In Figure~\ref{fig:spatial_variability}, decomposing the variables into static and fluctuating components in a time-averaged sense revealed some correlations between the temperature and flow statistics.
It is, therefore, reasonable to expect that $\Bar{\langle\textbf{u}\rangle} \Bar{\langle T\rangle}$ and $\langle\textbf{u}\rangle' \langle T\rangle'$ variations may provide insights into interactions between heat transfer and flow patterns. The time-averaged energy equation justifies the causal relationship between these variables:
\begin{gather}
\diffp{}{t} \left(\overline{{\langle \rho \rangle \langle h \rangle}} \right) + \nabla\cdot\left(\Bar{\langle \rho \rangle} \Bar{\langle h \rangle} \Bar{\langle \textbf{u} \rangle}\right) = -\nabla\cdot\left(\Bar{\langle \rho \rangle} \overline{\langle h\rangle' \langle\textbf{u}\rangle'}\right) 
+ \Bar{\dot Q} \;, 
\end{gather}
where $\overline{(\cdot)}$ denotes an ensemble averaging operation over time and $(\cdot)'$ a fluctuation from the ensemble average. 

Figures~\ref{fig:tpup} and \ref{fig:tp&up} show time series of $\overline{\langle \rho / \rho_0 \rangle} \Bar{\langle\textbf{u}^*\rangle} \overline{\langle T/T_s \rangle}$,  $\overline{\langle \rho/\rho_0\rangle} \langle\textbf{u}^*\rangle'\langle T/T_s\rangle'$, $\langle\textbf{u}^*\rangle'$, and $\langle T/T_s\rangle'$. As a plume rises, it entrains surrounding fluid into the plume core. This entrainment process can dilute the momentum of the rising fluid, slowing down the velocity and decreasing the temperature (depending on the distance from the plume core) which can lead to negative values of 
$\langle\textbf{u}\rangle' \langle T\rangle'$, and therefore, a decrease in the associated convective heat fluxes. Looking at cases A $\Bar{\langle \rho\rangle} \Bar{\langle u_y \rangle} \Bar{\langle T\rangle}$ is positive; however, it monotonically decreases up to a point near $t^*\approx 4.5$; this is modulated by the negative values of $\Bar{\langle \rho\rangle} \Bar{\langle u_x \rangle} \Bar{\langle T\rangle}$; combined with negative $\Bar{\langle \rho\rangle} \langle u_x \rangle' \langle T\rangle'$ and $\Bar{\langle \rho\rangle} \langle u_y \rangle' \langle T\rangle'$ in that time window, these effects explain the temperature drop observed at $t^*\approx 4.5$. 
Conversely, in case B, $\Bar{\langle \rho\rangle} \Bar{\langle u_y \rangle} \Bar{\langle T\rangle}$ shows a minimum at $t^*\approx 2.5$ but continuously increases after that point; similarly, $\Bar{\langle \rho\rangle} \langle u_y \rangle' \langle T\rangle'$ is negative until $t^*\approx 2.5$ but begins to increase with positive values after that point, leading to a rise in temperature to the critical value necessary for ignition. 
For case A, $\Bar{\langle \rho\rangle} \Bar{\langle\ u_x \rangle} \Bar{\langle T\rangle}$, $\Bar{\langle \rho\rangle} \langle u_x \rangle' \langle T\rangle'$ and $\Bar{\langle \rho\rangle} \langle u_y \rangle' \langle T\rangle'$ all become positive after $t^*\approx 17.5$, explaining the ignition at $t^*\approx 20$. 
Comparing cases C and B, with symmetric geometries, they show similar trends but both the rate of increase and the values of $\Bar{\langle \rho\rangle} \Bar{\langle u_y \rangle} \Bar{\langle T\rangle}$ and $\Bar{\langle \rho\rangle} \langle u_y \rangle' \langle T\rangle'$ are considerably lower in case C, explaining the absence of ignition. 
Another consideration for case C is the considerably lower values of $\dot Q$ (Figure~\ref{qdot}), which reduces the temperature growth at early times compared to case B, despite their similar temperatures at $t^*\approx 1$. 
This should result from lower mass fractions of the oxidizer for case C---the gas-phase reaction rates are functions of temperature and mass factions of fuel and oxidizer, with higher sensitivity to $Y_{\ce{O2}}$, as the reaction orders are higher with respect to \ce{O2}.

From another perspective, understanding the correlation between velocity and temperature has traditionally been a goal in compressible turbulence research, aiming to elucidate the complex nonlinear coupling between thermal and velocity fields, particularly in boundary layers. For example, Morkovin’s hypothesis postulates that high-speed turbulence structure in zero pressure-gradient turbulent boundary layers remains largely the same as its incompressible counterpart \cite{zhang2018direct}. A consequence of this hypothesis is the analogy between the temperature and velocity fields that leads to velocity-temperature relations such as the ones suggested by the Generalized Reynolds Analogy \cite{zhang2014generalized} and Huang's model \cite{huang1995compressible}. These reveal that the turbulence-induced wall-normal (or vertical) transfer of heat closely resembles that of momentum and suggest that the coefficient of correlation between temperature and velocity fluctuations ($R_{u_y' T'}$) is affected by the sign and magnitude of $\left(\Bar{\rho} \overline{T' u_y'}\right)/\left(\Bar{\rho} \overline{u_x' u_y'}\right)$ or equivalently $\partial\Bar{T}/\partial\Bar{u_y}$. For a rising plume, the ejections of the high-speed/high-temperature fluid and the sweeps of the low-speed/high-temperature fluid would result in opposite signs of $\left(\Bar{\rho} \overline{T' u_y'}\right)$ and $\left(\Bar{\rho} \overline{u_x' u_y'}\right)$, therefore a negative $R_{u_y' T'}$ (consistent with the results in Figure~\ref{fig:tpup} at early times). Additionally, the relatively constant distribution of $R_{u_y' T'}$ is due to small variations of $\Bar{u_y}$ while $\Bar{T}$ would reach maximum near the location where $R_{u_y' T'}$ (or $R_{u_y' u_x'}$)  has a sign transition. This is consistent with results in Figure~\ref{fig:spatial_variability} where the change in the sign of $\overline{u_y' u_x'}$ coincides with a maximum in temperature.

\begin{figure}
\centering
\includegraphics[width=0.5\linewidth]{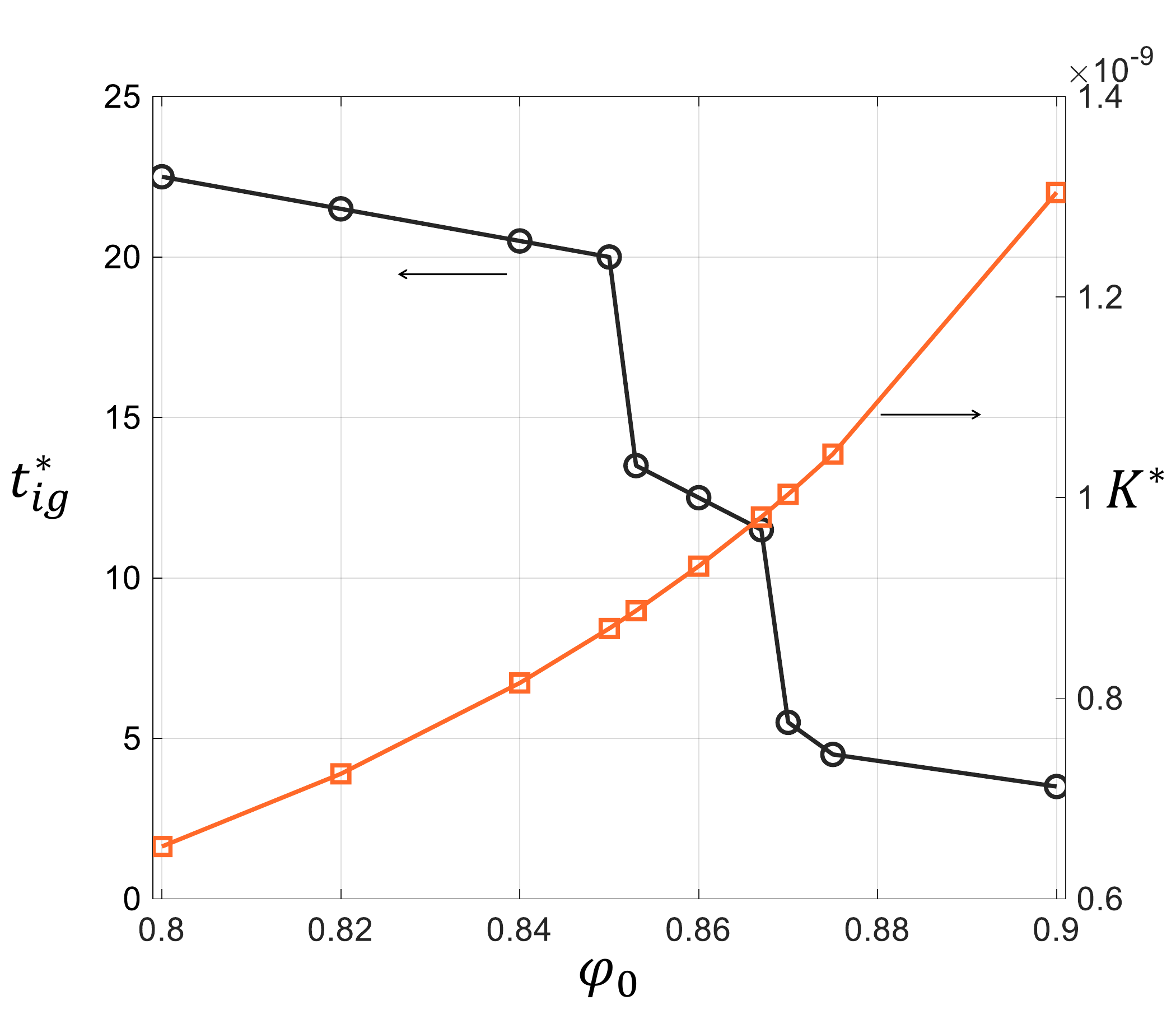}
\caption{\small \revise{Effect of initial porosity on nondimensional permeability ($K^*$) and ignition time ($t^*_{ig}$) for case B.}{revTwo}}
\label{fig:porosity}
\end{figure}

\begin{figure*}
\centering
\begin{subfigure}[b]{.32\textwidth}%
\centering\captionsetup{width=\linewidth}%
\includegraphics[width=\linewidth]{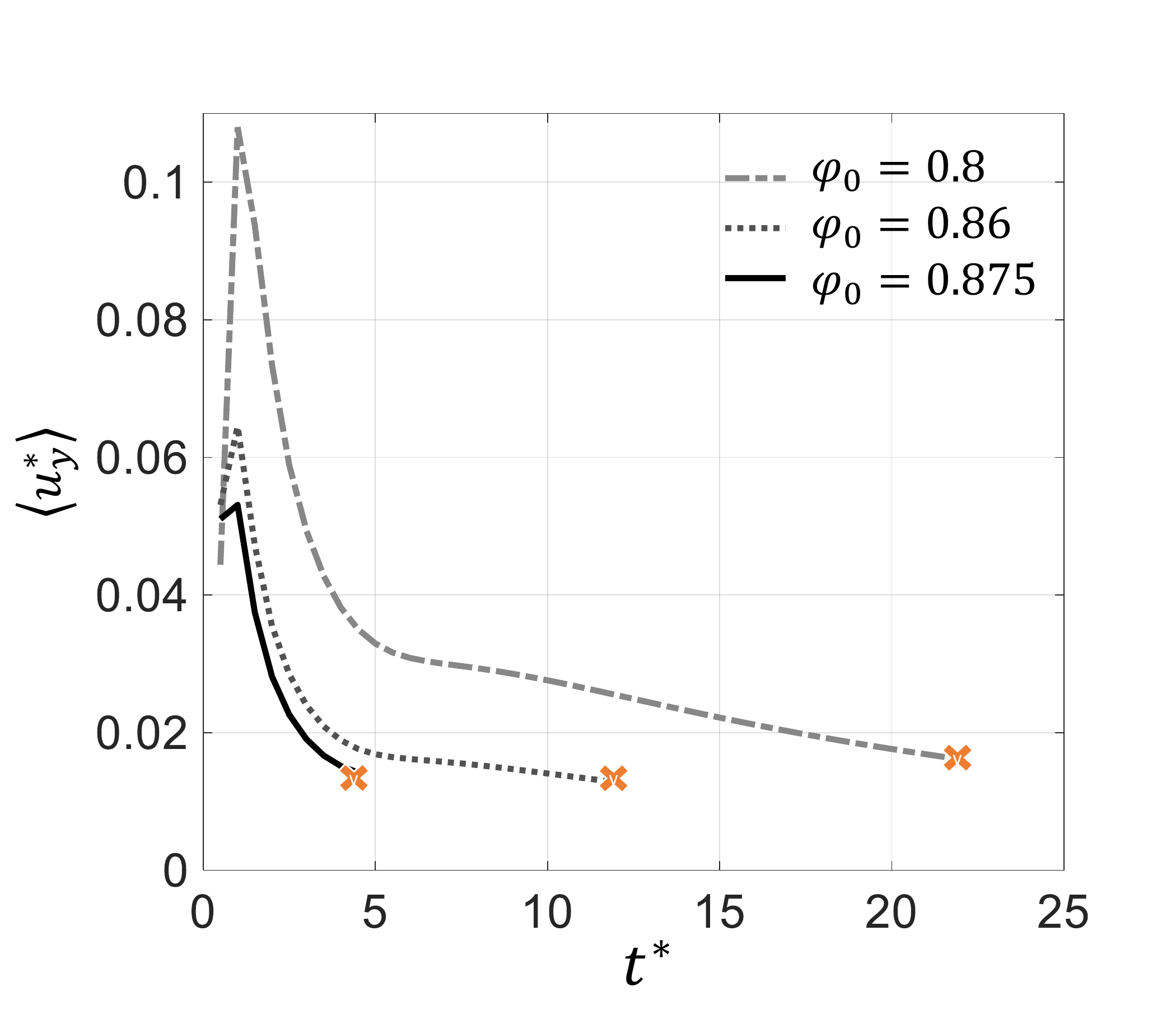}%
\caption{Vertical velocity}
\label{phi_v}
\end{subfigure} 
~
\begin{subfigure}[b]{.32\textwidth}%
    \centering\captionsetup{width=\linewidth}%
\includegraphics[width=\linewidth]{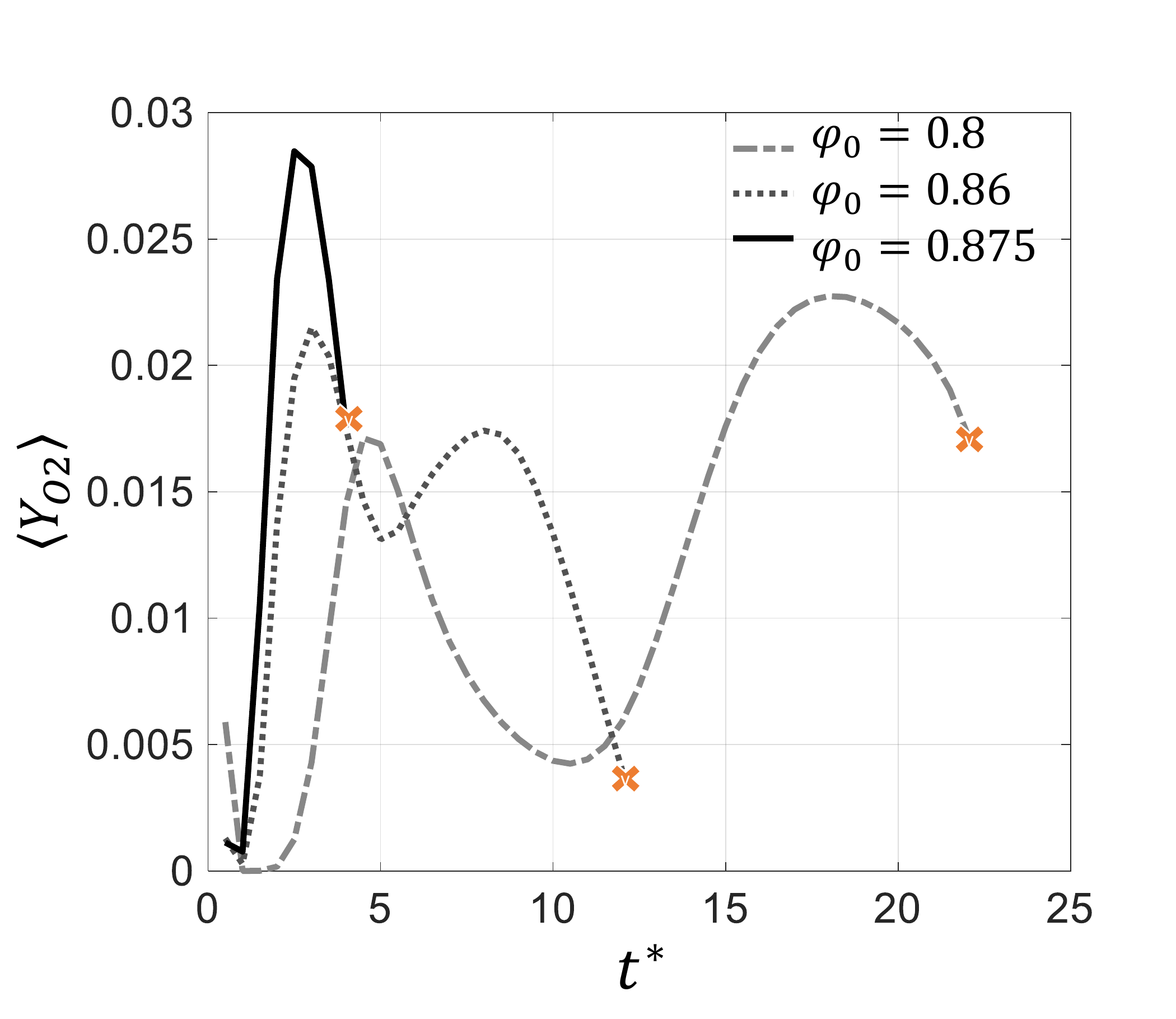}%
\caption{\ce{O2} mass fraction}
\label{phi_o2}
\end{subfigure}
~
\begin{subfigure}[b]{.32\textwidth}%
    \centering\captionsetup{width=\linewidth}%
\includegraphics[width=\linewidth]{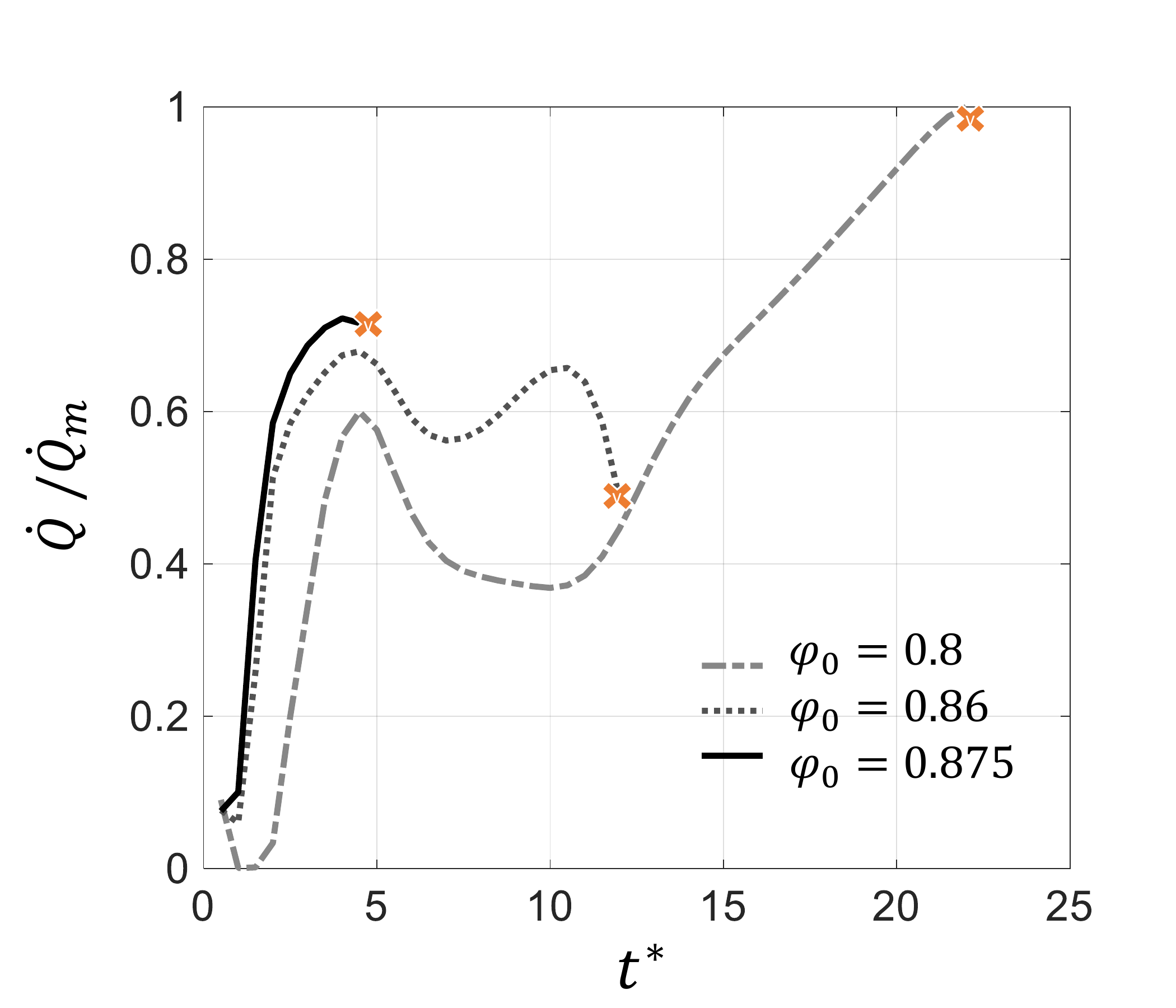}%
\caption{Chemical heat release rate}
\label{phi_qdot}
\end{subfigure}
\caption{\small \revise{Evolution of variables at a point on the interface ($x/L=0.5,\ y/L=H/3$) for cases B with different initial porosities. The cross sign indicates the ignition point.}{revTwo}}
\label{fig:phi_investigation}
\end{figure*}

\paragraph{\revise{Effect of initial porosity.}{revTwo}}\revise{Just as geometric characteristics affect ignition, physical properties of the solid material, such as porosity, can disrupt the fragile conditions necessary for ignition. 
Figure~\ref{fig:porosity} shows how changing the initial porosity ($\phi_0$) of the solid region impacts nondimensional permeability and ultimately ignition time, for case B.
Ignition time varies nonlinearly with initial porosity, with a sharp increase as porosity drops from the original value (0.875).}{revTwo}

\revise{The sharp rise in the ignition time occurring at the porosity values between 0.85 and 0.87 calls for additional investigation.
Figure~\ref{fig:phi_investigation} shows how vertical velocity ($\langle u_y \rangle$), \ce{O2} mass fraction ($\langle Y_{\ce{O2}} \rangle$), and normalized chemical heat release rate ($\dot{Q}/\dot{Q}_m$) vary with time for different porosity values in this range.
From Figure~\ref{phi_v}, vertical velocity increases nonlinearly as the material becomes less porous (which is more obvious at the peaks); this surge in velocity (for the lower $\phi_0$ values) results from a higher production rate of pyrolysis gases and temporarily depletes the surface region of the oxidizer (Figure~\ref{phi_o2}), whose variations correlate with those of chemical heat release rate (Figure~\ref{phi_qdot}). 
A competing effect is the decrease in permeability as the material becomes less porous, but since the porosities are relatively high in the entire range, this effect is less significant.
As mentioned before, the gas-phase reaction rates are sensitive to $Y_{\ce{O2}}$ and therefore ignition is delayed until this quantity reaches sufficiently high values.
}{revTwo}

\paragraph{Cavity (or gap) width effect.} As seen in Figure~\ref{fig:density_field}, doubling the width of the gap between the middle features increases the size of the plume in both symmetric and asymmetric cases, inflating it to approximately twice its size in some instances (e.g., case D $\rightarrow$ A at $t^* = 10$). This is also reflected in the velocity profiles in Figure~\ref{cavity_size}, where $\langle u_y^* \rangle$ in case A peaks at approximately twice its value in case D. Bres et al.~\cite{bres2008three} and others who studied the centrifugal instabilities in a cavity flow made similar observations that the recirculating vortical flow inside the cavity exhibits characteristics similar to a solid-body rotation away from the walls, with velocity and circulation along the streamlines increasing
linearly with the distance to the center of rotation. Furthermore, conservation of mass dictates a higher velocity in the plume core for the wider cavity as the larger hot boundary at the bottom supplies a higher influx of hot gases, which then concentrate in a narrower region as the instabilities develop. The variation trends, however, are similar for both cavity widths with the plume core moving away from the centerline and occasional fluctuations due to the previously mentioned flapping motion. The observations in this study are limited to the early time window up to the ignition time of case A; different flow behaviors may occur after that point. 

\section{Conclusions} 

In this study, we investigated a comprehensive model for coupling the processes in porous solid and surrounding fluid at different flow regimes, using the single domain approach. 
The model considers a set of spatially averaged conservation equations including the compressible momentum balance, two energy conservation equations for heat transfer in the solid matrix and interstitial fluid in the porous region, and multi-species transport along with detailed chemical kinetics. 
The capabilities of the model in merging the two regions allow more-accurately predicting variables associated with surface events such as ignition time and location. 

We verified the model's performance on non-reacting laminar and turbulent flow cases and showed that the resulting velocity profiles match well with direct numerical simulation data.
We further validated the model against a set of experimental data for the pyrolysis and subsequent ignition of wooden spheres in a combustion chamber in the presence of a forced flow. 
The predicted ignition times agree closely with the experimental measurements. 

Next, we applied the model to the emission of the buoyant reacting plumes from the surface of a heated solid fuel. 
We identified hints of Rayleigh--Taylor and Kelvin--Helmholtz flow instabilities and showed that the shear layer near the interface leads to viscous torque and vorticity generation, associated with Kelvin--Helmholtz instabilities. As expected, a combination of the baroclinic and viscous terms drives the growth of observed vortical structures.

The modeling approach also facilitates the study of plume and fire spread in more-complicated situations such as multi-object settings. To demonstrate this, we extended the study of buoyant reacting plumes and analyzed the effect of interface morphology on near-field momentum and energy transport, as well as the resulting ignition. As expected, the presence of geometrical features considerably impact the coupled flow and combustion behaviors. Ignition is suppressed in cases with narrower gap regions while the asymmetry in the cases with a wider gap alters the ignition time and location by causing the plume to lean to one side. Furthermore, asymmetric cases exhibit the most oscillatory behavior, whereas symmetric cases result in more uniform profiles. These oscillations initially cause some negative heat fluxes, which prevent the temperature from reaching the critical level necessary to trigger the release of high heat due to reactions. These behaviors have implications for supplying oxidizer to the fuel and fire propagation, and, on a larger scale, for the exchange of emissions between forests and the atmosphere, as well as the dispersion of pollutants in cities, where plant canopies and densely built-up urban areas can be modeled as porous media.

While this work focused on the early-time developments, different flow behaviors may occur beyond that point as the instabilities expand and potentially drive a flow regime transition. We have preliminary evidence of phenomena such as vortex shedding near the shorter feature in the asymmetric case, leading to the formation of a second plume branch later on. Additionally, further investigation is needed to understand how different variables are spatially correlated, such as the impact of roughness features upstream or downstream of the heated surface on velocity or temperature fields over time, which can be explored using techniques like two-point spatial correlations. Finally, cases with persistent flaming combustion coupled with flow instabilities warrant a dedicated study.

\begin{acknowledgments}

This project was funded by a contract from the Strategic Environmental Research and Development 
Program (SERDP), project number RC19-1092. 
The authors would also like to thank Sourabh Apte for providing guidance and helpful discussions. 

\end{acknowledgments}

\section*{Data Availability Statement}
The software implementation of the model is available openly via GitHub under the GNU General Public License version 3 (GPL-3.0), and the version used here was archived on Zenodo~\cite{software}.

\appendix
\section{Radiation source term} \label{radiation}
Radiation heat transfer is based on the radiative transfer equation (RTE) \cite{kaviany2012}, further simplified using the $P_1$ approximation approach \cite{modest2013radiative}. In this method, all radiating surfaces are considered diffuse, with the directional dependence in RTE integrated out. For porous media, there are two contributions to the RTE: the gas mixture, modeled here as gray gas, and the solid phase, assumed as gray body. RTE in the form of incident radiation, $G$, is:
\begin{equation}
    \nabla \cdot (\Gamma \nabla G) + 4\pi \left(a \frac{\sigma T^4}{\pi} + E_s \right) - (a+a_s)G = 0 \;,
\end{equation}
where $a$ is the gas mixture absorption coefficient, $E_s$ is the equivalent emission of the solid and $a_s$ is the equivalent absorption coefficient of the solid. $\Gamma$ is defined as:
\begin{equation}
    \Gamma = \frac{1}{3(a+a_s+\sigma_{\text{scatter}}) + C\sigma_{\text{scatter}}} \;,
\end{equation}
where $\sigma_{\text{scatter}}$ is the scattering coefficient and $C$ is the coefficient associated with the forward/backward scattering. $E_s$ is evaluated as (assuming the Kirchhoff law holds):
\begin{equation*}
    \Gamma = a_s \frac{\sigma T_s^4}{\pi} \;.
\end{equation*}
These result in the following source terms for the gas-phase and solid-phase energy equations:
\begin{align}
    S^{f, \text{radiation}} &= - 4\pi \left(a \frac{\sigma T^4}{\pi}\right) + aG\\
    S^{s, \text{radiation}} &= - 4\pi (E_s) + a_sG \;.
\end{align}
The model also assumes that $a_s$ is constant across the porous medium but its surface additionally interacts with the incident radiation within a certain distance, defined by the user.

\begin{figure}
\centering
\includegraphics[width=.5\linewidth]{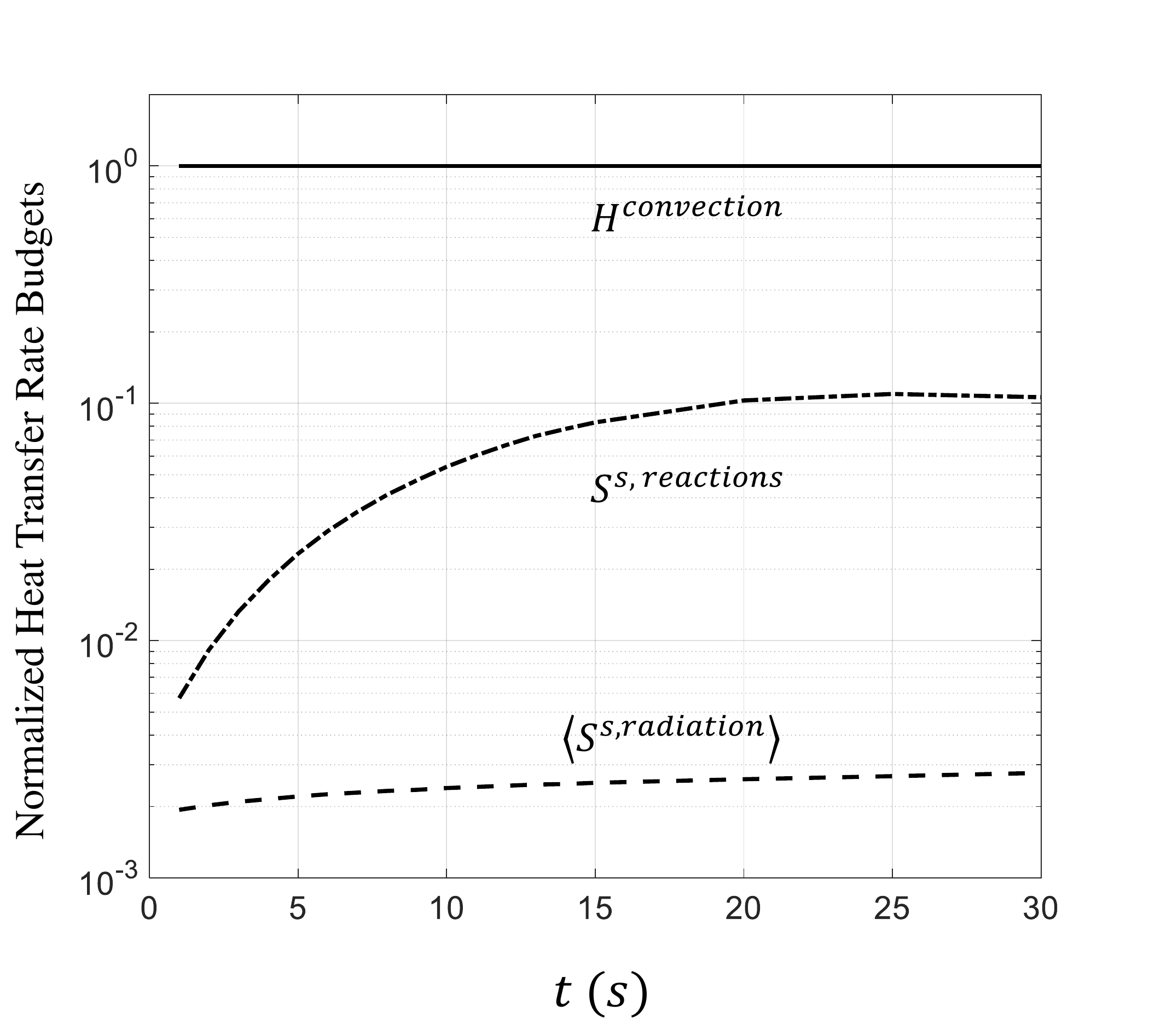}
\caption{\small \revise{Budgets in the solid-phase energy equation: $S^{\text{reactions}}$ and $H^{\text{convection}}$ indicate the second and third terms in Eq.~\eqref{eq:solid_energy}. The values are normalized by $H^{\text{convection}}$.}{revTwo}}
\label{fig:radiation}
\end{figure}

\revise{Figure~\ref{fig:radiation} demonstrates the relative contribution of radiation heat transfer compared with contributions from convection and chemical reactions in Eq.~\ref{eq:solid_energy}, based on simulation data for the oak sphere combustion case in Section~\ref{subsec:sphere}.
Radiation contributes nearly two orders of magnitude less than chemical reactions, and nearly three orders of magnitude less than convection. Thus, it plays a fairly minor role in these problems.}{revTwo}

\bibliography{aipsamp}

\providecommand{\noopsort}[1]{}\providecommand{\singleletter}[1]{#1}%
\begin{thebibliography}{44}%
\makeatletter
\providecommand \@ifxundefined [1]{%
 \@ifx{#1\undefined}
}%
\providecommand \@ifnum [1]{%
 \ifnum #1\expandafter \@firstoftwo
 \else \expandafter \@secondoftwo
 \fi
}%
\providecommand \@ifx [1]{%
 \ifx #1\expandafter \@firstoftwo
 \else \expandafter \@secondoftwo
 \fi
}%
\providecommand \natexlab [1]{#1}%
\providecommand \enquote  [1]{``#1''}%
\providecommand \bibnamefont  [1]{#1}%
\providecommand \bibfnamefont [1]{#1}%
\providecommand \citenamefont [1]{#1}%
\providecommand \href@noop [0]{\@secondoftwo}%
\providecommand \href [0]{\begingroup \@sanitize@url \@href}%
\providecommand \@href[1]{\@@startlink{#1}\@@href}%
\providecommand \@@href[1]{\endgroup#1\@@endlink}%
\providecommand \@sanitize@url [0]{\catcode `\\12\catcode `\$12\catcode
  `\&12\catcode `\#12\catcode `\^12\catcode `\_12\catcode `\%12\relax}%
\providecommand \@@startlink[1]{}%
\providecommand \@@endlink[0]{}%
\providecommand \url  [0]{\begingroup\@sanitize@url \@url }%
\providecommand \@url [1]{\endgroup\@href {#1}{\urlprefix }}%
\providecommand \urlprefix  [0]{URL }%
\providecommand \Eprint [0]{\href }%
\providecommand \doibase [0]{https://doi.org/}%
\providecommand \selectlanguage [0]{\@gobble}%
\providecommand \bibinfo  [0]{\@secondoftwo}%
\providecommand \bibfield  [0]{\@secondoftwo}%
\providecommand \translation [1]{[#1]}%
\providecommand \BibitemOpen [0]{}%
\providecommand \bibitemStop [0]{}%
\providecommand \bibitemNoStop [0]{.\EOS\space}%
\providecommand \EOS [0]{\spacefactor3000\relax}%
\providecommand \BibitemShut  [1]{\csname bibitem#1\endcsname}%
\let\auto@bib@innerbib\@empty
\bibitem [{\citenamefont {Savard}\ \emph {et~al.}(2015)\citenamefont {Savard},
  \citenamefont {Xuan}, \citenamefont {Bobbitt},\ and\ \citenamefont
  {Blanquart}}]{savard2015}%
  \BibitemOpen
  \bibfield  {author} {\bibinfo {author} {\bibfnamefont {B.}~\bibnamefont
  {Savard}}, \bibinfo {author} {\bibfnamefont {Y.}~\bibnamefont {Xuan}},
  \bibinfo {author} {\bibfnamefont {B.}~\bibnamefont {Bobbitt}},\ and\ \bibinfo
  {author} {\bibfnamefont {G.}~\bibnamefont {Blanquart}},\ }\bibfield  {title}
  {\enquote {\bibinfo {title} {A computationally-efficient, semi-implicit,
  iterative method for the time-integration of reacting flows with stiff
  chemistry},}\ }\href {https://doi.org/10.1016/j.jcp.2015.04.018} {\bibfield
  {journal} {\bibinfo  {journal} {Journal of Computational Physics}\ }\textbf
  {\bibinfo {volume} {295}},\ \bibinfo {pages} {740--769} (\bibinfo {year}
  {2015})}\BibitemShut {NoStop}%
\bibitem [{\citenamefont {Motheau}\ and\ \citenamefont
  {Abraham}(2016)}]{motheau2016}%
  \BibitemOpen
  \bibfield  {author} {\bibinfo {author} {\bibfnamefont {E.}~\bibnamefont
  {Motheau}}\ and\ \bibinfo {author} {\bibfnamefont {J.}~\bibnamefont
  {Abraham}},\ }\bibfield  {title} {\enquote {\bibinfo {title} {A high-order
  numerical algorithm for {DNS} of low-{Mach}-number reactive flows with
  detailed chemistry and quasi-spectral accuracy},}\ }\href
  {https://doi.org/10.1016/j.jcp.2016.02.059} {\bibfield  {journal} {\bibinfo
  {journal} {Journal of Computational Physics}\ }\textbf {\bibinfo {volume}
  {313}},\ \bibinfo {pages} {430--454} (\bibinfo {year} {2016})}\BibitemShut
  {NoStop}%
\bibitem [{\citenamefont {Fillo}\ \emph {et~al.}(2020)\citenamefont {Fillo},
  \citenamefont {Schlup}, \citenamefont {Beardsell}, \citenamefont
  {Blanquart},\ and\ \citenamefont {Niemeyer}}]{Fillo2020}%
  \BibitemOpen
  \bibfield  {author} {\bibinfo {author} {\bibfnamefont {A.~J.}\ \bibnamefont
  {Fillo}}, \bibinfo {author} {\bibfnamefont {J.}~\bibnamefont {Schlup}},
  \bibinfo {author} {\bibfnamefont {G.}~\bibnamefont {Beardsell}}, \bibinfo
  {author} {\bibfnamefont {G.}~\bibnamefont {Blanquart}},\ and\ \bibinfo
  {author} {\bibfnamefont {K.~E.}\ \bibnamefont {Niemeyer}},\ }\bibfield
  {title} {\enquote {\bibinfo {title} {A fast, low-memory, and stable algorithm
  for implementing multicomponent transport in direct numerical simulations},}\
  }\href {https://doi.org/10.1016/j.jcp.2019.109185} {\bibfield  {journal}
  {\bibinfo  {journal} {Journal of Computational Physics}\ }\textbf {\bibinfo
  {volume} {406}},\ \bibinfo {pages} {109185} (\bibinfo {year}
  {2020})}\BibitemShut {NoStop}%
\bibitem [{\citenamefont {Desai}\ \emph {et~al.}(2021)\citenamefont {Desai},
  \citenamefont {Kim}, \citenamefont {Song}, \citenamefont {Luong},
  \citenamefont {P{\'e}rez}, \citenamefont {Sankaran},\ and\ \citenamefont
  {Im}}]{desai2021}%
  \BibitemOpen
  \bibfield  {author} {\bibinfo {author} {\bibfnamefont {S.}~\bibnamefont
  {Desai}}, \bibinfo {author} {\bibfnamefont {Y.~J.}\ \bibnamefont {Kim}},
  \bibinfo {author} {\bibfnamefont {W.}~\bibnamefont {Song}}, \bibinfo {author}
  {\bibfnamefont {M.~B.}\ \bibnamefont {Luong}}, \bibinfo {author}
  {\bibfnamefont {F.~E.~H.}\ \bibnamefont {P{\'e}rez}}, \bibinfo {author}
  {\bibfnamefont {R.}~\bibnamefont {Sankaran}},\ and\ \bibinfo {author}
  {\bibfnamefont {H.~G.}\ \bibnamefont {Im}},\ }\bibfield  {title} {\enquote
  {\bibinfo {title} {Direct numerical simulations of turbulent reacting flows
  with shock waves and stiff chemistry using many-core/{GPU} acceleration},}\
  }\href {https://doi.org/10.1016/j.compfluid.2020.104787} {\bibfield
  {journal} {\bibinfo  {journal} {Computers \& Fluids}\ }\textbf {\bibinfo
  {volume} {215}},\ \bibinfo {pages} {104787} (\bibinfo {year}
  {2021})}\BibitemShut {NoStop}%
\bibitem [{\citenamefont {Gentile}\ \emph {et~al.}(2017)\citenamefont
  {Gentile}, \citenamefont {Debiagi}, \citenamefont {Cuoci}, \citenamefont
  {Frassoldati}, \citenamefont {Ranzi},\ and\ \citenamefont {Faravelli}}]{cfd}%
  \BibitemOpen
  \bibfield  {author} {\bibinfo {author} {\bibfnamefont {G.}~\bibnamefont
  {Gentile}}, \bibinfo {author} {\bibfnamefont {P.~E.~A.}\ \bibnamefont
  {Debiagi}}, \bibinfo {author} {\bibfnamefont {A.}~\bibnamefont {Cuoci}},
  \bibinfo {author} {\bibfnamefont {A.}~\bibnamefont {Frassoldati}}, \bibinfo
  {author} {\bibfnamefont {E.}~\bibnamefont {Ranzi}},\ and\ \bibinfo {author}
  {\bibfnamefont {T.}~\bibnamefont {Faravelli}},\ }\bibfield  {title} {\enquote
  {\bibinfo {title} {A computational framework for the pyrolysis of anisotropic
  biomass particles},}\ }\href {https://doi.org/10.1016/j.cej.2017.03.113}
  {\bibfield  {journal} {\bibinfo  {journal} {Chemical Engineering Journal}\
  }\textbf {\bibinfo {volume} {321}},\ \bibinfo {pages} {458--473} (\bibinfo
  {year} {2017})}\BibitemShut {NoStop}%
\bibitem [{\citenamefont {Lautenberger}\ and\ \citenamefont
  {Fernandez-Pello}(2009)}]{lautenberger2009}%
  \BibitemOpen
  \bibfield  {author} {\bibinfo {author} {\bibfnamefont {C.}~\bibnamefont
  {Lautenberger}}\ and\ \bibinfo {author} {\bibfnamefont {C.}~\bibnamefont
  {Fernandez-Pello}},\ }\bibfield  {title} {\enquote {\bibinfo {title}
  {Generalized pyrolysis model for combustible solids},}\ }\href
  {https://doi.org/10.1016/j.firesaf.2009.03.011} {\bibfield  {journal}
  {\bibinfo  {journal} {Fire Safety Journal}\ }\textbf {\bibinfo {volume}
  {44}},\ \bibinfo {pages} {819--839} (\bibinfo {year} {2009})}\BibitemShut
  {NoStop}%
\bibitem [{\citenamefont {Mahiques}\ \emph {et~al.}(2023)\citenamefont
  {Mahiques}, \citenamefont {Br\"{o}mmer}, \citenamefont {Wirtz}, \citenamefont
  {van Wachem},\ and\ \citenamefont {Scherer}}]{Mahiques2023}%
  \BibitemOpen
  \bibfield  {author} {\bibinfo {author} {\bibfnamefont {E.~I.}\ \bibnamefont
  {Mahiques}}, \bibinfo {author} {\bibfnamefont {M.}~\bibnamefont
  {Br\"{o}mmer}}, \bibinfo {author} {\bibfnamefont {S.}~\bibnamefont {Wirtz}},
  \bibinfo {author} {\bibfnamefont {B.}~\bibnamefont {van Wachem}},\ and\
  \bibinfo {author} {\bibfnamefont {V.}~\bibnamefont {Scherer}},\ }\bibfield
  {title} {\enquote {\bibinfo {title} {Simulation of reacting, moving granular
  assemblies of thermally thick particles by discrete element
  method/computational fluid dynamics},}\ }\href
  {https://doi.org/10.1002/ceat.202200520} {\bibfield  {journal} {\bibinfo
  {journal} {Chemical Engineering \& Technology}\ }\textbf {\bibinfo {volume}
  {46}},\ \bibinfo {pages} {1317--1332} (\bibinfo {year} {2023})}\BibitemShut
  {NoStop}%
\bibitem [{\citenamefont {Beavers}\ and\ \citenamefont
  {Joseph}(1967)}]{beavers1967}%
  \BibitemOpen
  \bibfield  {author} {\bibinfo {author} {\bibfnamefont {G.~S.}\ \bibnamefont
  {Beavers}}\ and\ \bibinfo {author} {\bibfnamefont {D.~D.}\ \bibnamefont
  {Joseph}},\ }\bibfield  {title} {\enquote {\bibinfo {title} {Boundary
  conditions at a naturally permeable wall},}\ }\href
  {https://doi.org/10.1017/S0022112067001375} {\bibfield  {journal} {\bibinfo
  {journal} {Journal of Fluid Mechanics}\ }\textbf {\bibinfo {volume} {30}},\
  \bibinfo {pages} {197--207} (\bibinfo {year} {1967})}\BibitemShut {NoStop}%
\bibitem [{\citenamefont {Hahn}, \citenamefont {Je},\ and\ \citenamefont
  {Choi}(2002)}]{hahn2002}%
  \BibitemOpen
  \bibfield  {author} {\bibinfo {author} {\bibfnamefont {S.}~\bibnamefont
  {Hahn}}, \bibinfo {author} {\bibfnamefont {J.}~\bibnamefont {Je}},\ and\
  \bibinfo {author} {\bibfnamefont {H.}~\bibnamefont {Choi}},\ }\bibfield
  {title} {\enquote {\bibinfo {title} {Direct numerical simulation of turbulent
  channel flow with permeable walls},}\ }\href
  {https://doi.org/10.1017/S0022112001006437} {\bibfield  {journal} {\bibinfo
  {journal} {Journal of Fluid Mechanics}\ }\textbf {\bibinfo {volume} {450}},\
  \bibinfo {pages} {259--285} (\bibinfo {year} {2002})}\BibitemShut {NoStop}%
\bibitem [{\citenamefont {Saffman}(1971)}]{saffman1971}%
  \BibitemOpen
  \bibfield  {author} {\bibinfo {author} {\bibfnamefont {P.~G.}\ \bibnamefont
  {Saffman}},\ }\bibfield  {title} {\enquote {\bibinfo {title} {On the boundary
  condition at the surface of a porous medium},}\ }\href
  {https://doi.org/10.1002/sapm197150293} {\bibfield  {journal} {\bibinfo
  {journal} {Studies in Applied Mathematics}\ }\textbf {\bibinfo {volume}
  {50}},\ \bibinfo {pages} {93--101} (\bibinfo {year} {1971})}\BibitemShut
  {NoStop}%
\bibitem [{\citenamefont {Brinkman}(1949)}]{brinkman1949calculation}%
  \BibitemOpen
  \bibfield  {author} {\bibinfo {author} {\bibfnamefont {H.~C.}\ \bibnamefont
  {Brinkman}},\ }\bibfield  {title} {\enquote {\bibinfo {title} {A calculation
  of the viscous force exerted by a flowing fluid on a dense swarm of
  particles},}\ }\href {https://doi.org/10.1007/BF02120313} {\bibfield
  {journal} {\bibinfo  {journal} {Flow, Turbulence and Combustion}\ }\textbf
  {\bibinfo {volume} {1}},\ \bibinfo {pages} {27--34} (\bibinfo {year}
  {1949})}\BibitemShut {NoStop}%
\bibitem [{\citenamefont {Lesieur}\ and\ \citenamefont
  {Metais}(1996)}]{lesieur1996}%
  \BibitemOpen
  \bibfield  {author} {\bibinfo {author} {\bibfnamefont {M.}~\bibnamefont
  {Lesieur}}\ and\ \bibinfo {author} {\bibfnamefont {O.}~\bibnamefont
  {Metais}},\ }\bibfield  {title} {\enquote {\bibinfo {title} {New trends in
  large-eddy simulations of turbulence},}\ }\href
  {https://doi.org/10.1146/annurev.fl.28.010196.000401} {\bibfield  {journal}
  {\bibinfo  {journal} {Annual Review of Fluid Mechanics}\ }\textbf {\bibinfo
  {volume} {28}},\ \bibinfo {pages} {45--82} (\bibinfo {year}
  {1996})}\BibitemShut {NoStop}%
\bibitem [{\citenamefont {Whitaker}(1998)}]{whitaker1998}%
  \BibitemOpen
  \bibfield  {author} {\bibinfo {author} {\bibfnamefont {S.}~\bibnamefont
  {Whitaker}},\ }\href@noop {} {\emph {\bibinfo {title} {The Method of Volume
  Averaging}}},\ Vol.~\bibinfo {volume} {13}\ (\bibinfo  {publisher} {Springer
  Science \& Business Media},\ \bibinfo {year} {1998})\BibitemShut {NoStop}%
\bibitem [{\citenamefont {Bae}\ and\ \citenamefont
  {Moon}(2011)}]{bae2011effect}%
  \BibitemOpen
  \bibfield  {author} {\bibinfo {author} {\bibfnamefont {Y.}~\bibnamefont
  {Bae}}\ and\ \bibinfo {author} {\bibfnamefont {Y.~J.}\ \bibnamefont {Moon}},\
  }\bibfield  {title} {\enquote {\bibinfo {title} {Effect of passive porous
  surface on the trailing-edge noise},}\ }\href
  {https://doi.org/10.1063/1.3662447} {\bibfield  {journal} {\bibinfo
  {journal} {Physics of Fluids}\ }\textbf {\bibinfo {volume} {23}} (\bibinfo
  {year} {2011}),\ 10.1063/1.3662447}\BibitemShut {NoStop}%
\bibitem [{\citenamefont {Sadowski}\ \emph {et~al.}(2023)\citenamefont
  {Sadowski}, \citenamefont {Sayyari}, \citenamefont {Di~Mare},\ and\
  \citenamefont {Marschall}}]{sadowski2023large}%
  \BibitemOpen
  \bibfield  {author} {\bibinfo {author} {\bibfnamefont {W.}~\bibnamefont
  {Sadowski}}, \bibinfo {author} {\bibfnamefont {M.}~\bibnamefont {Sayyari}},
  \bibinfo {author} {\bibfnamefont {F.}~\bibnamefont {Di~Mare}},\ and\ \bibinfo
  {author} {\bibfnamefont {H.}~\bibnamefont {Marschall}},\ }\bibfield  {title}
  {\enquote {\bibinfo {title} {Large eddy simulation of flow in porous media:
  Analysis of the commutation error of the double-averaged equations},}\ }\href
  {https://doi.org/10.1063/5.0148130} {\bibfield  {journal} {\bibinfo
  {journal} {Physics of Fluids}\ }\textbf {\bibinfo {volume} {35}} (\bibinfo
  {year} {2023}),\ 10.1063/5.0148130}\BibitemShut {NoStop}%
\bibitem [{\citenamefont {Zhou}, \citenamefont {Walther},\ and\ \citenamefont
  {Fernandez-Pello}(2002)}]{zhou2002}%
  \BibitemOpen
  \bibfield  {author} {\bibinfo {author} {\bibfnamefont {Y.}~\bibnamefont
  {Zhou}}, \bibinfo {author} {\bibfnamefont {D.}~\bibnamefont {Walther}},\ and\
  \bibinfo {author} {\bibfnamefont {A.}~\bibnamefont {Fernandez-Pello}},\
  }\bibfield  {title} {\enquote {\bibinfo {title} {Numerical analysis of
  piloted ignition of polymeric materials},}\ }\href
  {https://doi.org/10.1016/S0010-2180(02)00396-6} {\bibfield  {journal}
  {\bibinfo  {journal} {Combustion and Flame}\ }\textbf {\bibinfo {volume}
  {131}},\ \bibinfo {pages} {147--158} (\bibinfo {year} {2002})}\BibitemShut
  {NoStop}%
\bibitem [{\citenamefont {Jiang}\ and\ \citenamefont {Luo}(2000)}]{jiang2000}%
  \BibitemOpen
  \bibfield  {author} {\bibinfo {author} {\bibfnamefont {X.}~\bibnamefont
  {Jiang}}\ and\ \bibinfo {author} {\bibfnamefont {K.}~\bibnamefont {Luo}},\
  }\bibfield  {title} {\enquote {\bibinfo {title} {Combustion-induced buoyancy
  effects of an axisymmetric reactive plume},}\ }\href
  {https://doi.org/10.1016/S0082-0784(00)80605-0} {\bibfield  {journal}
  {\bibinfo  {journal} {Proceedings of the Combustion Institute}\ }\textbf
  {\bibinfo {volume} {28}},\ \bibinfo {pages} {1989--1995} (\bibinfo {year}
  {2000})}\BibitemShut {NoStop}%
\bibitem [{\citenamefont {Meehan}, \citenamefont {Wimer},\ and\ \citenamefont
  {Hamlington}(2022)}]{meehan2022}%
  \BibitemOpen
  \bibfield  {author} {\bibinfo {author} {\bibfnamefont {M.~A.}\ \bibnamefont
  {Meehan}}, \bibinfo {author} {\bibfnamefont {N.~T.}\ \bibnamefont {Wimer}},\
  and\ \bibinfo {author} {\bibfnamefont {P.~E.}\ \bibnamefont {Hamlington}},\
  }\bibfield  {title} {\enquote {\bibinfo {title} {Richardson and {Reynolds}
  number effects on the near field of buoyant plumes: temporal variability and
  puffing},}\ }\href {https://doi.org/10.1017/jfm.2022.788} {\bibfield
  {journal} {\bibinfo  {journal} {Journal of Fluid Mechanics}\ }\textbf
  {\bibinfo {volume} {950}},\ \bibinfo {pages} {A24} (\bibinfo {year}
  {2022})}\BibitemShut {NoStop}%
\bibitem [{\citenamefont {Bres}\ and\ \citenamefont
  {Colonius}(2008)}]{bres2008three}%
  \BibitemOpen
  \bibfield  {author} {\bibinfo {author} {\bibfnamefont {G.~A.}\ \bibnamefont
  {Bres}}\ and\ \bibinfo {author} {\bibfnamefont {T.}~\bibnamefont
  {Colonius}},\ }\bibfield  {title} {\enquote {\bibinfo {title}
  {Three-dimensional instabilities in compressible flow over open cavities},}\
  }\href {https://doi.org/10.1017/S0022112007009925} {\bibfield  {journal}
  {\bibinfo  {journal} {Journal of Fluid Mechanics}\ }\textbf {\bibinfo
  {volume} {599}},\ \bibinfo {pages} {309--339} (\bibinfo {year}
  {2008})}\BibitemShut {NoStop}%
\bibitem [{\citenamefont {Banerjee}\ \emph {et~al.}(2020)\citenamefont
  {Banerjee}, \citenamefont {Heilman}, \citenamefont {Goodrick}, \citenamefont
  {Hiers},\ and\ \citenamefont {Linn}}]{banerjee2020effects}%
  \BibitemOpen
  \bibfield  {author} {\bibinfo {author} {\bibfnamefont {T.}~\bibnamefont
  {Banerjee}}, \bibinfo {author} {\bibfnamefont {W.}~\bibnamefont {Heilman}},
  \bibinfo {author} {\bibfnamefont {S.}~\bibnamefont {Goodrick}}, \bibinfo
  {author} {\bibfnamefont {J.~K.}\ \bibnamefont {Hiers}},\ and\ \bibinfo
  {author} {\bibfnamefont {R.}~\bibnamefont {Linn}},\ }\bibfield  {title}
  {\enquote {\bibinfo {title} {Effects of canopy midstory management and fuel
  moisture on wildfire behavior},}\ }\href
  {https://doi.org/10.1038/s41598-020-74338-9} {\bibfield  {journal} {\bibinfo
  {journal} {Scientific Reports}\ }\textbf {\bibinfo {volume} {10}},\ \bibinfo
  {pages} {17312} (\bibinfo {year} {2020})}\BibitemShut {NoStop}%
\bibitem [{\citenamefont {Li}\ and\ \citenamefont
  {Giometto}(2024)}]{li2024structure}%
  \BibitemOpen
  \bibfield  {author} {\bibinfo {author} {\bibfnamefont {W.}~\bibnamefont
  {Li}}\ and\ \bibinfo {author} {\bibfnamefont {M.~G.}\ \bibnamefont
  {Giometto}},\ }\bibfield  {title} {\enquote {\bibinfo {title} {The structure
  of turbulence in unsteady flow over urban canopies},}\ }\href
  {https://doi.org/10.1017/jfm.2023.974} {\bibfield  {journal} {\bibinfo
  {journal} {Journal of Fluid Mechanics}\ }\textbf {\bibinfo {volume} {985}},\
  \bibinfo {pages} {A5} (\bibinfo {year} {2024})}\BibitemShut {NoStop}%
\bibitem [{\citenamefont {{\.Z}uk}\ \emph {et~al.}(2022)\citenamefont
  {{\.Z}uk}, \citenamefont {Tu{\.z}nik}, \citenamefont {Rymarz}, \citenamefont
  {Kwiatkowski}, \citenamefont {Dudy{\'n}ski}, \citenamefont {Galeazzo},\ and\
  \citenamefont {Krieger~Filho}}]{zuk2022}%
  \BibitemOpen
  \bibfield  {author} {\bibinfo {author} {\bibfnamefont {P.~J.}\ \bibnamefont
  {{\.Z}uk}}, \bibinfo {author} {\bibfnamefont {B.}~\bibnamefont {Tu{\.z}nik}},
  \bibinfo {author} {\bibfnamefont {T.}~\bibnamefont {Rymarz}}, \bibinfo
  {author} {\bibfnamefont {K.}~\bibnamefont {Kwiatkowski}}, \bibinfo {author}
  {\bibfnamefont {M.}~\bibnamefont {Dudy{\'n}ski}}, \bibinfo {author}
  {\bibfnamefont {F.~C.}\ \bibnamefont {Galeazzo}},\ and\ \bibinfo {author}
  {\bibfnamefont {G.~C.}\ \bibnamefont {Krieger~Filho}},\ }\bibfield  {title}
  {\enquote {\bibinfo {title} {{OpenFOAM} solver for thermal and chemical
  conversion in porous media},}\ }\href
  {https://doi.org/10.1016/j.cpc.2022.108407} {\bibfield  {journal} {\bibinfo
  {journal} {Computer Physics Communications}\ }\textbf {\bibinfo {volume}
  {278}},\ \bibinfo {pages} {108407} (\bibinfo {year} {2022})}\BibitemShut
  {NoStop}%
\bibitem [{\citenamefont {Kaviany}(2012)}]{kaviany2012}%
  \BibitemOpen
  \bibfield  {author} {\bibinfo {author} {\bibfnamefont {M.}~\bibnamefont
  {Kaviany}},\ }\href@noop {} {\emph {\bibinfo {title} {Principles of Heat
  Transfer in Porous Media}}}\ (\bibinfo  {publisher} {Springer Science \&
  Business Media},\ \bibinfo {year} {2012})\BibitemShut {NoStop}%
\bibitem [{\citenamefont {Le~Bars}\ and\ \citenamefont
  {Worster}(2006)}]{le2006}%
  \BibitemOpen
  \bibfield  {author} {\bibinfo {author} {\bibfnamefont {M.}~\bibnamefont
  {Le~Bars}}\ and\ \bibinfo {author} {\bibfnamefont {M.~G.}\ \bibnamefont
  {Worster}},\ }\bibfield  {title} {\enquote {\bibinfo {title} {Interfacial
  conditions between a pure fluid and a porous medium: implications for binary
  alloy solidification},}\ }\href {https://doi.org/10.1017/S0022112005007998}
  {\bibfield  {journal} {\bibinfo  {journal} {Journal of Fluid Mechanics}\
  }\textbf {\bibinfo {volume} {550}},\ \bibinfo {pages} {149--173} (\bibinfo
  {year} {2006})}\BibitemShut {NoStop}%
\bibitem [{\citenamefont {Sutherland}(1893)}]{sutherland1893}%
  \BibitemOpen
  \bibfield  {author} {\bibinfo {author} {\bibfnamefont {W.}~\bibnamefont
  {Sutherland}},\ }\bibfield  {title} {\enquote {\bibinfo {title} {{LII.} {The}
  viscosity of gases and molecular force},}\ }\href
  {https://doi.org/10.1080/14786449308620508} {\bibfield  {journal} {\bibinfo
  {journal} {The London, Edinburgh, and Dublin Philosophical Magazine and
  Journal of Science}\ }\textbf {\bibinfo {volume} {36}},\ \bibinfo {pages}
  {507--531} (\bibinfo {year} {1893})}\BibitemShut {NoStop}%
\bibitem [{\citenamefont {Breugem}, \citenamefont {Boersma},\ and\
  \citenamefont {Uittenbogaard}(2006)}]{breugem2006influence}%
  \BibitemOpen
  \bibfield  {author} {\bibinfo {author} {\bibfnamefont {W.-P.}\ \bibnamefont
  {Breugem}}, \bibinfo {author} {\bibfnamefont {B.-J.}\ \bibnamefont
  {Boersma}},\ and\ \bibinfo {author} {\bibfnamefont {R.~E.}\ \bibnamefont
  {Uittenbogaard}},\ }\bibfield  {title} {\enquote {\bibinfo {title} {The
  influence of wall permeability on turbulent channel flow},}\ }\href
  {https://doi.org/10.1017/S0022112006000887} {\bibfield  {journal} {\bibinfo
  {journal} {Journal of Fluid Mechanics}\ }\textbf {\bibinfo {volume} {562}},\
  \bibinfo {pages} {35--72} (\bibinfo {year} {2006})}\BibitemShut {NoStop}%
\bibitem [{\citenamefont {Wood}, \citenamefont {He},\ and\ \citenamefont
  {Apte}(2020)}]{wood2020}%
  \BibitemOpen
  \bibfield  {author} {\bibinfo {author} {\bibfnamefont {B.~D.}\ \bibnamefont
  {Wood}}, \bibinfo {author} {\bibfnamefont {X.}~\bibnamefont {He}},\ and\
  \bibinfo {author} {\bibfnamefont {S.~V.}\ \bibnamefont {Apte}},\ }\bibfield
  {title} {\enquote {\bibinfo {title} {Modeling turbulent flows in porous
  media},}\ }\href {https://doi.org/10.1146/annurev-fluid-010719-060317}
  {\bibfield  {journal} {\bibinfo  {journal} {Annual Review of Fluid
  Mechanics}\ }\textbf {\bibinfo {volume} {52}},\ \bibinfo {pages} {171--203}
  (\bibinfo {year} {2020})}\BibitemShut {NoStop}%
\bibitem [{\citenamefont {Yoshizawa}(1986)}]{yoshizawa1986}%
  \BibitemOpen
  \bibfield  {author} {\bibinfo {author} {\bibfnamefont {A.}~\bibnamefont
  {Yoshizawa}},\ }\bibfield  {title} {\enquote {\bibinfo {title} {Statistical
  theory for compressible turbulent shear flows, with the application to
  subgrid modeling},}\ }\href {https://doi.org/10.1063/1.865552} {\bibfield
  {journal} {\bibinfo  {journal} {Physics of Fluids}\ }\textbf {\bibinfo
  {volume} {29}},\ \bibinfo {pages} {2152--2164} (\bibinfo {year}
  {1986})}\BibitemShut {NoStop}%
\bibitem [{\citenamefont {Harten}(1997)}]{harten1997high}%
  \BibitemOpen
  \bibfield  {author} {\bibinfo {author} {\bibfnamefont {A.}~\bibnamefont
  {Harten}},\ }\bibfield  {title} {\enquote {\bibinfo {title} {High resolution
  schemes for hyperbolic conservation laws},}\ }\href
  {https://doi.org/10.1006/jcph.1997.5713} {\bibfield  {journal} {\bibinfo
  {journal} {Journal of Computational Physics}\ }\textbf {\bibinfo {volume}
  {135}},\ \bibinfo {pages} {260--278} (\bibinfo {year} {1997})}\BibitemShut
  {NoStop}%
\bibitem [{\citenamefont {{OpenFOAM Foundation}}(2020)}]{openfoam}%
  \BibitemOpen
  \bibfield  {author} {\bibinfo {author} {\bibnamefont {{OpenFOAM
  Foundation}}},\ }\href {https://openfoam.org/} {\enquote {\bibinfo {title}
  {{OpenFOAM}},}\ }\bibinfo {howpublished} {\url{https://openfoam.org/}}
  (\bibinfo {year} {2020}),\ \bibinfo {note} {version 8}\BibitemShut {NoStop}%
\bibitem [{\citenamefont {Hairer}\ and\ \citenamefont
  {Wanner}(1996)}]{hairer1991ii}%
  \BibitemOpen
  \bibfield  {author} {\bibinfo {author} {\bibfnamefont {E.}~\bibnamefont
  {Hairer}}\ and\ \bibinfo {author} {\bibfnamefont {G.}~\bibnamefont
  {Wanner}},\ }\href {https://doi.org/10.1007/978-3-642-05221-7} {\emph
  {\bibinfo {title} {Solving Ordinary Differential Equations {II}: Stiff and
  Differential-Algebraic Problems}}},\ Springer Series in Computational
  Mathematics\ (\bibinfo  {publisher} {Springer Berlin Heidelberg},\ \bibinfo
  {year} {1996})\BibitemShut {NoStop}%
\bibitem [{\citenamefont {Sandu}\ \emph {et~al.}(1997)\citenamefont {Sandu},
  \citenamefont {Verwer}, \citenamefont {Blom}, \citenamefont {Spee},
  \citenamefont {Carmichael},\ and\ \citenamefont
  {Potra}}]{sandu1997benchmarking}%
  \BibitemOpen
  \bibfield  {author} {\bibinfo {author} {\bibfnamefont {A.}~\bibnamefont
  {Sandu}}, \bibinfo {author} {\bibfnamefont {J.}~\bibnamefont {Verwer}},
  \bibinfo {author} {\bibfnamefont {J.}~\bibnamefont {Blom}}, \bibinfo {author}
  {\bibfnamefont {E.}~\bibnamefont {Spee}}, \bibinfo {author} {\bibfnamefont
  {G.}~\bibnamefont {Carmichael}},\ and\ \bibinfo {author} {\bibfnamefont
  {F.}~\bibnamefont {Potra}},\ }\bibfield  {title} {\enquote {\bibinfo {title}
  {Benchmarking stiff {ODE} solvers for atmospheric chemistry problems {II}:
  {Rosenbrock} solvers},}\ }\href
  {https://doi.org/10.1016/S1352-2310(97)83212-8} {\bibfield  {journal}
  {\bibinfo  {journal} {Atmospheric Environment}\ }\textbf {\bibinfo {volume}
  {31}},\ \bibinfo {pages} {3459--3472} (\bibinfo {year} {1997})}\BibitemShut
  {NoStop}%
\bibitem [{\citenamefont {Finnigan}(2000)}]{finnigan2000}%
  \BibitemOpen
  \bibfield  {author} {\bibinfo {author} {\bibfnamefont {J.}~\bibnamefont
  {Finnigan}},\ }\bibfield  {title} {\enquote {\bibinfo {title} {Turbulence in
  plant canopies},}\ }\href {https://doi.org/10.1146/annurev.fluid.32.1.519}
  {\bibfield  {journal} {\bibinfo  {journal} {Annual Review of Fluid
  Mechanics}\ }\textbf {\bibinfo {volume} {32}},\ \bibinfo {pages} {519--571}
  (\bibinfo {year} {2000})}\BibitemShut {NoStop}%
\bibitem [{\citenamefont {Kuo}\ and\ \citenamefont {Hsi}(2005)}]{sphere}%
  \BibitemOpen
  \bibfield  {author} {\bibinfo {author} {\bibfnamefont {J.~T.}\ \bibnamefont
  {Kuo}}\ and\ \bibinfo {author} {\bibfnamefont {C.-L.}\ \bibnamefont {Hsi}},\
  }\bibfield  {title} {\enquote {\bibinfo {title} {Pyrolysis and ignition of
  single wooden spheres heated in high-temperature streams of air},}\ }\href
  {https://doi.org/10.1016/j.combustflame.2005.04.002} {\bibfield  {journal}
  {\bibinfo  {journal} {Combustion and Flame}\ }\textbf {\bibinfo {volume}
  {142}},\ \bibinfo {pages} {401--412} (\bibinfo {year} {2005})}\BibitemShut
  {NoStop}%
\bibitem [{\citenamefont {Debiagi}\ \emph {et~al.}(2015)\citenamefont
  {Debiagi}, \citenamefont {Pecchi}, \citenamefont {Gentile}, \citenamefont
  {Frassoldati}, \citenamefont {Cuoci}, \citenamefont {Faravelli},\ and\
  \citenamefont {Ranzi}}]{debiagi}%
  \BibitemOpen
  \bibfield  {author} {\bibinfo {author} {\bibfnamefont {P.~E.~A.}\
  \bibnamefont {Debiagi}}, \bibinfo {author} {\bibfnamefont {C.}~\bibnamefont
  {Pecchi}}, \bibinfo {author} {\bibfnamefont {G.}~\bibnamefont {Gentile}},
  \bibinfo {author} {\bibfnamefont {A.}~\bibnamefont {Frassoldati}}, \bibinfo
  {author} {\bibfnamefont {A.}~\bibnamefont {Cuoci}}, \bibinfo {author}
  {\bibfnamefont {T.}~\bibnamefont {Faravelli}},\ and\ \bibinfo {author}
  {\bibfnamefont {E.}~\bibnamefont {Ranzi}},\ }\bibfield  {title} {\enquote
  {\bibinfo {title} {Extractives extend the applicability of multistep kinetic
  scheme of biomass pyrolysis},}\ }\href
  {https://doi.org/10.1021/acs.energyfuels.5b01753} {\bibfield  {journal}
  {\bibinfo  {journal} {Energy \& Fuels}\ }\textbf {\bibinfo {volume} {29}},\
  \bibinfo {pages} {6544--6555} (\bibinfo {year} {2015})}\BibitemShut {NoStop}%
\bibitem [{\citenamefont {Smith}\ \emph {et~al.}(2023)\citenamefont {Smith},
  \citenamefont {Golden}, \citenamefont {Frenklach}, \citenamefont {Moriarty},
  \citenamefont {Eiteneer}, \citenamefont {Goldenberg}, \citenamefont {Bowman},
  \citenamefont {Hanson}, \citenamefont {Song}, \citenamefont {Gardiner~Jr}
  \emph {et~al.}}]{gri}%
  \BibitemOpen
  \bibfield  {author} {\bibinfo {author} {\bibfnamefont {G.}~\bibnamefont
  {Smith}}, \bibinfo {author} {\bibfnamefont {D.}~\bibnamefont {Golden}},
  \bibinfo {author} {\bibfnamefont {M.}~\bibnamefont {Frenklach}}, \bibinfo
  {author} {\bibfnamefont {N.}~\bibnamefont {Moriarty}}, \bibinfo {author}
  {\bibfnamefont {B.}~\bibnamefont {Eiteneer}}, \bibinfo {author}
  {\bibfnamefont {M.}~\bibnamefont {Goldenberg}}, \bibinfo {author}
  {\bibfnamefont {C.}~\bibnamefont {Bowman}}, \bibinfo {author} {\bibfnamefont
  {R.}~\bibnamefont {Hanson}}, \bibinfo {author} {\bibfnamefont
  {S.}~\bibnamefont {Song}}, \bibinfo {author} {\bibfnamefont {W.}~\bibnamefont
  {Gardiner~Jr}}, \emph {et~al.},\ }\href@noop {} {\enquote {\bibinfo {title}
  {{GRI-Mech 3.0}},}\ }\bibinfo {howpublished}
  {\url{http://combustion.berkeley.edu/gri-mech/}} (\bibinfo {year} {1999;
  accessed Dec. 2023})\BibitemShut {NoStop}%
\bibitem [{\citenamefont {Fillo}, \citenamefont {Hamlington},\ and\
  \citenamefont {Niemeyer}(2022)}]{fillo2022assessing}%
  \BibitemOpen
  \bibfield  {author} {\bibinfo {author} {\bibfnamefont {A.~J.}\ \bibnamefont
  {Fillo}}, \bibinfo {author} {\bibfnamefont {P.~E.}\ \bibnamefont
  {Hamlington}},\ and\ \bibinfo {author} {\bibfnamefont {K.~E.}\ \bibnamefont
  {Niemeyer}},\ }\bibfield  {title} {\enquote {\bibinfo {title} {Assessing
  diffusion model impacts on enstrophy and flame structure in turbulent lean
  premixed flames},}\ }\href {https://doi.org/10.1080/13647830.2022.2049882}
  {\bibfield  {journal} {\bibinfo  {journal} {Combustion Theory and Modelling}\
  }\textbf {\bibinfo {volume} {26}},\ \bibinfo {pages} {712--727} (\bibinfo
  {year} {2022})}\BibitemShut {NoStop}%
\bibitem [{\citenamefont {Cetegen}(1997)}]{cetegen1997measurements}%
  \BibitemOpen
  \bibfield  {author} {\bibinfo {author} {\bibfnamefont {B.~M.}\ \bibnamefont
  {Cetegen}},\ }\bibfield  {title} {\enquote {\bibinfo {title} {Measurements of
  instantaneous velocity field of a non-reacting pulsating buoyant plume by
  particle image velocimetry},}\ }\href
  {https://doi.org/10.1080/00102209708935636} {\bibfield  {journal} {\bibinfo
  {journal} {Combustion Science and Technology}\ }\textbf {\bibinfo {volume}
  {123}},\ \bibinfo {pages} {377--387} (\bibinfo {year} {1997})}\BibitemShut
  {NoStop}%
\bibitem [{\citenamefont {Cetegen}\ and\ \citenamefont
  {Dong}(2000)}]{cetegen2000experiments}%
  \BibitemOpen
  \bibfield  {author} {\bibinfo {author} {\bibfnamefont {B.}~\bibnamefont
  {Cetegen}}\ and\ \bibinfo {author} {\bibfnamefont {Y.}~\bibnamefont {Dong}},\
  }\bibfield  {title} {\enquote {\bibinfo {title} {Experiments on the
  instability modes of buoyant diffusion flames and effects of ambient
  atmosphere on the instabilities},}\ }\href
  {https://doi.org/10.1007/s003480050415} {\bibfield  {journal} {\bibinfo
  {journal} {Experiments in Fluids}\ }\textbf {\bibinfo {volume} {28}},\
  \bibinfo {pages} {546--558} (\bibinfo {year} {2000})}\BibitemShut {NoStop}%
\bibitem [{\citenamefont {Zhang}, \citenamefont {Duan},\ and\ \citenamefont
  {Choudhari}(2018)}]{zhang2018direct}%
  \BibitemOpen
  \bibfield  {author} {\bibinfo {author} {\bibfnamefont {C.}~\bibnamefont
  {Zhang}}, \bibinfo {author} {\bibfnamefont {L.}~\bibnamefont {Duan}},\ and\
  \bibinfo {author} {\bibfnamefont {M.~M.}\ \bibnamefont {Choudhari}},\
  }\bibfield  {title} {\enquote {\bibinfo {title} {Direct numerical simulation
  database for supersonic and hypersonic turbulent boundary layers},}\ }\href
  {https://doi.org/10.2514/1.J057296} {\bibfield  {journal} {\bibinfo
  {journal} {AIAA Journal}\ }\textbf {\bibinfo {volume} {56}},\ \bibinfo
  {pages} {4297--4311} (\bibinfo {year} {2018})}\BibitemShut {NoStop}%
\bibitem [{\citenamefont {Zhang}\ \emph {et~al.}(2014)\citenamefont {Zhang},
  \citenamefont {Bi}, \citenamefont {Hussain},\ and\ \citenamefont
  {She}}]{zhang2014generalized}%
  \BibitemOpen
  \bibfield  {author} {\bibinfo {author} {\bibfnamefont {Y.-S.}\ \bibnamefont
  {Zhang}}, \bibinfo {author} {\bibfnamefont {W.-T.}\ \bibnamefont {Bi}},
  \bibinfo {author} {\bibfnamefont {F.}~\bibnamefont {Hussain}},\ and\ \bibinfo
  {author} {\bibfnamefont {Z.-S.}\ \bibnamefont {She}},\ }\bibfield  {title}
  {\enquote {\bibinfo {title} {A generalized {Reynolds} analogy for
  compressible wall-bounded turbulent flows},}\ }\href
  {https://doi.org/10.1017/jfm.2013.620} {\bibfield  {journal} {\bibinfo
  {journal} {Journal of Fluid Mechanics}\ }\textbf {\bibinfo {volume} {739}},\
  \bibinfo {pages} {392--420} (\bibinfo {year} {2014})}\BibitemShut {NoStop}%
\bibitem [{\citenamefont {Huang}, \citenamefont {Coleman},\ and\ \citenamefont
  {Bradshaw}(1995)}]{huang1995compressible}%
  \BibitemOpen
  \bibfield  {author} {\bibinfo {author} {\bibfnamefont {P.~G.}\ \bibnamefont
  {Huang}}, \bibinfo {author} {\bibfnamefont {G.~N.}\ \bibnamefont {Coleman}},\
  and\ \bibinfo {author} {\bibfnamefont {P.}~\bibnamefont {Bradshaw}},\
  }\bibfield  {title} {\enquote {\bibinfo {title} {Compressible turbulent
  channel flows: {DNS} results and modelling},}\ }\href
  {https://doi.org/10.1017/S0022112095004599} {\bibfield  {journal} {\bibinfo
  {journal} {Journal of Fluid Mechanics}\ }\textbf {\bibinfo {volume} {305}},\
  \bibinfo {pages} {185--218} (\bibinfo {year} {1995})}\BibitemShut {NoStop}%
\bibitem [{\citenamefont {Behnoudfar}(2024)}]{software}%
  \BibitemOpen
  \bibfield  {author} {\bibinfo {author} {\bibfnamefont {D.}~\bibnamefont
  {Behnoudfar}},\ }\href {https://doi.org/10.5281/zenodo.14166205} {\enquote
  {\bibinfo {title} {{porousGasificationFoam} v2024.11.14},}\ }\bibinfo
  {howpublished} {\url{https://doi.org/10.5281/zenodo.14166205},
  \url{https://github.com/Niemeyer-Research-Group/porousGasificationFoam}}
  (\bibinfo {year} {2024})\BibitemShut {NoStop}%
\bibitem [{\citenamefont {Modest}(2013)}]{modest2013radiative}%
  \BibitemOpen
  \bibfield  {author} {\bibinfo {author} {\bibfnamefont {M.~F.}\ \bibnamefont
  {Modest}},\ }\href@noop {} {\emph {\bibinfo {title} {Radiative Heat
  Transfer}}}\ (\bibinfo  {publisher} {Academic Press},\ \bibinfo {year}
  {2013})\BibitemShut {NoStop}%
\end{thebibliography}%

\end{document}